%% file: master8d.tex
\documentclass[twocolumn,showpacs,preprintnumbers,amsmath,amssymb]{revtex4}
\usepackage{graphicx}% Include figure files
\usepackage{dcolumn}% Align table columns on decimal point
\usepackage{bm}% bold math
\usepackage{mathrsfs}
\begin{document}

% Use /ltx/revtex4/sample/apssamp.tex as a template
%\preprint{APS/123-QED}

\title{Machian space quanta}

\author{David W. Essex}
 \email{D.W.Essex@damtp.cam.ac.uk}
 \affiliation{Clare College, Trinity Lane, Cambridge, CB2 1TL, England}
% \affiliation{Centre for Mathematical Sciences, Wilberforce Road, Cambridge, CB3 0WA, England}
% \homepage{http://www.homepage}
\date{\today}

\begin{abstract}
  A new model for space and matter is obtained by joining every pair of point charges in the 
observable universe by an ethereal string. Positive gravitational potential energy in each string gives an attractive 
gravitational force due to the action of an energy conservation constraint.
% The kinetic energy of a massive particle is part of the energy in the Machian strings connecting it to distant matter.
  Newton's laws of motion are derived and inertia is explained in accordance with Mach's principle. The Machian string 
model gives a surprisingly simple way to understand the expansion history of the Universe. The decelerating expansion in 
the radiation era and the matter era is explained without using General Relativity and the transition from deceleration 
to acceleration is explained without the need to introduce a separate \lq dark energy' component. The interaction 
between Machian strings gives a physical model for modified Newtonian dynamics (MOND) and is therefore an alternative 
to \lq dark matter'.

\vspace*{0.2cm}

\footnotesize

 \lq\lq \,I consider it entirely possible that physics cannot be based upon the field concept, that is on continuous structures. Then nothing will remain of my whole castle in the air, including the theory of gravitation, but also nothing of the rest of contemporary physics." (Albert Einstein)

% Valid PACS numbers may be entered using the \verb+\pacs{#1}+ command.
\end{abstract}
%\pacs{Valid PACS appear here}% PACS, the Physics and Astronomy
                              % Classification Scheme.
\maketitle

\section{Motivation}

 The purpose of the present paper is to suggest a new model for the structure of elementary particles and a new model for
 space and to offer a new approach to the problems of \lq dark matter' and \lq dark energy'.

One of the basic properties of elementary particles is that they have spin one-half. Quantum mechanically, a spin one-half particle ends up in a different state when rotated through an angle of $2\pi$ and only returns to its original state when rotated through an angle of $4\pi$. There is no analogue of such behaviour for a classical point particle but a mechanical model for spin one-half can be constructed when the particle is connected to its surroundings, for example by someone holding onto it or by strings connecting it to a rigid external frame~\cite{mtw,bpr}. Suppose that the strings are initially not entangled. A rotation of the particle through an angle of $4\pi$ results in a configuration that is equivalent to the initial configuration, in the sense that the strings can be disentangled, but a rotation through an angle of $2\pi$ results in a configuration that is not equivalent to the initial configuration.

%{\it The inferred mass distribution depends on the resulting force between two objects and also on the force law. It is assumed that the force between two point masses or two point charges is $1/r^2$, so when a $1/r^2$ force law is discovered between two masses or charges, the mass and charge is inferred to be pointlike, i.e. it's a circular argument. Are we not free to alter the mass or charge distribution arbitrarily provided we make an appropriate change in the force law\,? This might be worth a separate investigation on its own in the context of conventional electrodynamics or Newtonian gravity. The idea that elementary particles are pointlike is consistent with the fact that they appear to be very small, but this might be a circular argument again. An arbitrary mass distribution may be written as a superposition of point masses, where a point mass is defined as an object whose mass density is a delta function. The force law is, by definition, the force between two delta function mass distributions as a function of their separation. Whether or not an elementary particle is accurately represented by a delta function mass distribution is logically a separate question.}

In searching for a new model for space it is natural to reconsider Mach's principle, which was never fully
incorporated into General Relativity~\cite{mach1,sciama,brans}. Instead of defining inertial frames in terms of a local
 absolute space, Mach argued that the inertial forces experienced by an accelerated particle are due to some sort of 
interaction with the distant matter in the Universe. Perhaps a physical model for space can be constructed by 
defining space itself in terms of distant matter~\cite{mach2}.

% Since Mach's principle implies that every particle is in instantaneous contact with
% every other.
% The simplest possible model for space and matter consistent with Mach's principle is
% obtained by connecting every pair of elementary particles by a straight line and
% defining the physical space of the universe to be the set of points along all the
% interconnecting lines. Since every particle then has its own \lq particle space'
% defined by the set of points in all the lines connected to it, as illustrated in
% Figure~\ref{two}, the model may be referred to as the particle space model. All
% physical fields are confined to the network of interconnecting lines associated with
% physical space. In the particle space model, space is defined by the relative
% positions of all the particles in the universe and cannot exist without matter.

\begin{figure}[h]
\vspace*{-0.2cm}
\includegraphics[height=5cm,width=7cm]{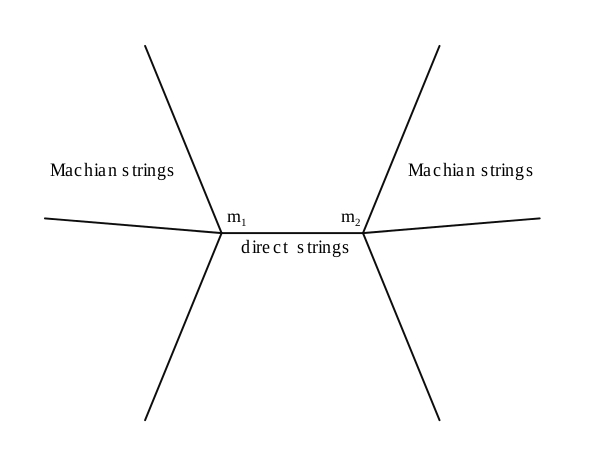}
\vspace*{-0.4cm} \caption{\label{strings} Schematic diagram of the direct strings and the Machian strings in the Machian
space quanta of two masses $m_1$ and $m_2$. The direct strings join the centres of the space quanta associated with the two
masses and the Machian strings connect the centres with the centres of all the other space quanta in the observable
universe.}
\end{figure}

In the model of Machian space quanta, the mechanical model for spin one-half and Mach's principle are incorporated by postulating that an elementary particle consists of a set of ethereal strings connecting a charged centre to the centre of every other elementary particle in the observable universe. The set of strings associated with a given elementary particle can be imagined as a quantum of space with the size of the observable universe and will be referred to as a Machian space quantum. The strings contain energy, and hence mass, but are not charged. The strings associated with the space quanta of two nearby particles of mass $m_1$ and $m_2$ are illustrated in Figure~\ref{strings}. The strings connecting two such masses will be referred to as direct strings and strings connected to distant matter will be referred to as Machian strings.

 The idea that an elementary particle could have the size of the observable universe appears, at first sight, to be completely contrary to experience. Indeed, high energy scattering experiments have failed to find any spatial extent at all in the charge distribution of either quarks or electrons. However, experiments cannot probe an entire space quantum because the only part of a space quantum that can interact directly with another space quantum is the centre. The only conclusion that may therefore be drawn from scattering experiments is that the charge distribution at the centre of a space quantum is pointlike.
% It is not possible to infer the mass distribution at the centre of an individual space quantum from experimental tests of the gravitational interaction due to the very large number of elementary particles involved.

 The gravitational interaction between the two masses $m_1$ and $m_2$ illustrated in Figure~\ref{strings} is defined by specifying the total energy, $Gm_1m_2/r$, in the direct strings joining them, where $r$ is the distance between the centres of the two masses. To a first approximation, the gravitational masses $m_1$ and $m_2$ are simply the total masses in the associated space quanta and it makes no difference at all whether all the mass of a space quantum is at the centre or whether the mass is distributed throughout the strings. For the further development of the theory, however, the fact that the mass is distributed throughout the strings is very significant.
%In Section~\ref{mmg}, a model of Newtonian gravity is obtained by postulating that the direct strings joining the centres of two particles of mass $m_1$ and $m_2$, separated by a distance $r$, have an energy $Gm_1m_2/r$. Having specified the interaction energy, it makes no difference whatsoever whether the masses are deemed to be at a point or whether they are
% deemed to be spread throughout their Machian strings. For the further development of the theory, however, the difference is enormous. Apart from the conceptual difference,
 The fundamental variables in the problem are the strings themselves, rather than positions and velocities of the particle centres. The energies in the strings may be calculated explicitly, including the dependence on the velocities of the particles, and it is then possible to derive the properties of particle inertia and kinetic energy from first principles instead of putting them in by hand. When the contribution to the energies of the strings from the expansion of the Universe is taken into account, the string model is found to give an accelerating expansion of the Universe without the need to introduce any mysterious \lq dark energy'.

It might be thought, at first sight, that Machian strings give a very simple solution to the problem of \lq dark matter'. It is well known that flat galaxy rotation curves can be accounted for by assuming that galaxies lie within a dark matter halo whose density at a distance $r$ from the centre of the galaxy is proportional to $1/r^2$~\cite{binney}. For strings with uniform mass per unit length, the mass in the strings enclosed within a spherical shell of thickness $dr$ at any radius is proportional to $dr$ and it follows that the density in the strings is proportional to $1/r^2$, which appears to be exactly what is required. This observation was, in fact, the original motivation for the Machian string model. Unfortunately the idea doesn't work because the only gravitational interaction between space quanta is the interaction due to the energies in their common direct strings. Unlike a conventional mass distribution, the mass elements within a Machian string are invisible to other space quanta and do not, therefore, act as additional point sources for the gravitational field. In any case, such a model would give an additional gravitational acceleration proportional to $M/r$, whereas the additional acceleration required to account for galaxy rotation curves is proportional to $\sqrt M/r$. A different model for \lq dark matter', based on a contact interaction between the strings of different space quanta, is introduced in Section~\ref{dm}.

%Machian space quanta provide a physical representation for the wavefunction of Quantum Mechanics and Schr\"{o}dinger's
%$3N$-\,dimensional configuration space may just be a superposition of $N$ space quanta.

%After explaining why we adopt a model for a massive particle in which the mass is extended but the charge is not, it remains to consider a photon because a new model for gravity must deal with the electromagnetic interaction as well as the
%gravitational interaction.

%{\it What does our choice to keep the charge at a point tell us about the photon\,? There's obviously a link here because the photon mediates the electromagnetic interaction between elementary charges and the properties of a
%photon therefore depend on their properties. If an elementary charge was extended rather than pointlike, what would the
%effect be on the Maxwell equations\,? Recall that Feynman discussed an extended charge theory of Bopp which was motivated
%by the desire to make the electromagnetic interaction finite, at least at the classical level. But the extended charge distributions in this context are real charge distributions in the sense that all points within in act as centres of Coulomb attraction.}

Whereas the quantisation of matter is automatically present at the classical level a photon is, by definition, a quantum mechanical object. The strange quantum mechanical behaviour of the photon was described by Eddington in 1928 as follows.
\begin{quotation}\small\noindent
The pursuit of the quantum leads to many surprises; but probably none is more outrageous to our preconceptions than the
regathering of light and other radiant energy into $h$-units, when all the classical pictures show it to be dispersing more and more. Consider the light waves which are emitted as a result of a single emission on the star Sirius. These bear away
a certain amount of energy endowed with a certain period, and the product is $h$. The period is carried by the waves without change, but the energy spreads out in an ever-widening circle. Eight years and nine months after the emission the wave-front is due to reach the earth. A few minutes before the arrival some person takes it into his head to go out and admire the glories of the heavens and - in short - to stick his eye in the way. The light waves when they started out could have had no notion of what they were going to hit; for all they knew they would be bound on a journey through endless space, as most of their colleagues were. Their energy would seem to be dissipated beyond recovery over a sphere of $50$ billion miles' radius. And yet if that energy is ever to enter matter again, if it is to work those chemical changes in the retina which give rise to the sensation of light, it must enter as a single quantum of action $h$~\cite{eddington3}.
\end{quotation}
 How can a photon possibly manage to keep itself all together as a single tiny quantum when it is so incredibly spread out across the Universe\,? The answer, perhaps, is that a photon is also a Machian space quantum. A photon space quantum still has a centre, defined as the point of intersection of all the photon strings, but the centre has no mass and no charge.

 Machian space quanta may also help to understand the invariance of the speed of light. Special Relativity is based on the postulate that the speed of a light wave, $c$, is independent of the motion of the light source relative to the observer. Is there any way to understand the invariance of $c$ with some sort of physical model or must the postulate simply be accepted\,? In the old ether model, the speed of light was assumed to be equal to $c$ in the rest frame of an \lq ether' that pervades all space. The invariance of the speed of light waves with respect to motion of the source is then obvious, just as the speed of sound waves in air and water waves in water are also independent of the motion of the source. However, the invariance of $c$ with respect to motion of the observer was much harder to explain and led to the conclusion that motion through the ether somehow causes the time dilation of clocks and the length contraction of metre sticks. The Machian string model offers an alternative point of view because the strings of the observer space quantum define an ether that is always at rest relative to the observer. Suppose the speed $c$ is a characteristic of the interaction between a photon space quantum and the space quantum of an observer. The invariance of $c$ with respect to motion of the observer then follows automatically because, when the observer starts moving, the space quantum of the observer moves with it.

% The Machian string model is a discrete model of matter in the sense that a particle consists of a discrete set of strings.
 Although the centre of a space quantum can be anywhere in the continuous three-dimensional coordinate space, physical space is defined as the space occupied by the strings and is therefore quantised.
% Physical space can be thought of as the superposition of the space quanta associated with all the particles.
 The space quantum of an elementary particle is an extended object and Quantum Mechanics also requires that every elementary particle is associated with an extended object, namely the wavefunction.
 % Space quanta may therefore provide a physical picture of the actual, quantum mechanical, universe.
 Perhaps Schrodinger's 3N-dimensional configuration space should not be regarded as an abstract mathematical space but should instead be identified with the physical space defined by the superposition of N space quanta.

% Machian space quanta may also help to understand the invariance of the speed of light. According to Special Relativity, the speed of light is independent of the motion of the source relative to the observer. Although it is easy to imagine how the speed of light can be independent of the motion of the source, since the independence of the speed of sound waves and water waves with respect to motion of the source is familiar from everyday experience, it is much harder to imagine how the speed of light can be independent of the motion of the observer. The Machian string model offers a solution, because the strings of the observer define an \lq ether' that is always at rest relative to the observer. When the observer starts moving, the space quantum of the observer moves with it. It is easy to see that if a photon always propagates at speed $c$ through the space quantum defined by the strings of the observer then the invariance of $c$ with respect to motion of the observer follows automatically.

 The inertial mass constraint (IMC) for space quanta is introduced in Section~\ref{imc} and, in Section~\ref{mmg}, the IMC is used to give a physical model for Newtonian gravity. When relativistic effects are included and the experimental constraints are applied, the equations of motion are found to be the same as in General Relativity. Inertia and kinetic energy are explained in terms of the energies in the Machian strings and a Machian formula is given for $G$, the constant of gravitation. Possible solutions to the dark energy and dark matter problems are given in Sections~\ref{de} and~\ref{dm}, respectively.

% The Machian string model is a discrete model of matter in the sense that a particle consists of a discrete set of strings. Although the centre of a space quantum can be anywhere in the continuous three-dimensional coordinate space, physical space is defined as the space occupied by the strings and is therefore also discrete. Physical space can be thought of as the superposition of the space quanta associated with all the particles.

% Machian space quanta define a quantisation of space and explain the double-rotation property of elementary particles. The space quantum of an elementary particle is an extended object and Quantum Mechanics also requires that every elementary particle is associated with an extended object, namely the wavefunction.
% % Space quanta may therefore provide a physical picture of the actual, quantum mechanical, universe.
% Perhaps Schrodinger's 3N-dimensional configuration space should not be regarded as an abstract mathematical space but should instead be identified with the physical space defined by the superposition of N space quanta.

\section{The inertial mass constraint (IMC)}\label{imc}
 Inertial mass is defined as the total energy in the space quantum of a massive particle in its instantaneous rest frame
 and the inertial mass constraint (IMC) states that the inertial mass of every massive particle is constant. The IMC allows an attractive gravitational force to be obtained with positive potential energies in the strings, as explained in Section~\ref{mmg}. The IMC is also used in Section~\ref{de} on \lq dark energy' and Section~\ref{dm} on \lq dark
 matter'.

\section{A Machian model for gravity}\label{mmg}
 The equations of motion for a Machian model of gravity are derived in the Appendix. The model is a direct interaction theory, as opposed to a field theory, and is defined by the energies in the strings connecting all pairs of particles in the observable universe. The energy in the strings has the form of positive Newtonian gravitational potential energy, with additional terms depending on the velocities of the particles.

\subsection{Inertia and Newton's laws of motion}\label{inertia}
 According to Newton's laws of motion, a free particle in an \lq inertial frame' moves with uniform velocity
 and a force $m{\bf a}$ is required to give a mass $m$ an acceleration ${\bf a}$. In General Relativity, acceleration is defined relative to the local \lq spacetime' around the particle and it has been suggested that an interaction with the quantum vacuum is responsible for the existence of inertial forces~\cite{haisch,rueda}. In the Machian string model, the energy of a particle is not concentrated at a point but is instead distributed throughout the Machian strings. Inertial frames are identified with frames moving at uniform velocity relative to distant matter, in accordance with Mach's principle. Appendix~\ref{appmg4} shows that the kinetic energy of a particle is part of the gravitational interaction energy in the Machian strings. Appendix~\ref{applaws4} shows that distant matter exerts a force $-m{\bf a}$ on a particle that has acceleration ${\bf a}$ relative to distant matter, so that an external force $m{\bf a}$ is needed to maintain the acceleration.

%From the Machian point of view, the force $m{\bf a}$ is due to the change in energy in the field strings due to the
%acceleration relative to distant matter. Mach's principle suggests that inertial frames should be identified with frames
% moving at uniform velocity relative to distant matter. According to Newton's second law, a particle moving with
%acceleration ${\bf a}$ in an inertial frame requires a force $m{\bf a}$.

\subsection{Newtonian gravity and the constant of gravitation}\label{newtonian}
 In the Machian string model, Newton's point masses and negative gravitational potential energies are both
 replaced by positive gravitational potential energy in the strings. In the simplest model of Machian space quanta, the
 direct strings in Figure~\ref{strings} joining the centres of the two masses $m_1$ and $m_2$, a distance $r$ apart, have
 energy $+Gm_1m_2/r$, where $G$ is the constant of gravitation. Due to the action of the IMC, the total energy at the
 centre and in the Machian strings connected to $m_1$ is $m_1c^2-Gm_1m_2/r$, and similarly for $m_2$. The total energy in
 the centres and all the strings of $m_1$ and $m_2$ is therefore  $m_1c^2 + m_2c^2 - Gm_1m_2/r$, so the interaction energy
 is the same as in Newtonian gravity. The idea that the gravitational mass energy is positive, with a compensating
 reduction in the energy of each mass, was suggested many years ago~\cite{abraham,ohanian}.

 There is a well-known Machian relation for $G$~\cite{brans}, namely
\begin{eqnarray}\label{bd}
 G\,\sim\,c^2\Big(\frac{M_U}{R_U}\Big)^{\!-1}\,,
\end{eqnarray}
 where $M_U$ and $R_U$ are the mass and radius of the observable universe, respectively. In the string model, the
 relation~(\ref{bd}) follows directly from the fact that the inertial mass, $mc^2$, is the same order of magnitude as the
 total gravitational potential energy in the strings. Since the average length of a string is of order $R_U$, the total
 gravitational potential energy in the strings connecting a mass $m$ to distant matter of total mass $M_U$ is of order
 $GmM_U/R_U$ so $mc^2\sim GmM_U/R_U$.
% The coefficient of proportionality in~(\ref{bd}) is derived in Appendix~\ref{appmg2}.

\subsection{Inertial and gravitational mass}\label{ig}
 The inertial mass in the space quantum of the $i^{th}$ particle, $m_i$, is defined as the mass at the centre of the space
 quantum plus the mass in all the strings connected to it. The gravitational mass of the $i^{th}$ particle, $\tilde m_i$,
 is defined as the mass at the centre plus the mass in the (very large) subset of strings connected to distant matter only.
 When all the space quanta are superposed, each pair of space quanta has two strings connecting their centres. The total
 energy in the two strings is $G\tilde m_i\tilde m_j/r_{ij}$ so the energy in each of the two strings is
 $G\tilde m_i\tilde m_j/2r_{ij}$. Since the difference between inertial mass and gravitational mass is due to the mass in
 the strings connected to nearby matter, it follows that inertial and gravitational mass are related by the equation
\begin{eqnarray}\label{mg1}
 && \hspace*{-1.8cm}m_ic^2=\tilde m_i c^2 \,+\!\!\!\sum_{j\in S_n,j\ne i}\!\!\!\frac{G\tilde m_i\tilde m_j}{2r_{ij}}\,,
\end{eqnarray}
 where $S_n$ denotes the set of nearby particles in the system under consideration.

\subsection{Experimental constraints}\label{exp}
\subsubsection{The two body and three body problems}
 Relativistic corrections for the motion of two gravitating masses and the acceleration of the Earth and Moon towards the Sun in the three-body Earth-Moon-Sun system are calculated in Appendix~\ref{apprelgrav}. After applying the experimental constraints, the equations of motion are found to be the same as in General Relativity. In the corresponding Machian string model, approximately $8\%$ of the mass of a space quantum is in the Machian strings and the remaining $92\%$ of the mass is at the centre. The effect of a gravitational field on the precession of a gyroscope is considered in Appendix~\ref{appprecess} and the usual results for the rates of geodetic and gravitomagnetic precession are recovered.
\subsubsection{Gravitational redshift and the propagation of light}
 Gravitational redshift and the propagation of light in a gravitational field are discussed in Appendix~\ref{applight2}. Photons propagate with constant energy in a gravitational field because a massive particle propagates with constant energy and a photon may be thought of as a massive particle with vanishingly small rest mass. From the relation $E=hf$, it follows that the frequency of a photon propagating in a gravitational field is constant. To explain the gravitational redshift, consider an atom of mass $m$ at rest at a distance $r$ from a mass $M$. The energy in the strings joining the two masses is $GMm/r$ and the IMC implies that the energy at the centre of the atom is $mc^2-GMm/r$, so the atomic energy levels are reduced by a factor of $1-GM/rc^2$. The frequency of a photon corresponding to a given atomic transition is therefore lower when the photon is emitted in a gravitational field. The frequency remains unchanged as the photon propagates out of the gravitational field, so the photon emerges with a lower frequency and therefore appears to be redshifted. Appendix~\ref{applight2} also shows that the speed of light is reduced in a gravitational field and Fermat's
 principle is used to calculate the deflection of a light beam and the time delay of radar signals.
\subsubsection{Gravitational radiation}
 The calculations in Appendix~\ref{appdarkmatter} show that the Machian strings around a massive body are deformed by the presence of a nearby mass. When two masses orbit one another, the deformation in the strings varies periodically in time and waves in the Machian strings are expected. A formula for the rate at which energy is radiated is derived in Appendix~\ref{appgravrad} for the case of circular orbits. Comparison with the experimentally verified formula for the rate of emission of gravitational radiation in General Relativity suggests that gravitational radiation may be attributed to the generation of gravitational waves in the strings.

\section{The expansion of the universe and \lq Dark energy'}\label{de}
 At first sight, the Machian relation~(\ref{bd}) implies that $G$ is a function of time since $M_U$ and $R_U$ vary with time as the Universe expands. It is shown in Appendix~\ref{appke3}, however, that the relation~(\ref{bd}) is not only consistent with the expansion history of the Universe but it can, in fact, be used to calculate the expansion history of the Universe.
% The expansion rates in the radiation and matter eras are found to be the same as in the conventional model based on the solution of the Friedmann equations and, moreover, the observed transition from deceleration to acceleration may be predicted without the need to introduce \lq dark energy'.

 On cosmological time scales, the expansion of the Universe has a significant effect on the number of Machian
 strings in the Universe and on the energy in each string. In all space quanta, the number
 of Machian strings increases with time because the number of particles in the observable universe increases
 as the Universe expands. In addition, the mass associated with the potential energy in each string has a
 kinetic energy associated with its recessional velocity. In the early universe, it turns out that the kinetic
 energy of the potential energy is small compared with the potential energy so, as the Universe expands, the
 total energy in every Machian string decreases. The IMC requires the effects of the increasing number of strings and the decreasing energy of each string to balance and the result is that the
 expansion of the Universe decelerates. The expansion rates in the radiation era and the matter era are the same as in the conventional $\Lambda$CDM model.

\begin{figure}[h]
\vspace*{0.5cm}
\includegraphics[height=5.5cm,width=7.5cm]{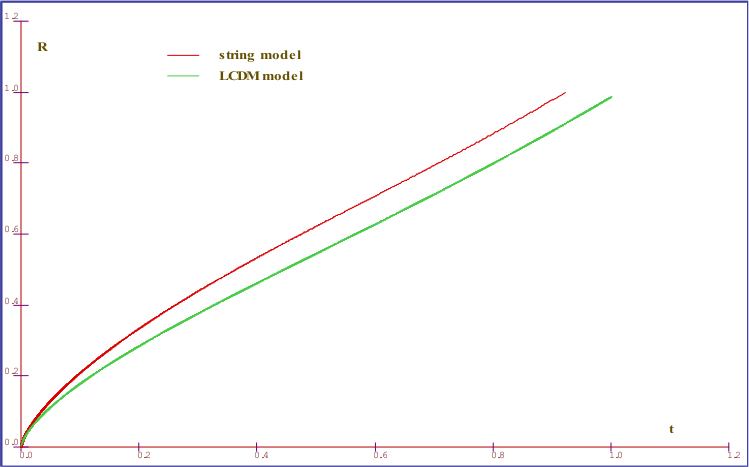}
\caption{\label{scalefactor} Scale factor, $R(t)$, in the string model (red) and in the $\Lambda$CDM model (green)
as a function of the time, $t$. In both models, the present time is at $R=1$. The two scale factors are almost indistinguishable experimentally, as discussed in Appendix~\ref{appke3}.}
\end{figure}

 In the later universe, the kinetic energy in each string becomes larger than the potential energy and the IMC then requires an accelerating expansion to keep the total kinetic energy in the space quantum of a massive particle constant. The time evolution of the scale factor in the string model calculated in Appendix~\ref{appke3} is shown in Figure~\ref{scalefactor} and compared to the time evolution in the conventional $\Lambda$CDM model.

\section{Galaxy rotation curves and \lq Dark matter'}\label{dm}
 It is known that Newton's law of gravity, when applied to the observed mass distribution, does not give the correct dynamics on the scale of galaxies and galaxy clusters. Either there is a very large amount of still unseen dark matter or Newton's law is incorrect. In modified Newtonian dynamics (MOND)~\cite{milgrom}, the Newtonian $1/r^2$ force law is replaced by a $1/r$ force law when the Newtonian acceleration is less than a characteristic acceleration, $a_0$. It has been shown~\cite{mcgaugh2} that MOND gives a very good fit to the observed galaxy rotation curves for $a_0\approx 1.2\times 10^{-10}$m/s$^2$.
% the required modification to Newton's law occurs below a characteristic acceleration $a_0\approx 1.2\times 10^{-10}$m/s$^2$, which is approximately the Newtonian acceleration at the visible edge of the galaxy [ref to Freeman's law]. A good fit to the observed rotation curves is found if the Newtonian $1/r^2$ law is replaced by a $1/r$ law when the Newtonian acceleration is smaller than $a_0$, but no physical reason for such a modification of the Newtonian force law is given.

 In the Machian string model, an acceleration scale of order $a_0$ is a natural consequence of a contact interaction between Machian strings. Consider a test mass in the vicinity of a larger mass $M$. In addition to the Machian strings of $M$, the test mass is immersed within the background space defined by the Machian strings joining all the pairs of particles in the Universe. It is therefore reasonable to suppose that the effect of the Machian strings of $M$ on the Machian strings of the test mass depends on the ratio of the density of the Machian strings of $M$ to the density of background strings. In Appendix~\ref{appdarkmatter}, it is shown that the ratio of densities is equivalent to the ratio of the Newtonian gravitational acceleration to $a_0$.

 Physically, the additional acceleration is due to the forces exerted on the centre of the test mass by the Machian strings connected to it. Due to the action of the IMC, the force between every pair of space quanta is attractive so the strings are in tension. In the absence of the mass $M$, the distribution of Machian strings around the test mass would be spherically symmetrical and all the tension forces on the centre would cancel out. The interaction with the Machian strings of $M$ causes the distribution of Machian strings around the test mass to become asymmetrical, so the forces exerted on the centres no longer cancel out and the test mass experiences an additional acceleration.
% Thus, in addition to the Newtonian gravitational acceleration associated with the direct strings, there is also a gravitational acceleration due to the net force exerted by the Machian strings on the centres.
% The net forces exerted by the strings on the centres give an additional gravitational acceleration, in addition to the Newtonian gravitational acceleration associated with the direct strings.

 To estimate the magnitude of the additional acceleration, consider the total energy in the Machian strings connected to a test mass $m$. As noted in Section~\ref{exp}, about $8\%$ of the rest mass, $0.08 mc^2$, is in the Machian strings. Dividing the total energy in the Machian strings by the average length of a Machian string, $R_U/2$, gives a force of about $0.16 mc^2/R_U$. In the limiting case of maximum asymmetry, with all the Machian strings pointing in the same direction at the centre, an additional acceleration of $0.16 c^2/R_U$ would be expected. The acceleration $a_0$ is approximately $0.4\,c^2/R_U$, where $R_U= 3.6\times 10^{26}$\,m is the radius of the observable universe found in Appendix~\ref{appke3}, so the maximum additional acceleration is about $0.4 a_0$.

 In contrast to MOND, which simply postulates a change in Newton's law, the additional acceleration in the string model is added to the Newtonian acceleration. Thus, although the additional acceleration is significantly less than $a_0$, it is still sufficient to explain the MOND phenomenology. A specific model for the interaction between Machian strings is considered in Appendix~\ref{appdarkmatter} and the maximum additional acceleration is found to be about $0.2 a_0$. A plot of the Newtonian gravitational acceleration and the additional acceleration calculated in Appendix~\ref{appdarkmatter} is shown in Figure~\ref{atotal}, plotted in units of $a_0$.
\begin{figure}[h]
%\vspace*{7cm}
%\vspace*{0.5cm}
\includegraphics[height=5cm,width=7cm]{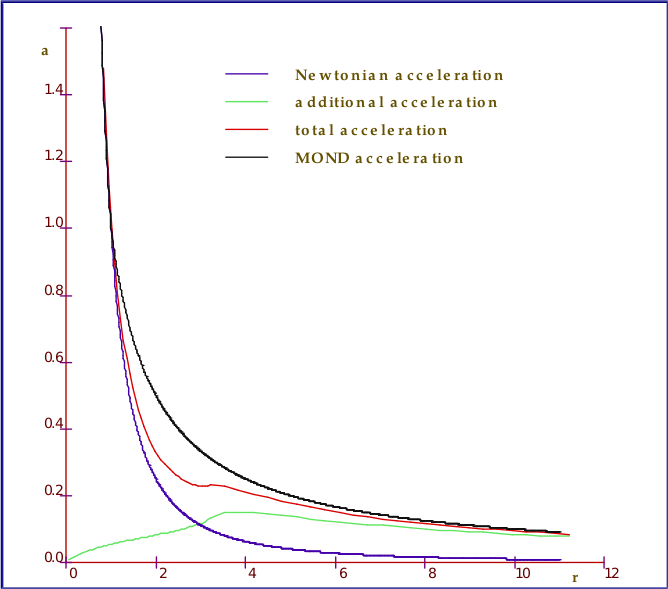}
%\vspace*{-0.8cm}
\caption{\label{atotal} The Newtonian acceleration (red curve) and the
additional acceleration due to the interaction between Machian strings
(green curve) around a point mass $M$ in the Machian string model for the model considered in 
Appendix~\ref{appdarkmatter}. The acceleration is plotted in units of the MOND acceleration scale, $a_0$, and the 
radial distance $r$ is plotted in units of the radius $R_g$, defined as the radius at which the Newtonian gravitational 
acceleration is equal to $a_0$, so the MOND acceleration is equal to $1/r^2$ for $r<1$ and $1/r$ for $r>1$.}
%With $r$ in units of $R_g$, the
%Newtonian acceleration, $-GM/r^2$, becomes approximately $-0.1/r^2$
%since $GM_U\approx 0.1\,R_Uc^2$ (taking $M_U=10^{22}$ Solar masses and
%$R_U=10^{26}$\,m). The additional acceleration from the field strings
%is approximately $-0.4/r$ for $r\gg 1$.
\end{figure}
 The total acceleration in the string model is seen to be very close to the MOND acceleration when the acceleration is 
much less than $a_0$, so the string model is expected to give a good fit to the observed galaxy rotation curves.

\section{Summary}
 In the Machian string model, direct strings give Newtonian gravity, Machian strings give inertia and the expansion of the Universe and dark matter effects are due to the interaction between Machian strings.

 A formula for the gravitational potential energy in the strings is derived from a relativistic string action 
in Appendix~\ref{applorentz2}. The usual formula for the Newtonian kinetic energy of a massive particle can be derived 
by a suitable choice of the fraction, $\kappa$, of the mass of a space quantum that lies in the strings. It is 
shown in Appendix~\ref{appmg4} that the relativistic correction to the kinetic energy and all the experimental 
results for the relativistic two-body problem, the Nordtvedt effect and gyroscope precession are obtained if the free 
parameters $a_1$, $a_2$, $b_1$ and $b_2$ are suitably
chosen. The theory gives the standard results for gravitational
redshift, light bending and radar time delay and the electromagnetic interaction between charged particles can also be 
included in the formalism, as explained in Appendix~\ref{applight2}. A physical model for the observed rate of 
emission of gravitational radiation from binary pulsars is discussed in Appendix~\ref{appgravrad}.

 The expansion history of the universe can be calculated by considering the total energy in all the Machian strings of 
a space quantum. The usual results for the decelerating expansion of the Universe in the radiation and matter eras are 
obtained if the mass fraction $\kappa$ is assumed to be constant. By including the effect of the Hubble flow on the 
energy of a Machian string, a good fit to the experimental data 
is obtained without introducing \lq dark energy'.

 The interaction between Machian strings gives effects usually attributed to \lq dark matter'. The strength of the 
interaction depends on the density of space defined by the Machian strings. Since the ratio of the density of space 
around a mass to the background density of space is equivalent to the ratio of the Newtonian gravitational acceleration 
to the acceleration scale $a_0$, the MOND phenomenology is explained.

% Machian space quanta define a quantisation of space and explain the double-rotation property of elementary particles. The space quantum of an elementary particle is an extended object and Quantum Mechanics also requires that every elementary particle is associated with an extended object, namely the wavefunction.
% % Space quanta may therefore provide a physical picture of the actual, quantum mechanical, universe.
% Perhaps Schrodinger's 3N-dimensional configuration space should not be regarded as an abstract mathematical space but should instead be identified with the physical space defined by the superposition of N space quanta.

\section{Acknowledgements}
 I would like to thank both my parents for all their help and support over the years. They gave me the time and space in which to work and the courage to think along different lines. The financial support of Trinity College, Cambridge, is also gratefully acknowledged. The programming language used for the numerical calculations is available online~\cite{raphael}.

\begin{appendix}
 \input{applorentz2.tex}

 \input{appmg4.tex}

 \input{appmotion4.tex}
 \input{applaws4.tex}

 \input{apprelgravb.tex}
 \input{appprecess.tex}
 \input{applight2.tex}

 \input{appke3.tex}

 \input{appdarkmatter.tex}

 \input{appgravrad.tex}

\end{appendix}

\bibliography{master}

\end{document}

%% file: applorentz2.tex
\section{Lorentz invariance and the Lagrangian for a string}\label{applorentz2}

 Let $X^\mu(\tau,\sigma)$ be the spacetime coordinates of a classical string joining the centres of the $i^{th}$ and $j^{th}$ particles, where $\tau$ is a time-like coordinate and $\sigma\in [0,1]$ parameterises the position along the string~\cite{zwiebach}. An action for the string may be written in the form
\begin{eqnarray}\label{as1}
 S~=~\int \!{\mathcal L}~d\tau d\sigma\,,
\end{eqnarray}
 where the Lagrangian density ${\mathcal L}$ is some scalar function of the four-vectors $\partial X^\mu/\partial\tau$ and $\partial X^\mu/\partial\sigma$. The most general form of a Lagrangian density with dimensions $r^n$ is
\begin{eqnarray}\label{ld1}
 {\mathcal L}\,=\,\Big[\frac{A(s)}{c^2}\Big(\frac{\partial X^\mu}{\partial \tau}\frac{\partial X_\mu}{\partial \sigma}\Big)^{\!2}
 -B(s)\Big(\frac{\partial X^\mu}{\partial \sigma}\Big)^{\!2}\Big]^{n/2},
\end{eqnarray}
 where $A(s)$ and $B(s)$ are arbitrary functions of $s=(\partial X^\mu/\partial \tau)^2/c^2$. The Nambu-Goto Lagrangian~\cite{zwiebach}, for example, has $A=1$, $B=s$ and $n=1$. Let the centre of mass frame of the string, ${\mathcal S}_{cm}$, be defined as the frame in which the velocities of the ends of the string are equal and opposite. If the parameter $\tau$ is chosen to be the time coordinate in ${\mathcal S}_{cm}$ then
\begin{eqnarray}\label{ac}
  S\,= \! \int L_{cm} ~d\tau\,,
\end{eqnarray}
 where the Lagrangian in ${\mathcal S}_{cm}$ is given by
\begin{eqnarray}\label{sl}
  L_{cm} \,= \! \int {\mathcal L} ~d\sigma\,.
\end{eqnarray}
 If ${\bf x}$ denotes the spatial coordinates in ${\mathcal S}_{cm}$ then $X^\mu=(c\tau,{\bf x})$ and
\begin{eqnarray}\label{as2}
 \frac{\partial X^\mu}{\partial \tau}\,=\,(c,\dot{\bf x}) \hspace*{0.5cm}\mbox{and}\hspace*{0.5cm}\frac{\partial X^\mu}{\partial \sigma}\,=\,(0,{\bf x}^\prime)\,.
\end{eqnarray}

 It is desired to find the Lagrangian $L$ in a general frame, ${\mathcal S}$, in which the velocities of the $i^{th}$ and $j^{th}$ particles are ${\bf v}_i$ and ${\bf v}_j$, respectively. In the special case $B(s)=sA(s)$, the action is invariant under reparameterisation of the $\tau$ coordinate of the form $\tau\rightarrow \tau^\prime(\tau,\sigma)$ and $\tau$ can then be chosen to be the time coordinate in the general frame ${\mathcal S}$ as well as in ${\mathcal S}_{cm}$~\cite{chodos}. More generally, the Lagrangian must first be evaluated in ${\mathcal S}_{cm}$, using~(\ref{as2}), and a Lorentz transformation must then be applied to find the Lagrangian in ${\mathcal S}$.

 In the Machian string model, the energy in a string of length $r$ is proportional to $1/r$, so $n=-1$. Consider a straight string in the centre of mass frame. Then ${\bf x}^\prime= {\pmb \xi}$, where ${\pmb \xi}$ is the instantaneous separation of the two centres in ${\mathcal S}_{cm}$, so $(\partial X^\mu/\partial \tau)(\partial X_\mu/\partial \sigma)= -{\pmb \xi}.\dot{\bf x}$, $(\partial X^\mu/\partial \sigma)^2= -\xi^2$ and $s= 1-\dot x^2/c^2$. It is convenient to rewrite the Lagrangian density~(\ref{ld1}) in the form
 \begin{eqnarray}
 {\mathcal L}&=&f(z)\Big[\frac{g(z)}{c^2}\Big(\frac{\partial X^\mu}{\partial \tau}\frac{\partial X^\mu}{\partial \sigma}\Big)^{\!2}-\Big(\frac{\partial X^\mu}{\partial \sigma}\Big)^{\!2}\Big]^{-1/2}~~~~\label{ldd}\\
 &=& \frac{f(z)}{\xi}\Big[1 + g(z)\frac{(\widehat{\pmb \xi}.\dot{\bf x})^2}{c^2}\Big]^{-1/2}\,,\label{ld2}
 \end{eqnarray}
 where $f(z)$ and $g(z)$ are arbitrary functions of $z= \dot x^2/c^2$ defined by $A(s)= g(z)/f(z)^2$ and $B(s)= 1/f(z)^2$ and $\widehat{\pmb \xi}$ denotes the unit vector along ${\pmb \xi}$. Expanding $f$ and $g$ in powers of $z$, the Lagrangian~(\ref{sl}) has the form
%\begin{eqnarray}\label{sl}
%  && \hspace*{-0.5cm} S\, =\, \! \int L_{cm} \,dt_{cm}\, \\
%  \hspace*{-3cm}\mbox{where}\hspace*{1cm}&& L_{cm} \,= \! \int {\mathcal L} ~d\sigma\,.
%\end{eqnarray}
% The Lagrangian~(\ref{sl}) may be evaluated using~(\ref{as2}) to give
%\begin{eqnarray}\label{ld3}
%  && \hspace*{-0.8cm}L_{cm} = \frac{\Gamma}{\xi}\int_0^1 \!d\sigma\,\Big[ab\Big(1-\frac{\dot x^2}{c^2}\Big)^{\!p} \frac{({\pmb \xi}.\dot{\bf x})^2}{c^2}
%   + b\Big(1-\frac{\dot x^2}{c^2}\Big)^{\!m} \xi^2 \Big]^{-1/2}\hspace*{-0.5cm}\nonumber\\
%  && = \, \frac{1}{\xi} \int_0^1 \!d\sigma \,\Big[b\Big(1-m\frac{\dot x^2}{c^2}\Big) + ab \frac{(\widehat{\pmb \xi}.\dot{\bf x})^2}{c^2} + \dots\Big]^{-1/2}.\nonumber\\
%\end{eqnarray}
\begin{eqnarray}\label{ld3}
  && \hspace*{-0.8cm}L_{cm} = \frac{\Gamma}{\xi}\int_0^1 \!d\sigma\,(1 + a_1z + a_2z^2 +\dots)\nonumber\\
  && \hspace*{1.2cm}\times \Big[1 + (b_1 + b_2 z + \dots)\frac{(\widehat{\pmb \xi}.\dot{\bf x})^2}{c^2}\Big]^{-1/2}\!,~~
\end{eqnarray}
  for some constants $\Gamma$, $a_1$, $a_2$, $b_1$ and $b_2$. Let $(ct,{\bf r})$ denote the coordinates in ${\mathcal S}$. To lowest order, the relative positions and velocities of the ends of the string are Lorentz invariant, so $\widehat{\pmb \xi}$ may be replaced by ${\widehat {\bf r}}_{ij}$, the unit vector along the string in ${\mathcal S}$, and the velocity of the string as a function of $\sigma$ may be written as $\dot{\bf x}= (1/2-\sigma){\bf v}_{ij}$. Then, to ${\mathcal O}(v^2/c^2)$,~(\ref{ld3}) gives
\begin{eqnarray}\label{ld4}
  L_{cm}\,=\,\frac{\Gamma}{\xi}\Big[1 + \frac{a_1}{12}\frac{v_{ij}^2}{c^2} - \frac{b_1}{24}\frac{({\widehat {\bf r}}_{ij}.{\bf v}_{ij})^2}{c^2} + \dots \Big]\,.
\end{eqnarray}
 If ${\mathcal S}_{cm}$ moves with velocity ${\bf w}$ relative to ${\mathcal S}$ then $d\tau= dt/\gamma(w)$. From~(\ref{ac}), $L=L_{cm}/\gamma(w)$, so
\begin{eqnarray}\label{sl2}
 L\,=\,\frac{\Gamma}{\gamma(w)\xi}\Big[1 + \frac{a_1}{12}\frac{v_{ij}^2}{c^2} - \frac{b_1}{24}\frac{({\widehat {\bf r}}_{ij}.{\bf v}_{ij})^2}{c^2} + \dots \Big]\,.
\end{eqnarray}

\begin{figure}[h]
\vspace*{-0.1cm}
\includegraphics[height=7cm,width=7cm]{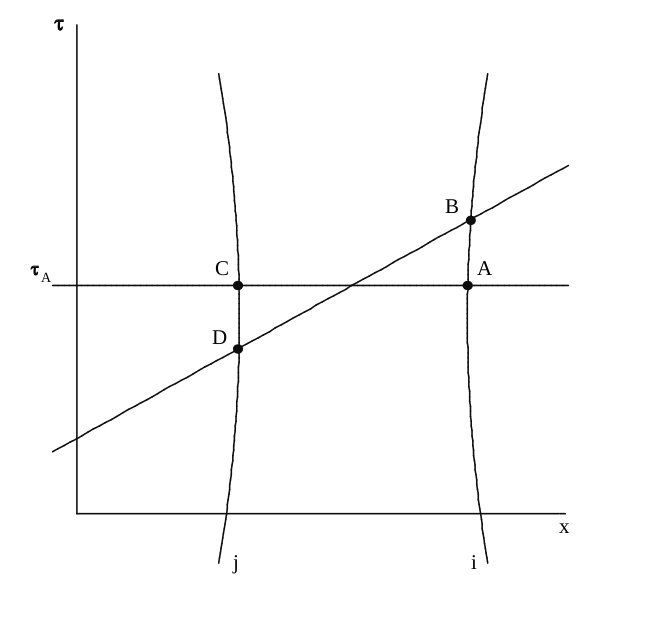}
\vspace*{-0.5cm} \caption{\label{times} Diagram showing the worldlines of the $i^{th}$ and $j^{th}$ particles in the centre of mass frame, ${\mathcal S}_{cm}$. The line $AC$ is a line of simultaneity in ${\mathcal S}_{cm}$ and the line $BD$ is a line of simultaneity in the frame ${\mathcal S}$.}
\vspace*{0.3cm}
\end{figure}

 The quantity $1/\gamma(w)\xi$ must now be rewritten in terms of the positions and velocities at a given time in frame ${\mathcal S}$. The instantaneous separation ${\pmb \xi}$ of the two centres in frame ${\mathcal S}_{cm}$ can be related to the instantaneous separation ${\bf r}_{ij}$ in frame ${\mathcal S}$ by a modification of the method in~\cite{kennedy} that is symmetrical with respect to the two centres. Consider the two events $A$ and $C$ in Fig~\ref{times}, which are simultaneous in ${\mathcal S}_{cm}$ at time $\tau_A$ and separated by the distance ${\bf \xi}$. In ${\mathcal S}$, the events $A$ and $C$ are not simultaneous. A line of simultaneity in ${\mathcal S}$ is shown in Fig~\ref{times} connecting the events $B$ and $D$. The events $B$ and $D$ are defined to be symmetrical relative to the line of simultaneity in ${\mathcal S}_{cm}$, so that $\tau_B-\tau_A= \tau_A-\tau_D$. The Lorentz transformations relating the positions of events $B$ and $D$ in the two frames are
\begin{eqnarray}\label{lt1}
 \hspace*{-1cm}&& {\bf x}_B\,=\,{\bf r}_B + (\gamma-1)\widehat {\bf w}(\widehat {\bf w}.{\bf r}_B) -\gamma{\bf w}t_B \\\hspace*{-1cm}
 \mbox{and}\hspace*{0.2cm}&& {\bf x}_D\,=\,{\bf r}_D + (\gamma-1)\widehat {\bf w}(\widehat {\bf w}.{\bf r}_D) -\gamma{\bf w}t_D\,. \label{lt2}
\end{eqnarray}
 Since the velocities of the centres are equal and opposite in ${\mathcal S}_{cm}$ and $\tau_B-\tau_A= -(\tau_D-\tau_C)$, it follows that ${\bf x}_B-{\bf x}_A={\bf x}_D-{\bf x}_C$. Then
\begin{eqnarray}\label{lt3}
 && \hspace*{-1.8cm}{\pmb \xi}\,=\,{\bf x}_A-{\bf x}_C\,=\,{\bf x}_B-{\bf x}_D\,=\,{\bf r}_{ij} + (\gamma-1)\widehat {\bf w}(\widehat {\bf w}.{\bf r}_{ij})\nonumber\\
 \hspace*{1cm}\Rightarrow\hspace*{0.5cm}&& {\pmb \xi}= {\bf r}_{ij} + \frac{{\bf w}({\bf w}.{\bf r}_{ij})}{2c^2} +\dots\nonumber\\
 \hspace*{1cm}\Rightarrow\hspace*{0.5cm}&& \xi^2\,=\,r_{ij}^2\,+\,\frac{({\bf r}_{ij}.{\bf w})^2}{c^2} + \dots\,. \label{k1}
\end{eqnarray}

 For interactions with nearby matter, the centre of mass velocity is ${\bf w}= ({\bf v}_i+{\bf v}_j)/2$ so
\begin{eqnarray}\label{gx}
 \hspace*{-0.3cm}\frac{1}{\gamma(w)\xi}&=&\Big(1-\frac{w^2}{2c^2}+\dots \Big)\Big[r_{ij}^2 + \frac{({\bf r}_{ij}.{\bf w})^2}{c^2}\Big]^{-1/2}\hspace*{0.3cm} \\
 &=& \frac{1}{r_{ij}}\Big[1-\frac{w^2}{2c^2}-\frac{(\widehat{\bf r}_{ij}.{\bf w})^2}{2c^2} +\dots \Big] \nonumber\\
 &=& \frac{1}{r_{ij}}\Big[1-\frac{v_{ij}^2}{8c^2}-\frac{{\bf v}_i.{\bf v}_j}{2c^2}-\frac{(\widehat{\bf r}_{ij}.{\bf v}_{ij})^2}{8c^2} \nonumber\\
 && \hspace*{1.2cm}- ~\frac{(\widehat{\bf r}_{ij}.{\bf v}_i)(\widehat{\bf r}_{ij}.{\bf v}_j)}{2c^2} +\dots \Big]
 \label{gx2}
\end{eqnarray}
 and the Lagrangian~(\ref{sl2}) becomes
\begin{eqnarray}\label{ld5}
  \hspace*{-1.3cm}&& L\,=\,\frac{\Gamma}{r_{ij}}\Big[1 + \Big(\frac{a_1}{12}-\frac{1}{8}\Big)\frac{v_{ij}^2}{c^2} - \frac{{\bf v}_i.{\bf v}_j}{2c^2}\nonumber\\
  &&\hspace*{0.3cm} - \Big(\frac{b_1}{24}+\frac{1}{8}\Big)\frac{({\widehat {\bf r}}_{ij}.{\bf v}_{ij})^2}{c^2}
  - \frac{({\widehat {\bf r}}_{ij}.{\bf v}_i)({\widehat {\bf r}}_{ij}.{\bf v}_j)}{2c^2} + \dots \Big].\nonumber\\
\end{eqnarray}
 For interactions with distant matter, the centre of mass frame of $m_i$ and particles of distant matter is effectively the rest frame of the distant matter. The calculation of relativistic corrections to kinetic energy in Appendix~\ref{appmg4} requires ${\mathcal O}(v^4/c^4)$ terms proportional to $v_i^4/c^4$. The velocity of $m_i$ in ${\mathcal S}_{cm}$ is ${\bf v}_{ij}/(1-{\bf v}_i.{\bf v}_j/c^2)$, but the ${\mathcal O}(v^2/c^2)$ correction to ${\bf v}_{ij}$ does not lead to any terms in the Lagrangian proportional to $v_i^4/c^4$ and is therefore neglected. The velocity of the string as a function of $\sigma$ may therefore be written as $\dot{\bf x}= \sigma{\bf v}_{ij}$. Similarly, it is sufficient to use the approximation $\widehat{\pmb \xi}= \widehat{\bf r}_{ij}$ in the evaluation of~(\ref{ld3}) and expansion to ${\mathcal O}(v^4/c^4)$ then gives, instead of~(\ref{sl2}),
 \begin{eqnarray}\label{sl3}
 && \hspace*{-0.7cm}L\,=\,\frac{\Gamma}{\gamma(w)\xi}\Big[1 + \frac{a_1}{3}\frac{v_{ij}^2}{c^2} - \frac{b_1}{6}\frac{({\widehat {\bf r}}_{ij}.{\bf v}_{ij})^2}{c^2} + \frac{a_2}{5}\frac{v_{ij}^4}{c^4}\nonumber\\
 && \hspace*{-0.1cm} +~\frac{3b_1^2}{40}\frac{({\widehat {\bf r}}_{ij}.{\bf v}_{ij})^4}{c^4}-\frac{b_2+a_1b_1}{10}\frac{({\bf v}_{ij})^2({\widehat {\bf r}}_{ij}.{\bf v}_{ij})^2}{c^4}\Big]\,.~~\nonumber\\
\end{eqnarray}
 Evaluating the prefactor~(\ref{gx}) with ${{\bf w}=\bf v}_j$ instead of ${\bf w}=({\bf v}_i+{\bf v}_j)/2$ gives,
 instead of~(\ref{gx2}),
\begin{eqnarray}\label{gx3}
  \frac{1}{\gamma(w)\xi}\,=\,\frac{1}{r_{ij}}\Big[1-\frac{v_j^2}{2c^2}-\frac{(\widehat{\bf r}_{ij}.{\bf v}_j)^2}{2c^2} +\dots \Big]\,,
\end{eqnarray}
 with no ${\mathcal O}(v^4/c^4)$ terms proportional to $v_i^4/c^4$. Substituting into~(\ref{sl3}) then gives, instead of~(\ref{ld5}),
\begin{eqnarray}\label{ld6}
 && \hspace*{-0.5cm}L\,=\,\frac{\Gamma}{r_{ij}}\Big[1 -\frac{v_j^2}{2c^2} - \frac{({\widehat {\bf r}}_{ij}.{\bf v}_j)^2}{2c^2} + \frac{a_1}{3}\frac{v_{ij}^2}{c^2} - \frac{b_1}{6}\frac{({\widehat {\bf r}}_{ij}.{\bf v}_{ij})^2}{c^2} \nonumber\\
 && \hspace*{-0.4cm}+~ \frac{a_2}{5}\frac{v_{ij}^4}{c^4} +\frac{3b_1^2}{40}\frac{({\widehat {\bf r}}_{ij}.{\bf v}_{ij})^4}{c^4}
 -\frac{b_2 + a_1b_1}{10}\frac{({\bf v}_{ij})^2({\widehat {\bf r}}_{ij}.{\bf v}_{ij})^2}{c^4}\Big].
 \hspace*{-0.5cm}\nonumber\\
\end{eqnarray}
%\begin{eqnarray}\label{ld6}
%  && \hspace*{-1cm}L\,=\,\frac{\Gamma}{r_{ij}}\Big[1 -\frac{v_j^2}{2c^2} - \frac{({\widehat {\bf r}}_{ij}.{\bf v}_j)^2}{2c^2}\nonumber\\ &&\hspace*{0.5cm}
%  - ~\frac{a_1}{6}\frac{v_{ij}^2}{c^2} + \frac{b_0}{12}\frac{({\widehat {\bf r}}_{ij}.{\bf v}_{ij})^2}{c^2}
% + \dots \Big]\,.\nonumber\\
%\end{eqnarray}

%% file: appmg4.tex
\section{The total Lagrangian and experimental constraints}\label{appmg4}
 To define the total Lagrangian for a system of particles interacting with each other and with the distant universe, it is necessary to define the gravitational mass of the $i^{th}$ particle, $\tilde m_i$, for each interaction. For interactions between a particle in the system and a distant particle, the gravitational mass of each particle is equal to the inertial mass, $m_i$. For interactions between two particles in the system, the gravitational masses are defined by equation~(\ref{mg1}). The total Lagrangian for a system of nearby particles interacting with distant matter, using~(\ref{ld4}) for interactions between pairs of particles within the system and~(\ref{ld5}) for interactions with distant matter, is therefore
\begin{eqnarray}\label{lag1}
 &&\hspace*{-0.5cm} L\,= \!\!\sum_{i,j \in S_n, i<j}\!\!\frac{G\tilde m_i\tilde m_j}{r_{ij}}
 \Big[1 + \Big(\frac{a_1}{12}-\frac{1}{8}\Big)\frac{v_{ij}^2}{c^2} - \frac{{\bf v}_i.{\bf v}_j}{2c^2}\nonumber\\
  &&\hspace*{0.3cm} - \Big(\frac{b_1}{24}+\frac{1}{8}\Big)\frac{({\widehat {\bf r}}_{ij}.{\bf v}_{ij})^2}{c^2}
  - \frac{({\widehat {\bf r}}_{ij}.{\bf v}_i)({\widehat {\bf r}}_{ij}.{\bf v}_j)}{2c^2} + \dots \Big]\nonumber\\
 && \hspace*{-0.1cm}+ \!\!\sum_{i\in S_n, j\in S_d}\!\!\frac{Gm_im_j}{r_{ij}}
 \Big[1 -\frac{v_j^2}{2c^2} - \frac{({\widehat {\bf r}}_{ij}.{\bf v}_j)^2}{2c^2}\nonumber\\ &&\hspace*{0.5cm}
  +~\frac{a_1}{3}\frac{v_{ij}^2}{c^2} - \frac{b_1}{6}\frac{({\widehat {\bf r}}_{ij}.{\bf v}_{ij})^2}{c^2} + \frac{a_2}{5}\frac{v_{ij}^4}{c^4} +\frac{3b_1^2}{40}\frac{({\widehat {\bf r}}_{ij}.{\bf v}_{ij})^4}{c^4}
 \nonumber\\ && \hspace*{1.2cm}-~\frac{b_2 + a_1b_1}{10}\frac{({\bf v}_{ij})^2({\widehat {\bf r}}_{ij}.{\bf v}_{ij})^2}{c^4} + \dots \Big]\,,
\end{eqnarray}
 where $S_n$ denotes the set of nearby particles in the system and $S_d$ denotes the set of particles in the distant universe.
 Let the gravitational potential due to distant matter, $\kappa$, be defined by
\begin{eqnarray}\label{kd}
 \sum_{j \in S_d }\frac{Gm_j}{r_{ij}c^2}\,=\,\kappa\,,
\end{eqnarray}
 where $r_{ij}$ is the distance from a distant particle of mass $m_j$ to some nearby mass $m_i$, and suppose that, in frame ${\mathcal S}$, all distant matter particles have velocity ${\bf v}_d$. The average values of $(\widehat {\bf r}_{ij}.{\bf n})^2$ and $(\widehat {\bf r}_{ij}.{\bf n})^4$ when summed over distant matter are $1/3$ and $1/5$, respectively, for any fixed unit vector ${\bf n}$, so~(\ref{lag1}) becomes
% The average velocity of distant matter, ${\bf v}_d$, may be defined by the equation
%\begin{eqnarray}
% \sum_{j \in S_d }\frac{Gm_j}{r_{ij}c^2}{\bf v}_j\,=\,\kappa{\bf v}_d
%\end{eqnarray}
% and the Lagrangian
\begin{eqnarray}\label{lag3}
 &&\hspace*{-0.5cm} L\,= \sum_{i\in S_n}\kappa m_i c^2\Big[1 -\frac{2v_d^2}{3c^2} +\frac{1}{3} \Big(a_1-\frac{b_1}{6}\Big) \frac{v_{id}^2}{c^2} + \delta\frac{v_{id}^4}{c^4}\dots \Big] \nonumber\\ && \hspace*{0.1cm}+ \!\!\sum_{i,j \in S_n, i<j}\!\!\frac{G\tilde m_i\tilde m_j}{r_{ij}}
 \Big[1 + \Big(\frac{a_1}{12}-\frac{1}{8}\Big)\frac{v_{ij}^2}{c^2} - \frac{{\bf v}_i.{\bf v}_j}{2c^2}\nonumber\\
  &&\hspace*{0.3cm} - \Big(\frac{b_1}{24}+\frac{1}{8}\Big)\frac{({\widehat {\bf r}}_{ij}.{\bf v}_{ij})^2}{c^2}
  - \frac{({\widehat {\bf r}}_{ij}.{\bf v}_i)({\widehat {\bf r}}_{ij}.{\bf v}_j)}{2c^2} + \dots \Big],\nonumber\\
\end{eqnarray}
 where
\begin{eqnarray}\label{deldef}
 \delta\,=\, \frac{1}{5}\Big(a_2 + \frac{3b_1^2}{40} -\frac{b_2+a_1b_1}{6}\Big)\,.
\end{eqnarray}
 From equation~(\ref{mg1}), the gravitational mass $\tilde m_i$ is given in terms of the inertial mass $m_i$ by
\begin{eqnarray}\label{mg3}
  \hspace*{-0.3cm}\tilde m_ic^2&=&m_i c^2 \,-\!\!\sum_{j\in S_n,j\ne i}\!\!\frac{G\tilde m_i\tilde m_j}{2r_{ij}}
 \nonumber\\ &=& m_ic^2 \,-\!\!\sum_{j\in S_n,j\ne i}\!\!\frac{Gm_im_j}{2r_{ij}} \,+\, \dots\,.
\end{eqnarray}
 Substitution into~(\ref{lag3}) gives
%\begin{eqnarray}\label{lag4}
% && \hspace*{-0.5cm}L\,=\sum_{i,j\in S_n,i<j}\!\!\frac{Gm_im_j}{r_{ij}}
% \Big\{ \Big[1 - \!\!\!\sum_{k\in S_n,k\ne i}\!\frac{Gm_k}{r_{ik}c^2} + \dots\Big] \nonumber\\
% && \times \Big[1 - \!\!\!\sum_{k\in S_n,k\ne j}\!\frac{G\tilde m_k}{r_{jk}c^2} + \dots\Big]\Big[1 + \Big(\frac{m}{24}-\frac{1}{8}\Big)\frac{v_{ij}^2}{c^2} + \dots \Big]\Big\}\hspace*{-0.5cm}\nonumber\\
% && \hspace*{0.5cm}+~\sum_{i\in S_n,j\in S_d}\frac{Gm_im_j}{r_{ij}}\Big[1 -\frac{{v_j}^2}{c^2} \dots \Big].
%\end{eqnarray}
\begin{eqnarray}\label{lag4}
 &&\hspace*{-0.5cm} L\,= \sum_{i\in S_n}\kappa m_i c^2\Big[1 -\frac{2v_d^2}{3c^2} +\frac{1}{3} \Big(a_1-\frac{b_1}{6}\Big) \frac{v_{id}^2}{c^2} + \delta \frac{v_{id}^4}{c^4} + \dots \Big] \nonumber\\ && \hspace*{0.1cm}+ \!\!\sum_{i,j \in S_n, i<j}\!\!\frac{Gm_im_j}{r_{ij}}
 \Big[1 + \Big(\frac{a_1}{12}-\frac{1}{8}\Big)\frac{v_{ij}^2}{c^2} - \frac{{\bf v}_i.{\bf v}_j}{2c^2}\nonumber\\
  &&\hspace*{0.3cm} - \Big(\frac{b_1}{24}+\frac{1}{8}\Big)\frac{({\widehat {\bf r}}_{ij}.{\bf v}_{ij})^2}{c^2}
  - \frac{({\widehat {\bf r}}_{ij}.{\bf v}_i)({\widehat {\bf r}}_{ij}.{\bf v}_j)}{2c^2} + \dots \Big]\nonumber\\
  && \hspace*{1cm}-~\sum_{i,j,k \in S_n,j\ne i,k\ne j}\hspace*{-0.2cm}\frac{G^2m_im_jm_k}{2r_{ij}r_{jk}c^2}\,+\,\dots\,.
\end{eqnarray}
 If the constant $\kappa$ satisfies the relation
\begin{eqnarray}\label{kdef}
 \frac{\kappa}{3}\Big(a_1-\frac{b_1}{6}\Big)\,=\,\frac{1}{2}
\end{eqnarray}
 and the parameters $a_2$ and $b_2$ in~(\ref{deldef}) are such that
\begin{eqnarray}\label{ddef}
 \kappa\delta\,=\,\frac{1}{8}
\end{eqnarray}
 then, after dropping the first two terms in~(\ref{lag4}), which do not contribute to the equations of motion, the Lagrangian has the form
\begin{eqnarray}\label{lag5}
 &&\hspace*{-0.8cm} L\,= \sum_{i\in S_n}\Big[\frac{1}{2}m_iv_{id}^2 + \frac{1}{8}m_i\frac{v_{id}^4}{c^2}\Big] \nonumber\\ && \hspace*{0.1cm}+ \!\!\sum_{i,j \in S_n, i<j}\!\!\frac{Gm_im_j}{r_{ij}}
\Big[1+ \beta\frac{(v_i^2+v_j^2)}{c^2} +\gamma\frac{{\bf v}_i.{\bf v}_j}{c^2} \nonumber\\
 && \hspace*{0.4cm}+\,\mu\frac{({\widehat {\bf r}}_{ij}.{\bf v}_i)^2 + ({\widehat {\bf r}}_{ij}.{\bf v}_j)^2}{c^2} + \nu\frac{({\widehat {\bf r}}_{ij}.{\bf v}_i)({\widehat {\bf r}}_{ij}.{\bf v}_j)}{c^2}\Big]\nonumber\\
 % \Big[1 + \Big(\frac{m}{24}-\frac{1}{8}\Big)\frac{v_{ij}^2}{c^2} - \frac{{\bf v}_i.{\bf v}_j}{2c^2}\nonumber\\
%  &&\hspace*{0.3cm} - \Big(\frac{a}{24}+\frac{1}{8}\Big)\frac{({\widehat {\bf r}}_{ij}.{\bf v}_{ij})^2}{c^2}
%  - \frac{({\widehat {\bf r}}_{ij}.{\bf v}_i)({\widehat {\bf r}}_{ij}.{\bf v}_j)}{2c^2} + \dots \Big]\nonumber\\
  && \hspace*{0.7cm}+\hspace*{0.3cm}\epsilon\hspace*{-0.3cm}\sum_{i,j,k \in S_n,j\ne i,k\ne j}\hspace*{-0.2cm}\frac{G^2m_im_jm_k}{r_{ij}r_{jk}c^2}\,+\,\dots\,,
\end{eqnarray}
 where the parameters $\beta$, $\gamma$, $\mu$, $\nu$ and $\epsilon$ are given by
\begin{eqnarray}\label{param}
 \beta\,=\,\frac{a_1}{12}-\frac{1}{8}\,, \hspace*{0.3cm} \gamma\,=\,-\frac{a_1}{6}-\frac{1}{4}\,,\hspace*{0.3cm}
 \mu\,=\,-\frac{b_1}{24}-\frac{1}{8}\,,\nonumber\\
  \nu\,=\,\frac{b_1}{12}-\frac{1}{4}\hspace*{0.6cm}\mbox{and}\hspace*{0.6cm} \epsilon\,=\,-\frac{1}{2}\,.\hspace*{1.4cm}
\end{eqnarray}
 The derivation of the Lagrangian~(\ref{lag5}) shows that the kinetic energy of a mass is due to the velocity-dependence of the energy in the Machian strings connected to it.

 The equations of motion are presented in Appendix~\ref{appmotion4} and relativistic effects are calculated in Appendix~\ref{apprelgrav}. The experimental constraints~(\ref{aa1}) and~(\ref{aa2}) from the two-body problem and~(\ref{nc1}),~(\ref{nc2}) and~(\ref{nc3}) from the Nordtvedt effect have the unique solution
\begin{eqnarray}\label{psol}
 ~\beta\,=\,\frac{3}{2},~~\gamma\,=\,-\frac{7}{2},~~\mu\,=\,0,~~\nu\,=\,-\frac{1}{2}
 ~~\mbox{and}~~\epsilon\,=\,-\frac{1}{2}\,.\hspace*{-0.5cm}\nonumber\\
\end{eqnarray}
 With the parameters in~(\ref{psol}) and in the special case where the reference frame is at rest relative to distant matter, the Lagrangian simplifies to
\begin{eqnarray}\label{leih}
 &&\hspace*{-0.8cm} L\,= \sum_{i\in S_n}\Big[\frac{1}{2}m_iv_i^2 + \frac{1}{8}m_i\frac{v_i^4}{c^2}\Big] \nonumber\\ && \hspace*{0.1cm}+ \!\!\sum_{i,j \in S_n, i<j}\!\!\frac{Gm_im_j}{r_{ij}}
\Big[1+ \frac{3(v_i^2+v_j^2)}{2c^2} - \frac{7{\bf v}_i.{\bf v}_j}{2c^2} \nonumber\\
 && \hspace*{3cm}-\,\frac{({\widehat {\bf r}}_{ij}.{\bf v}_i)({\widehat {\bf r}}_{ij}.{\bf v}_j)}{2c^2}\Big]\nonumber\\
 % \Big[1 + \Big(\frac{m}{24}-\frac{1}{8}\Big)\frac{v_{ij}^2}{c^2} - \frac{{\bf v}_i.{\bf v}_j}{2c^2}\nonumber\\
%  &&\hspace*{0.3cm} - \Big(\frac{a}{24}+\frac{1}{8}\Big)\frac{({\widehat {\bf r}}_{ij}.{\bf v}_{ij})^2}{c^2}
%  - \frac{({\widehat {\bf r}}_{ij}.{\bf v}_i)({\widehat {\bf r}}_{ij}.{\bf v}_j)}{2c^2} + \dots \Big]\nonumber\\
  && \hspace*{0.7cm}-\hspace*{0.3cm}\hspace*{-0.3cm}\sum_{i,j,k \in S_n,j\ne i,k\ne j}\hspace*{-0.2cm}\frac{G^2m_im_jm_k}{2r_{ij}r_{jk}c^2}\,+\,\dots\,,
\end{eqnarray}
 which is the same as the Lagrangian derived by Einstein, Infeld and Hoffmann~\cite{eih}. The Lagrangian~(\ref{ld3}) describing the motion of the string in its centre of mass frame is therefore consistent with experiment if
\begin{eqnarray}\label{psol2}
 a_1\,=\,\frac{39}{2}\hspace*{0.5cm}\mbox{and}\hspace*{0.5cm}b_1\,=\,-3\,.
\end{eqnarray}
 Equation~(\ref{kdef}) then gives
\begin{eqnarray}\label{kapsol}
 \kappa\,=\,\frac{3}{40}
\end{eqnarray}
 and~(\ref{ddef}) requires $a_2$ and $b_2$ to satisfy
\begin{eqnarray}\label{ab}
 a_2\,-\,\frac{b_2}{6}\,=\,-\frac{251}{120}\,.
\end{eqnarray}
 Equation~(\ref{kapsol}) implies that the total energy in the Machian strings of a space quantum is
\begin{eqnarray}\label{ffrac}
 \sum_{j\in S_d}\frac{Gm_im_j}{r_{ij}}\,=\,\frac{3}{40}m_ic^2\,,
\end{eqnarray}
 so that approximately $8\%$ of the energy of a space quantum is in the Machian strings and the remaining $92\%$ is at the centre.

%% file: appmotion4.tex
\section{Equations of motion}\label{appmotion4}
 Variation of the Lagrangian~(\ref{lag5}) with respect to ${\bf r}_i$ gives the equation of motion for $m_i$,
\begin{eqnarray}\label{em1}
 \hspace*{-0.1cm}\frac{\partial L}{\partial {\bf r}_i}\,-\,\frac{d}{dt}\Big(\frac{\partial L}{\partial {\bf v}_i}\Big)\,=\,0\,.
\end{eqnarray}
% In the following equations, sums are understood to refer to nearby particles only.
 For the mass $m_1$, the equation of motion is
\begin{eqnarray}\label{em2}
 && \hspace*{-0.7cm}-\sum_{i\in S_n}\frac{Gm_1m_i}{{r_{1i}}^3}{\bf r}_{1i}\Bigg\{1 +\beta\frac{(v_1^2+v_i^2)}{c^2}+\gamma\frac{{\bf v}_1.{\bf v}_i}{c^2}\nonumber\\
 && \hspace*{0.3cm}+~3\mu\frac{({\widehat {\bf r}}_{1i}.{\bf v}_1)^2 + ({\widehat {\bf r}}_{1i}.{\bf v}_i)^2}{c^2}
 +3\nu\frac{({\widehat {\bf r}}_{1i}.{\bf v}_1)({\widehat {\bf r}}_{1i}.{\bf v}_i)}{c^2}\Bigg\} \nonumber\\
 && \hspace*{-0.3cm}+~\sum_{i\in S_n}\frac{Gm_1m_i}{r_{1i}c^2}\Big[2\mu \frac{({\widehat {\bf r}}_{1i}.{\bf v}_1){\bf v}_1 + ({\widehat {\bf r}}_{1i}.{\bf v}_i){\bf v}_i}{r_{1i}}\nonumber\\
 && \hspace*{3cm}+~\nu\frac{({\widehat {\bf r}}_{1i}.{\bf v}_i){\bf v}_1 + ({\widehat {\bf r}}_{1i}.{\bf v}_1){\bf v}_i}{r_{1i}}\Big] \nonumber\\
 && \hspace*{1cm}+~\frac{\partial}{\partial {\bf r}_1}\left[\hspace*{0.2cm}\epsilon\hspace*{-0.3cm}\sum_{i,j,k\in S_n,j\ne i,k\ne j}\hspace*{-0.2cm}\frac{G^2m_im_jm_k}{r_{ij}r_{jk}c^2}\right] \nonumber\\
 && \hspace*{-0.5cm}- \frac{d}{dt}\Bigg\{
  m_1\Big(1 + \frac{v_{1d}^2}{2c^2}\Big){\bf v}_{1d} \nonumber\\
 && \hspace*{0.1cm} +~ \sum_{i\in S_n}\frac{Gm_1m_i}{r_{1i}c^2}\Big[2\beta{\bf v}_1 + \gamma{\bf v}_i + 2\mu ({\widehat {\bf r}}_{1i}.{\bf v}_1){\widehat {\bf r}}_{1i} \nonumber\\ && \hspace*{3.3cm} + ~\nu({\widehat {\bf r}}_{1i}.{\bf v}_i){\widehat {\bf r}}_{1i}\Big]\Bigg\}\,=\,0\,.
\end{eqnarray}

%% file: applaws4.tex
\section{Inertia and Newton's laws of motion}\label{applaws4}

 Suppose that the masses $m_1$ and $m_2$ are in a reference frame accelerating with respect to distant matter with acceleration ${\bf a}$. In the accelerated frame, distant matter has an acceleration $-{\bf a}$, so  $\ddot{\bf r}_d  =-{\bf a}$. Equation~(\ref{em2}) then has the form
\begin{eqnarray}\label{em5}
 m_1\ddot{\bf r}_1= -m_1{\bf a}-\frac{Gm_1m_2}{r^3}{\bf r}\,+\,\dots\,.~~
\end{eqnarray}
 Equation~(\ref{em5}) defines the total force acting on the mass $m_1$ in an accelerated reference frame. When $m_2=0$, 
the additional force that has to be added to the right hand side to give $\ddot{\bf r}_1=0$ is $m_1{\bf a}$. A force 
${\bf F}\,=\,m_1{\bf a}$ is therefore needed to maintain an acceleration ${\bf a}$ relative to distant matter,
 which is Newton's second law of motion.
%\begin{eqnarray}\label{uf}
% {\bf F}\,=\,-\frac{3\gamma+\nu}{6\beta+2\mu}m_1{\bf a}\,.
%\end{eqnarray}
% If inertial frames are defined relative to distant matter then, in an inertial frame, a particle requires an external force~(\ref{uf}) to maintain an acceleration ${\bf a}$. Newton's second law of motion, ${\bf F}=m{\bf a}$, is recovered if the parameters satisfy the equation
%\begin{eqnarray}\label{p1}
% 6\beta\,+\,3\gamma\,+\,2\mu\,+\,\nu\,=\,0\,.
%\end{eqnarray}
 Newton's first law of motion, that a particle moves with $\ddot{\bf r}_1=0$ in an inertial frame when there are no external forces, corresponds to the special case ${\bf a}=0$.

%% file: apprelgravb.tex
\section{Relativistic gravity}\label{apprelgrav}

\subsection{The two body problem}

 Consider the motion of two masses, $m_1$ and $m_2$, in an inertial frame. Setting $\ddot{\bf r}_d=0$ and ${\bf v}_d= 0$, the equation of motion~(\ref{em2}) for $m_1$ is, to first order in $v^2/c^2$,
\begin{eqnarray}\label{mem1}
 &&\ddot{\bf r}_1\, +\, \frac{1}{2}\Big(\ddot{\bf r}_1\frac{v_1^2}{c^2} + 2{\bf v}_1\frac{{\bf v}_1.\ddot{\bf r}_1}{c^2}\Big)\nonumber\\
&& \hspace*{-0.1cm}=\,-\frac{Gm_2}{r^2}{\widehat {\bf r}}\Big[1 +\beta\frac{(v_1^2+v_2^2)}{c^2}+\gamma\frac{{\bf v}_1.{\bf v}_2}{c^2}\nonumber\\
 && \hspace*{1.3cm}+~3\mu\frac{({\widehat {\bf r}}.{\bf v}_1)^2 + ({\widehat {\bf r}}.{\bf v}_2)^2}{c^2}
 +3\nu\frac{({\widehat {\bf r}}.{\bf v}_1)({\widehat {\bf r}}.{\bf v}_2)}{c^2}\Big]\nonumber\\
 && +~\frac{Gm_2}{rc^2}\Big[ 2\mu \frac{({\widehat {\bf r}}.{\bf v}_1){\bf v}_1 + ({\widehat {\bf r}}.{\bf v}_2){\bf v}_2}{r}\nonumber\\
 && \hspace*{3cm}+~\nu\frac{({\widehat {\bf r}}.{\bf v}_2){\bf v}_1 + ({\widehat {\bf r}}.{\bf v}_1){\bf v}_2}{r}\Big] \nonumber\\
 && +~\frac{Gm_2}{r^3c^2}({\bf r}.{\bf v})\Big[2\beta{\bf v}_1+
 \gamma{\bf v}_2 + 6\mu ({\widehat {\bf r}}.{\bf v}_1){\widehat {\bf r}}+3\nu({\widehat {\bf r}}.{\bf v}_2){\widehat {\bf r}}\Big]\nonumber\\
 && \hspace*{0.1cm}-\frac{Gm_2}{rc^2}\Big[2\beta\ddot{\bf r}_1+\gamma\ddot{\bf r}_2+
  2\mu\frac{({\bf v}.{\bf v}_1)}{r}{\widehat {\bf r}} +
  2\mu \frac{({\widehat {\bf r}}.{\bf v}_1)}{r}{\bf v} \nonumber\\
  && \hspace*{0.5cm}+~2\mu ({\widehat {\bf r}}.\ddot{\bf r}_1){\widehat {\bf r}} +\nu\frac{({\widehat {\bf r}}.{\bf v}_2)}{r}{\bf v}+\nu\frac{({\bf v}.{\bf v}_i)}{r}{\widehat {\bf r}}+\nu({\widehat {\bf r}}.\ddot{\bf r}_2){\widehat {\bf r}}\Big]\nonumber\\
 && \hspace*{1.3cm}-~\frac{2\epsilon G^2}{r^3c^2}(m_1m_2+m_2^2){\widehat {\bf r}}\,.
\end{eqnarray}
 After substituting the Newtonian approximations $\ddot{\bf r}_1= -Gm_2{\bf r}/r^3$ and $\ddot{\bf r}_2= Gm_1{\bf r}/r^3$, equation~(\ref{mem1}) reduces to
\begin{eqnarray}\label{mem2}
 && \hspace*{1cm}\ddot{\bf r}_1\,=\,A_1\,{\widehat {\bf r}}~ + ~B_1\,{\bf v}_1~+~B_2{\bf v}_2\,, \\
  \label{mem3} && \hspace*{-0.4cm}\mbox{with} \nonumber\\
 && \hspace*{-0.2cm} A_1\,=\,-\frac{Gm_2}{r^2} + \frac{G^2m_2}{r^3c^2}\Big[(2\beta+2\mu) m_2 -(\gamma+\nu)m_1\Big] \nonumber\\ && +~ \frac{Gm_2}{r^2c^2}\Big[-\beta(v_1^2+v_2^2)-\gamma{\bf v}_1.{\bf v}_2\nonumber\\
 && -~3\mu\left\{({\widehat {\bf r}}.{\bf v}_1)^2 + ({\widehat {\bf r}}.{\bf v}_2)^2\right\}
    + 6\mu({\widehat {\bf r}}.{\bf v})({\widehat {\bf r}}.{\bf v}_1) - 2\mu{\bf v}.{\bf v}_1\nonumber\\
 && -~3\nu({\widehat {\bf r}}.{\bf v}_1)({\widehat {\bf r}}.{\bf v}_2)-\nu({\bf v}.{\bf v}_2) + 3\nu({\widehat {\bf r}}.{\bf v})({\widehat {\bf r}}.{\bf v}_2)\Big]\nonumber\\ && \hspace*{2.5cm}+~\frac{Gm_2}{2r^2c^2}v_1^2 - \frac{2\epsilon G^2}{r^3c^2}(m_1m_2+m_2^2)
 \,,\nonumber\\
 && \hspace*{0.5cm}B_1\,=\,\frac{Gm_2}{r^2c^2}\Big[{\widehat {\bf r}}.{\bf v}_1 +2\beta{\widehat {\bf r}}.{\bf v}\Big]\nonumber\\
 && \hspace*{-0.4cm}\mbox{and} \hspace*{0.4cm} B_2\,=\,\frac{Gm_2}{r^2c^2}\Big[(\nu+2\mu)({\widehat {\bf r}}.{\bf v}_1 + {\widehat {\bf r}}.{\bf v}_2)+ \gamma{\widehat {\bf r}}.{\bf v}\Big].\nonumber\\
\end{eqnarray}
 The equation of motion for ${\bf r}$ in the centre of mass frame may be found by subtracting the corresponding equation for $m_2$ and substituting ${\bf v}_1= m_2{\bf v}/M$ and ${\bf v}_2= -m_1{\bf v}/M$, where $M= m_1+m_2$, which gives
\begin{eqnarray}\label{em2b}
 && \hspace*{1.5cm}\ddot{\bf r}\,=\,A\,{\widehat {\bf r}}~ + ~B\,({\widehat {\bf r}}.{\bf v}){\bf v}\,, \\
 && \hspace*{-0.4cm}\mbox{with} \hspace*{0.5cm} A\,=\,-\frac{GM}{r^2} + \frac{G^2}{r^3c^2}\Big[(2\beta+2\mu-2\epsilon) (m_1^2+m_2^2) \nonumber\\ && \hspace*{3.8cm}- \,2(\gamma + \nu + 2\epsilon)m_1m_2\Big] \nonumber\\
 &&\hspace*{0.2cm}+~ \frac{G}{Mr^2c^2}\Big[\{
 3\mu(m_1^2+m_2^2)-3\nu m_1m_2\}({\widehat{\bf r}}.{\bf v})^2 \nonumber\\
 &&\hspace*{1.8cm} - \left\{\!\Big(\beta +2\mu- \frac{1}{2}\Big)(m_1^2+m_2^2) \right.\nonumber\\ && \hspace*{2.8cm} \left.- \,\Big(2\nu +\gamma - \frac{1}{2}\Big)m_1m_2\right\}v^2\Big] \hspace*{-0.3cm}\nonumber\\
 && \hspace*{-0.4cm}\mbox{and} \hspace*{0.5cm} B\,=\,\frac{G}{Mr^2c^2}\Big[(1+2\beta)(m_1^2+m_2^2)
  \nonumber\\ &&  \hspace*{4cm} -~(1+2\gamma) m_1m_2 \Big]\,.
\end{eqnarray}
 Introducing polar coordinates for ${\bf r}$, the angular component of~(\ref{em2b}) may be integrated to give
 $r^2\dot\theta=h\exp[\int^r\! B\, dr^\prime]$, for some constant $h$, and the radial component becomes
\begin{eqnarray}\label{r}
 &&\hspace*{-0.3cm}\frac{d^2u}{d\theta^2}\,+\,u\,=\,-\frac{A}{h^2u^2}\,e^{-2\!\int^r\!B\,dr^\prime}\nonumber\\
 &&\hspace*{0.9cm}= \frac{GM}{h^2} + \frac{G^2u}{h^2c^2}\Big[(2+2\beta-2\mu+2\epsilon)(m_1^2+m_2^2)\nonumber\\
 && \hspace*{2.5cm}-~(2+2\gamma-2\nu-4\epsilon)m_1m_2\Big] \nonumber\\
 &&\hspace*{1.7cm}+~\frac{3G[\nu m_1m_2-\mu(m_1^2+m_2^2)]}{Mh^2c^2}\dot r^2 \nonumber\\ &&\hspace*{0.3cm}+\,\frac{G}{Mh^2c^2}\Big[\Big(\beta+2\mu-\frac{1}{2}\Big)(m_1^2+m_2^2)\nonumber\\ && \hspace*{2.1cm}-~\Big(\gamma+2\nu-\frac{1}{2}\Big)m_1m_2\Big]v^2\,,
\end{eqnarray}
 where $u= 1/r$. After using the energy conservation equation $v^2=2GMu + 2EM/m_1m_2$ to substitute for
 $v^2$, the rate of periastron precession is found from the coefficient of $u$ on the right hand side to be
\begin{eqnarray}\label{rate}
 && \hspace*{-1.2cm}\Delta\theta\,=\,\frac{\pi G^2}{h^2c^2}\Big[(1+4\beta+2\mu+2\epsilon)(m_1^2+m_2^2)\nonumber\\
 && \hspace*{1.1cm}-~(1+4\gamma+2\nu-4\epsilon)m_1m_2\Big]
\end{eqnarray}
 per orbit. The prediction of General Relativity is $\Delta\theta= 6\pi G^2M^2/h^2c^2$, so agreement with the
 experimental data for the precession of the perihelion of Mercury and the periastron of a binary pulsar
 requires
\begin{eqnarray}\label{aa1}
 && \hspace*{0.5cm}4\beta\,+\,2\mu\,+\,2\epsilon=\,5 \\ \hspace*{-0.6cm}
 \mbox{and}&&\hspace*{0.5cm}4\gamma\,+\,2\nu\,-\,4\epsilon=\,-13\,.\hspace*{1cm}\label{aa2}
\end{eqnarray}
% The bending of light by a body of mass $M$ may be calculated by considering the limit $m_1/m_2\rightarrow
% 0$ for which~(\ref{r}) gives, to first order,
%\begin{eqnarray}\label{bend}
% \hspace*{-0.3cm}\frac{d^2u}{d\theta^2}\,+\,u\,=\, \frac{GM}{h^2}\Big[1 + \Big(\beta+2\mu-\frac{1}{2}\Big)\frac{v^2}{c^2}\Big].
%\end{eqnarray}
% The experimental result that the bending of light is twice that for a non-relativistic particle is recovered
% if
%\begin{eqnarray}\label{aa3}
% \beta\,+\,2\mu\,=\,\frac{3}{2}\,,
%\end{eqnarray}
% since the right hand side of~(\ref{bend}) is then twice as large in the limit $v\rightarrow c$ as in the
% limit $v\rightarrow 0$.

\input{apprelgrav2.tex}

% The solution of equations~(\ref{p1}),~(\ref{p2})~(\ref{aa1}),~(\ref{aa2}) and~(\ref{aa3}) is
%\begin{eqnarray}\label{esol}
% \alpha= \frac{1}{6},~~ \beta= \frac{4}{3},~~ \gamma= -\frac{7}{6},~~ \mu= -\frac{9}{2}, ~~\nu= \frac{11}{36}~~~
%\end{eqnarray}
% and the resulting Lagrangian is
%\begin{eqnarray}\label{result}
% && \hspace*{-0.5cm}L=\sum_{i<j}\frac{m_im_j}{r_{ij}}\Big[1+\frac{1}{6}\frac{{\bf r}_{ij}.{\bf a}_{ij}}{c^2} + \frac{4}{3}\frac{(v_i^2+v_j^2)}{c^2} -\frac{7}{6}\frac{{\bf v}_i.{\bf v}_j}{c^2}\nonumber\\
% && \hspace*{0.8cm}-~\frac{9}{2}\frac{({\widehat {\bf r}}_{ij}.{\bf v}_i)({\widehat {\bf r}}_{ij}.{\bf v}_j)}{c^2}+\frac{11}{36}\frac{(v_i^4+v_j^4)}{c^4}\Big]\,.
%\end{eqnarray}
% The first three terms in the square brackets may be absorbed into the redefinition of the gravitational mass $m_g$ given in equation~(\ref{mgrav}).

%% file: apprelgrav2.tex
\subsection{The Nordtvedt effect}

 The absence of a Nordtvedt effect in the Earth-Moon-Sun system~\cite{shapiro} implies that the accelerations of the Earth and Moon towards the Sun are the same, so that the acceleration of a composite body towards the Sun is independent of its gravitational self-interaction energy. For a composite body consisting of two point masses, $m_i$ and $m_j$, separated by a distance of $r_{ij}$, the acceleration of the centre of mass towards a massive external body must therefore be independent of $Gm_im_j/r_{ij}$.

 Consider the motion of the two masses $m_i$ and $m_j$ in the presence of an external mass $m_e$. The acceleration of the centre of mass is given by
\begin{eqnarray}\label{a11}
 {\bf a}&=&\frac{m_i\ddot{\bf r}_i + m_j\ddot{\bf r}_j}{M} \nonumber\\
        &=& \frac{m_i}{M}\ddot{\bf r}_i ~~ + ~~ \Big\{ i \leftrightarrow j \Big\}\,,\label{ecm}
\end{eqnarray}
 where $M= m_i+m_j$. The acceleration $\ddot{\bf r}_i$ of $m_i$ in the presence of $m_j$ and $m_e$ may be found by extending equation~(\ref{mem1}) to the case of two nearby masses. The required generalisation of~(\ref{mem1}) may be written in the form
\begin{eqnarray}\label{n1}
 &&\hspace*{-0.4cm}\ddot{\bf r}_i\,=\, -\frac{1}{2c^2}(v_i^2\ddot{\bf r}_i + {\bf v}_i.\ddot{\bf r}_i{\bf v}_i)\nonumber\\
 && -~\frac{Gm_j}{r_{ij}c^2}\Big[2\beta\ddot{\bf r}_i+\gamma\ddot{\bf r}_j
  +2\mu ({\widehat {\bf r}}_{ij}.\ddot{\bf r}_i){\widehat {\bf r}}_{ij} +\nu({\widehat {\bf r}}_{ij}.\ddot{\bf r}_j){\widehat {\bf r}}_{ij}\Big]\nonumber\\
  && -~\frac{Gm_e}{r_{ie}c^2}\Big[2\beta\ddot{\bf r}_i+\gamma\ddot{\bf r}_e
  +2\mu ({\widehat {\bf r}}_{ie}.\ddot{\bf r}_i){\widehat {\bf r}}_{ie} +\nu({\widehat {\bf r}}_{ie}.\ddot{\bf r}_e){\widehat {\bf r}}_{ie}\Big]\nonumber\\
 && -\frac{2\epsilon G^2}{c^2}\Big\{\Big[\frac{m_im_j+m_j^2}{r_{ij}^4} + \frac{m_jm_e}{r_{ij^3}}\Big(\frac{1}{r_{ie}}+\frac{1}{r_{je}}\Big)\Big]{\bf r}_{ij}\nonumber\\
 && \hspace*{0.7cm}+~\Big[\frac{m_im_e+m_e^2}{r_{ie}^4} + \frac{m_jm_e}{r_{ie^3}}\Big(\frac{1}{r_{ij}}+\frac{1}{r_{je}}\Big)\Big]{\bf r}_{ie}\Big]\Big\}\nonumber\\
  && +~\,{\bf X}_{ij}~+~{\bf X}_{ie}\,,
\end{eqnarray}
 where ${\bf X}_{ij}$ denotes the remaining acceleration-independent terms of the right hand side of~(\ref{mem1}) corresponding to the interaction of $m_i$ and $m_j$, i.e.
\begin{eqnarray}\label{n2}
 && \hspace*{-0.3cm}{\bf X}_{ij}\,=\, -\frac{Gm_j}{r_{ij}^3}{\bf r}_{ij}\Big[1 +\beta\frac{(v_i^2+v_j^2)}{c^2}+\gamma\frac{{\bf v}_i.{\bf v}_j}{c^2}\nonumber\\
 && \hspace*{0.8cm}+~3\mu\frac{({\widehat {\bf r}}_{ij}.{\bf v}_i)^2 + ({\widehat {\bf r}}_{ij}.{\bf v}_j)^2}{c^2}
 +3\nu\frac{({\widehat {\bf r}}_{ij}.{\bf v}_i)({\widehat {\bf r}}_{ij}.{\bf v}_j)}{c^2}\Big]\nonumber\\
 && +~\frac{Gm_j}{r_{ij}c^2}\Big[2\mu \frac{({\widehat {\bf r}}_{ij}.{\bf v}_i){\bf v}_i + ({\widehat {\bf r}}_{ij}.{\bf v}_j){\bf v}_j}{r}\nonumber\\
 && \hspace*{3cm}+~\nu\frac{({\widehat {\bf r}}_{ij}.{\bf v}_j){\bf v}_i + ({\widehat {\bf r}}_{ij}.{\bf v}_i){\bf v}_j}{r_{ij}}\Big] \nonumber\\
 && +~\frac{Gm_j}{r_{ij}^3c^2}({\bf r}_{ij}.{\bf v}_{ij})\Big[2\beta{\bf v}_i+
 \gamma{\bf v}_j + 6\mu ({\widehat {\bf r}}_{ij}.{\bf v}_i){\widehat {\bf r}}_{ij}\nonumber\\
 && \hspace*{4.8cm}+~3\nu({\widehat {\bf r}}_{ij}.{\bf v}_j){\widehat {\bf r}}_{ij}\Big]\nonumber\\
 && \hspace*{0.1cm}-~\frac{Gm_j}{r_{ij}c^2}\Big[2\mu \frac{({\bf v}_{ij}.{\bf v}_i)}{r_{ij}}{\widehat {\bf r}}_{ij} + 2\mu \frac{({\widehat {\bf r}}_{ij}.{\bf v}_i)}{r_{ij}}{\bf v}_{ij} \nonumber\\
  && \hspace*{2cm}+~\nu\frac{({\widehat {\bf r}}_{ij}.{\bf v}_j)}{r_{ij}}{\bf v}_{ij}+\nu\frac{({\bf v}_{ij}.{\bf v}_j)}{r_{ij}}{\widehat {\bf r}}_{ij}\Big].
\end{eqnarray}
 The acceleration-dependent terms in~(\ref{n1}) may be evaluated by substituting the Newtonian approximations
\begin{eqnarray}\label{newt}
 &&\hspace*{-0.3cm}\ddot{\bf r}_i\,=\,-\frac{Gm_j}{r_{ij}^2}{\widehat {\bf r}}_{ij}-\frac{Gm_e}{r_{ie}^2}{\widehat {\bf r}}_{ie}\,, \nonumber\\
 &&\hspace*{-0.3cm}\ddot{\bf r}_j\,=\,\frac{Gm_i}{r_{ij}^2}{\widehat {\bf r}}_{ij}-\frac{Gm_e}{r_{je}^2}{\widehat {\bf r}}_{je}\,, \nonumber\\
  &&\hspace*{-1.6cm}\mbox{and}\hspace*{0.7cm}\ddot{\bf r}_e\,=\,\frac{Gm_i}{r_{ie}^2}{\widehat {\bf r}}_{ie}+\frac{Gm_j}{r_{je}^2}{\widehat {\bf r}}_{je}\,.
\end{eqnarray}
 Substituting the Newtonian approximation for $\ddot{\bf r}_i$ into the first line of~(\ref{n1}) gives
\begin{eqnarray}\label{ar1}
 \ddot{\bf r}_i&=&{\bf Y}_{ij}~+~{\bf Y}_{ie}\nonumber\\
  && \hspace*{-0.2cm}-\frac{2\epsilon G^2}{c^2}\Big\{\Big[\frac{m_im_j+m_j^2}{r_{ij}^4} + \frac{m_jm_e}{r_{ij^3}}\Big(\frac{1}{r_{ie}}+\frac{1}{r_{je}}\Big)\Big]{\bf r}_{ij}~~~~\nonumber\\
 && \hspace*{-0.3cm}+~\Big[\frac{m_im_e+m_e^2}{r_{ie}^4} + \frac{m_jm_e}{r_{ie^3}}\Big(\frac{1}{r_{ij}}+\frac{1}{r_{je}}\Big)\Big]{\bf r}_{ie}\Big]\Big\}\,,~
\end{eqnarray}
 where
\begin{eqnarray}\label{n3}
 && \hspace*{-0.2cm}{\bf Y}_{ij}\,=\, {\bf X}_{ij}\, + \frac{Gm_j}{2r_{ij}^2c^2}\left\{v_i^2{\widehat {\bf r}}_{ij} + ({\bf v}_i.{\widehat {\bf r}}_{ij}){\bf v}_i\right\} \nonumber\\
 && -~\frac{Gm_j}{r_{ij}c^2}\Big[2\beta\ddot{\bf r}_i+\gamma\ddot{\bf r}_j
  +2\mu ({\widehat {\bf r}}_{ij}.\ddot{\bf r}_i){\widehat {\bf r}}_{ij} +\nu({\widehat {\bf r}}_{ij}.\ddot{\bf r}_j){\widehat {\bf r}}_{ij}\Big].\nonumber\\
\end{eqnarray}
 To evaluate ${\bf Y}_{ij}$, it may be noted that if the parts of $\ddot{\bf r}_i$ and $\ddot{\bf r}_j$ proportional to ${\widehat {\bf r}}_{ij}$ in~(\ref{newt}) are substituted into~(\ref{n3}) then~(\ref{n3}) has the same form
\begin{eqnarray}
  A_1^j\,{\widehat {\bf r}}_{ij}~ + ~B_1^j\,{\bf v}_i~+~B_2^j{\bf v}_j
\end{eqnarray}
 as given in~(\ref{mem2}), where $A_1^j$, $B_1^j$ and $B_2^j$ are of the form corresponding to~(\ref{mem3}) for the interaction of $m_i$ and $m_j$, i.e.
\begin{eqnarray}\label{mem4}
 && \hspace*{-0.2cm} A_1^j\,=\,-\frac{Gm_j}{r_{ij}^2}+ \frac{G^2m_j}{r_{ij}^3c^2}\Big[(2\beta+2\mu) m_j  -(\gamma+\nu)m_i\Big] \nonumber\\ && +~ \frac{Gm_j}{r_{ij}^2c^2}\Big[-\beta(v_i^2+v_j^2)-\gamma{\bf v}_i.{\bf v}_j\nonumber\\
 && \hspace*{0.3cm}-~3\mu\left\{({\widehat {\bf r}}_{ij}.{\bf v}_i)^2 + ({\widehat {\bf r}}_{ij}.{\bf v}_j)^2\right\}
    + 6\mu ({\widehat {\bf r}}_{ij}.{\bf v}_{ij})({\widehat {\bf r}}_{ij}.{\bf v}_i) \nonumber\\
    && \hspace*{0.3cm}-~ 2\mu{\bf v}_{ij}.{\bf v}_i - 3\nu({\widehat {\bf r}}_{ij}.{\bf v}_i)({\widehat {\bf r}}_{ij}.{\bf v}_j)-\nu({\bf v}_{ij}.{\bf v}_j) \nonumber\\
    && \hspace*{2.7cm} +~ 3\nu({\widehat {\bf r}}_{ij}.{\bf v}_{ij})({\widehat {\bf r}}_{ij}.{\bf v}_j)\Big]\,+\,\frac{Gm_j}{2r_{ij}^2c^2}v_i^2\,,\nonumber\\
 && \hspace*{0.5cm}B_1^j\,=\,\frac{Gm_j}{r_{ij}^2c^2}\Big[{\widehat {\bf r}}_{ij}.{\bf v}_i +2\beta{\widehat {\bf r}}_{ij}.{\bf v}_{ij}\Big]\nonumber\\
 && \hspace*{-0.4cm}\mbox{and} \hspace*{0.4cm} B_2^j\,=\,\frac{Gm_j}{r_{ij}^2c^2}\Big[(\nu+2\mu)({\widehat {\bf r}}_{ij}.{\bf v}_i + {\widehat {\bf r}}_{ij}.{\bf v}_j) \nonumber\\
 && \hspace*{4.1cm}+~ \gamma {\widehat {\bf r}}_{ij}.{\bf v}_{ij}\Big].
\end{eqnarray}
 Adding the additional contributions to ${\bf Y}_{ij}$ from the parts of $\ddot{\bf r}_i$ and $\ddot{\bf r}_j$ in~(\ref{newt}) proportional to ${\widehat {\bf r}}_{ie}$ and ${\widehat {\bf r}}_{je}$, respectively, gives
\begin{eqnarray}\label{yij}
 && \hspace*{-0.8cm}{\bf Y}_{ij}\,=\, A_1^j\,{\widehat {\bf r}}_{ij}\, + \,B_1^j\,{\bf v}_i\,+\,B_2^j{\bf v}_j \nonumber\\
 && +~\frac{Gm_j}{r_{ij}c^2}\left\{\frac{Gm_e}{r_{ie}^2}\Big[2\beta{\widehat {\bf r}}_{ie}+2\mu({\widehat {\bf r}}_{ij}.{\widehat {\bf r}}_{ie}){\widehat {\bf r}}_{ij}\Big]\right.\nonumber\\
 && \left. \hspace*{1.2cm}+~\frac{Gm_e}{r_{je}^2}\Big[\gamma{\widehat {\bf r}}_{je}+\nu({\widehat {\bf r}}_{ij}.{\widehat {\bf r}}_{je}){\widehat {\bf r}}_{ij}\Big]\right\}.
\end{eqnarray}
 Similarly,
\begin{eqnarray}\label{yie}
 && \hspace*{-0.8cm}{\bf Y}_{ie}\,=\, A_1^e\,{\widehat {\bf r}}_{ie}\, + \,B_1^e\,{\bf v}_i\,+\,B_2^e{\bf v}_j \nonumber\\
 && +~\frac{Gm_e}{r_{ie}c^2}\left\{\frac{Gm_j}{r_{ij}^2}\Big[2\beta{\widehat {\bf r}}_{ij}+2\mu({\widehat {\bf r}}_{ie}.{\widehat {\bf r}}_{ij}){\widehat {\bf r}}_{ie}\Big]\right.\nonumber\\
 && \left. \hspace*{1.2cm}+~\frac{Gm_j}{r_{je}^2}\Big[\gamma{\widehat {\bf r}}_{je}-\nu({\widehat {\bf r}}_{ie}.{\widehat {\bf r}}_{je}){\widehat {\bf r}}_{ie}\Big]\right\}.
\end{eqnarray}

 To lowest order, the centre of mass acceleration~(\ref{ecm}) is equal to $-Gm_e\widehat{\bf r}/r^2$ where here, and in the remainder of this section, ${\bf r}$ denotes the position vector from $m_e$ to the centre of mass of $m_i$ and $m_j$. The Nordtvedt effect concerns corrections to $Gm_e/r^2$ that depend on the binding energy, $-Gm_im_j/r_{ij}$, of $m_i$ and $m_j$. Consider, therefore, the contributions to~(\ref{ecm}) proportional to $Gm_e/r^2$. The contribution to~(\ref{a11}) from the ${\mathcal O}(G^2)$ terms in~(\ref{ar1}) that are proportional to $m_e$ is
\begin{eqnarray}
  && \hspace*{-0.2cm}-\frac{2\epsilon G^2m_i}{Mc^2}\Big\{\Big[ \frac{m_jm_e}{r_{ij^3}}\Big(\frac{1}{r_{ie}}+\frac{1}{r_{je}}\Big)\Big]{\bf r}_{ij}\nonumber\\
 && \hspace*{-0.3cm}+~\Big[\frac{m_im_e}{r_{ie}^4} + \frac{m_jm_e}{r_{ie^3}}\Big(\frac{1}{r_{ij}}+\frac{1}{r_{je}}\Big)\Big]{\bf r}_{ie}\Big]\Big\}\,,\nonumber\\
 && \hspace*{1cm} + ~~ \Big\{ i \leftrightarrow j \Big\}
\end{eqnarray}
 and the $r_{ij}-$dependent part proportional to $Gm_e/r^2$ is
\begin{eqnarray}\label{cc4}
  && \hspace*{-0.2cm}-\frac{2\epsilon G^2m_im_jm_e}{Mr_{ij}c^2}\Big(\frac{{\bf r}_{ie}}{r_{ie}^3} + \frac{{\bf r}_{je}}{r_{je}^3}\Big) \nonumber\\
  &&\approx ~ -\frac{4\epsilon G^2m_im_jm_e}{Mr_{ij}r^2c^2}{\widehat {\bf r}}\,.
\end{eqnarray}
 The contribution to ${\bf Y}_{ij}$ proportional to $Gm_e/r^2$ is, from~(\ref{yij}),
\begin{eqnarray}
 \frac{Gm_jm_e}{r_{ij}r^2c^2}\Big[(2\beta+\gamma)\widehat{\bf r} + (2\mu+\nu)({\widehat {\bf r}}_{ij}.{\widehat {\bf r}}){\widehat {\bf r}}_{ij}\Big]
\end{eqnarray}
 and the corresponding contribution to~(\ref{ecm}) is
\begin{eqnarray}\label{cc1}
 \frac{2Gm_im_jm_e}{Mr_{ij}r^2c^2}\Big[(2\beta+\gamma)\widehat{\bf r} + (2\mu+\nu)({\widehat {\bf r}}_{ij}.{\widehat {\bf r}}){\widehat {\bf r}}_{ij}\Big]\,.~
\end{eqnarray}
 The expression~(\ref{yie}) for ${\bf Y}_{ie}$ contains contributions proportional to $Gm_e/r^2$ 
from $A_1^e\,{\widehat {\bf r}}_{ie}+ B_1^e\,{\bf v}_i+B_2^e{\bf v}_j$ and, from the other terms, contributions 
proportional to $Gm_e/r$. The contributions proportional to $Gm_e/r^2$ are
\begin{eqnarray}\label{c2a}
 && \hspace*{-0.2cm}\left\{-\frac{Gm_e}{r^2} + \frac{Gm_e}{r^2c^2}\Big[-\beta(v_i^2+v_e^2)\right.\nonumber\\
 && \hspace*{0.1cm}-~\gamma{\bf v}_i.{\bf v}_e-3\mu\left\{({\widehat {\bf r}}.{\bf v}_i)^2 + ({\widehat {\bf r}}.{\bf v}_e)^2\right\}
    + 6\mu ({\widehat {\bf r}}.{\bf v}_{ie})({\widehat {\bf r}}.{\bf v}_i) \nonumber\\
    && \hspace*{0.1cm}-~ 2\mu{\bf v}_{ie}.{\bf v}_i - 3\nu({\widehat {\bf r}}.{\bf v}_i)({\widehat {\bf r}}.{\bf v}_e)-\nu({\bf v}_{ie}.{\bf v}_e) \nonumber\\
    && \left.\hspace*{2.7cm} +~ 3\nu({\widehat {\bf r}}.{\bf v}_{ie})({\widehat {\bf r}}.{\bf v}_e)\Big]\,+\,\frac{Gm_e}{2r^2c^2}v_i^2 \right\}{\widehat {\bf r}}  \nonumber\\
 && \hspace*{0.2cm}+~\frac{Gm_e}{r^2c^2}\Big[{\widehat {\bf r}}.{\bf v}_i +2\beta{\widehat {\bf r}}.{\bf v}_{ie}\Big]{\bf v}_i\nonumber\\
 && \hspace*{0.2cm}+~\frac{Gm_e}{r^2c^2}\Big[(\nu+2\mu)({\widehat {\bf r}}.{\bf v}_i + {\widehat {\bf r}}.{\bf v}_e)+ \gamma{\widehat {\bf r}}.{\bf v}_{ie}\Big]{\bf v}_e.
\end{eqnarray}
 Suppose, for simplicity, that the masses $m_i$ and $m_j$ are in a circular orbit. Then $v_i^2$ is constant and equal to $Gm_j^2/r_{ij}$. The time averaged values of ${\widehat {\bf r}}.{\bf v}_i$ and
 ${\bf v}_i.{\bf v}_e$ are zero, so the time-averaged part of~(\ref{c2a}) that depends on $r_{ij}$ is
\begin{eqnarray}\label{c2b}
 && \hspace*{-0.8cm}-\frac{Gm_e}{r^2}{\widehat {\bf r}} \,+\, \frac{G^2m_j^2m_e}{Mr_{ij}r^2c^2}\Big[\,\frac{1}{2}-\beta -2\mu \nonumber\\ &&
 \hspace*{3cm}+~\,3\mu\langle({\widehat {\bf r}}.{\widehat{\bf v}}_i)^2\rangle \Big]\,{\widehat {\bf r}}\nonumber\\
 && \hspace*{0.2cm}+~\frac{G^2m_j^2m_e}{Mr_{ij}r^2c^2}(1+2\beta)\langle({\widehat {\bf r}}.{\widehat
 {\bf v}}_i){\widehat {\bf v}}_i\rangle\,,
\end{eqnarray}
 where the angled brackets denote the time average. After replacing ${\widehat{\bf v}}_i$ by ${\widehat{\bf v}}_{ij}$, the contribution of~(\ref{c2b}) to~(\ref{ecm}) is
\begin{eqnarray}\label{cc2}
 && \hspace*{-0.8cm}-\frac{Gm_e}{r^2}{\widehat {\bf r}} \,+\, \frac{G^2m_im_jm_e}{Mr_{ij}r^2c^2}\Big[\,\frac{1}{2} -\beta -2\mu \nonumber\\ &&
 \hspace*{2.8cm}+~\,3\mu\langle({\widehat {\bf r}}.{\widehat{\bf v}}_{ij})^2\rangle \Big]\,{\widehat {\bf r}}\nonumber\\
 && +~\frac{G^2m_im_jm_e}{Mr_{ij}r^2c^2}(1+2\beta)\langle({\widehat {\bf r}}.{\widehat
 {\bf v}}_{ij}){\widehat {\bf v}}_{ij}\rangle\,,
\end{eqnarray}
 so the contribution to ${\bf Y}_{ie}$ proportional to $Gm_e/r^2$ gives a contribution to~(\ref{ecm}) proportional to $Gm_e/r^2$. The contribution to ${\bf Y}_{ie}$ proportional to $Gm_e/r$ is
\begin{eqnarray}
 \frac{G^2m_jm_e}{r_{ij}^2r_{ie}c^2}\Big[2\beta{\widehat {\bf r}}_{ij} + 2\mu({\widehat {\bf r}}_{ie}.{\widehat {\bf r}}_{ij}){\widehat {\bf r}}_{ie}\Big]\nonumber
\end{eqnarray}
 which gives a contribution to~(\ref{ecm}) of
\begin{eqnarray}
 &&\hspace*{-0.5cm}\frac{G^2m_em_im_j}{Mr_{ij}^2c^2}\Big[2\beta\Big(\frac{1}{r_{ie}}-\frac{1}{r_{je}}\Big){\widehat {\bf r}}_{ij}\nonumber\\ \label{c3a}
 && + ~2\mu\Big\{\frac{({\widehat {\bf r}}_{ie}.{\widehat {\bf r}}_{ij})}{r_{ie}}{\widehat {\bf r}}_{ie} - \frac{({\widehat {\bf r}}_{je}.{\widehat {\bf r}}_{ij})}{r_{je}}{\widehat {\bf r}}_{je}\Big\}\Big]
 \nonumber\\ && =~\frac{G^2m_em_im_j}{Mr_{ij}^2c^2}\Big[f({\bf r}_i)\,-\,f({\bf r}_j)\Big]\,,  \\
 && \hspace*{-0.6cm}\mbox{where}\hspace*{0.5cm}f({\bf x})\,=\,2\beta\frac{{\widehat {\bf r}}_{ij}}{|{\bf x}-{\bf x}_e|} \nonumber\\ && \hspace*{0.3cm}+~ 2\mu\frac{({\bf x}-{\bf x}_e).{\widehat {\bf r}}_{ij}}{|{\bf x}-{\bf x}_e|^3}({\bf x}-{\bf x}_e).
\end{eqnarray}
 With the approximation $f({\bf r}_i)-f({\bf r}_j)=({\bf r}_i-{\bf r}_j).{\pmb{\nabla}}f({\bf r}_i)$, the time-averaged contribution to~(\ref{ecm}) from~(\ref{c3a}) becomes
\begin{eqnarray}\label{cc3}
 && \hspace*{-0.4cm}\frac{G^2m_em_im_j}{Mr_{ij}r^2c^2}\Big[2\mu{\widehat {\bf r}} + (2\mu-2\beta)\langle({\widehat {\bf r}}.{\widehat {\bf r}}_{ij}){\widehat {\bf r}}_{ij}\rangle\nonumber\\
 &&\hspace*{3cm} -~6\mu\langle({\widehat {\bf r}}.{\widehat {\bf r}}_{ij})^2\rangle{\widehat {\bf r}}\Big]\,,
\end{eqnarray}
 so the contribution to ${\bf Y}_{ie}$ proportional to $Gm_e/r$ also gives a contribution to~(\ref{ecm}) proportional to $Gm_e/r^2$. The time averages of $({\widehat {\bf r}}.{\widehat{\bf r}}_{ij})^2$ and $({\widehat {\bf r}}.{\widehat{\bf v}}_{ij})^2$ are equal to $(\sin^2\theta)/2$, where $\theta$ is the angle between ${\widehat {\bf r}}$ and the normal to the orbital plane of $m_i$ and $m_j$, and the time averages of $({\widehat {\bf r}}.{\widehat {\bf r}}_{ij}){\widehat {\bf r}}_{ij}$ and $({\widehat {\bf r}}.{\widehat {\bf v}}_{ij}){\widehat {\bf v}}_{ij}$ are equal to $(\sin^2\theta\,{\widehat {\bf r}}+\sin\theta\cos\theta \,{\bf n})/2$, where ${\bf n}$ is a unit vector perpendicular to ${\widehat {\bf r}}$ in the plane containing ${\widehat {\bf r}}$ and the normal to the orbital plane. The sum of~(\ref{cc4}),~(\ref{cc1}),~(\ref{cc2}) and~(\ref{cc3}) gives, for the total $r_{ij}-$dependent centre of mass acceleration proportional to $Gm_e/r^2$,
\begin{eqnarray}\label{ct}
 && \hspace*{-0.8cm}-\frac{Gm_e}{r^2}{\widehat {\bf r}} \,+\, \frac{G^2m_im_jm_e}{Mr_{ij}r^2c^2}\Big[\Big\{\frac{1}{2} +3\beta +2\gamma-4\epsilon \nonumber\\ &&
 \hspace*{4cm}-\,\frac{3}{2}\mu\sin^2\theta \Big\}\,{\widehat {\bf r}}\nonumber\\
 && +~\frac{1}{2}(1+6\mu+2\nu)(\sin^2\theta\,{\widehat {\bf r}}+\sin\theta\cos\theta \,{\bf n})\Big]\,.\nonumber\\
\end{eqnarray}
 The constraints on the parameters needed to ensure the absence of the Nordtvedt effect are therefore
\begin{eqnarray}\label{nc1}
 && \hspace*{0.2cm}3\beta\,+\,2\gamma\,-\,4\epsilon\,=\,-\frac{1}{2}\,, \\ \hspace*{-0.6cm}\label{nc2}
 && \hspace*{1cm}\mu\,=\,0 \\ \hspace*{-0.6cm}
 \mbox{and}&&\hspace*{1cm}\nu\,=\,-\frac{1}{2}\,.\hspace*{1cm}\label{nc3}
\end{eqnarray} 

%% file: appprecess.tex
\section{The precession of gyroscopes}\label{appprecess}
 The precession of a gyroscope in orbit around the Earth may be calculated using the EIH Lagrangian~(\ref{leih}). The method is described in~\cite{landau} and explicit calculations are given here for completeness.
\subsection{Geodetic precession}\label{geo}
  Consider a gyroscope rotating with angular velocity ${\pmb \omega}$ and moving with velocity ${\bf V}$ relative to the Earth. The Lagrangian for the gyroscope may be obtained by integrating the Lagrangian~(\ref{llight}) for a point particle over the mass distribution of the gyroscope. A point with position vector ${\bf x}$ relative to the centre of mass of the gyroscope has velocity $v({\bf x})= {\bf V}+ {\pmb \omega}\times{\bf x}$ relative to the Earth. The geodetic precession comes from the change in the Lagrangian that is proportional to the angular velocity of the gyroscope and the mass $M$ of the Earth. After substituting into~(\ref{llight}), the last term in~(\ref{llight}) becomes
\begin{eqnarray}
 \frac{3GM}{2c^2}\int d^3{\bf x}\,
 \frac{\rho({\bf x})v({\bf x})^2}{|{\bf r}+{\bf x}|}\,,
\end{eqnarray}
 where $\rho({\bf x})$ is the mass density of the gyroscope and ${\bf r}$ is the position vector of the centre of the gyroscope relative to the centre of the Earth. The required change in the Lagrangian is therefore
\begin{eqnarray}\label{lin}
 \delta L &=& \frac{3GM}{c^2}\int d^3{\bf x}\,
 \frac{\rho({\bf x}){\bf V}.({\pmb \omega}\times
 {\bf x})}{|{\bf r}+{\bf x}|}\nonumber\\
 &=& \frac{3GM}{c^2}V_i\epsilon_{ijk}\omega_j\int d^3{\bf x}\,
 \rho({\bf x})x_k
  \left(\frac{1}{r}-\frac{{\bf r}.{\bf x}}{r^3}+\dots\right).\nonumber\\
\end{eqnarray}
 The first term vanishes by definition of the centre of mass and, if the gyroscope is assumed to be a sphere,
\begin{eqnarray}\label{sec}
 &&\hspace*{-0.2cm}\int d^3{\bf x}\,\rho({\bf x})x_ix_j = \frac{1}{2}I \delta_{ij}\,,\hspace*{1cm}\\
 \mbox{where}\hspace*{0.5cm}&& I= \frac{2}{3}\int d^3{\bf x}\,\rho({\bf x})x^2
\end{eqnarray}
 is the moment of inertia.
%\begin{eqnarray}
% && I_{ij}=\int d^3{\bf x}\,\rho({\bf x})(x^2\delta_{ij}-x_ix_j)=
% I\,\delta_{ij}, \nonumber\\\mbox{where}\hspace*{0.2cm}&& I=
%  \frac{2}{3}\int d^3{\bf x}\,\rho({\bf x})x^2,
%\end{eqnarray}
 Equation~(\ref{lin}) then gives
\begin{eqnarray}\label{lin2}
 \delta L &=& -\frac{3IGM}{2r^3c^2}V_i\epsilon_{ijk}\omega_jx_k\nonumber\\
 &=& -\frac{3GM}{2r^3c^2}{\bf S}.({\bf r}\times{\bf V}),
\end{eqnarray}
 where ${\bf S}$ is the angular momentum of the gyroscope. The additional
term~(\ref{lin2}) in the Lagrangian gives a precession of the gyroscope with angular velocity
\begin{eqnarray}
 {\bf \Omega}_{geo}= \frac{3GM}{2r^3c^2}({\bf r}\times{\bf V}),
\end{eqnarray}
 in agreement with the experimental value.

\subsection{Gravitomagnetic precession}\label{gm}
 When the rotation of the Earth is taken into account there is an additional contribution to the rate of gyroscope precession, known as gravitomagnetic precession, associated with the angular momentum of the Earth. Let the angular velocity of the Earth be ${\pmb \omega}_E$ and a consider mass element $m_1$ in the gyroscope with position vector ${\bf x}$ relative to the centre of the gyroscope and a mass element $m_2$ in the Earth with position vector ${\bf x}_E$ relative to the centre of the Earth. The velocities of the two mass elements are $v_1({\bf x})= {\bf V}+ {\pmb \omega}\times{\bf x}$ and $v_2({\bf x}_E)= {\pmb \omega}_E\times{\bf x}_E$, respectively, and their relative position vector is ${\bf r}_{12}= {\bf r}+{\bf x}-{\bf x}_E$. The terms in the Lagrangian~(\ref{leih}) that are proportional to $Gm_1m_2/r$ and quadratic in the velocities ${\bf v}_1$ and ${\bf v}_2$ are
\begin{eqnarray}
 \frac{Gm_1m_2}{r_{12}c^2}\Big[\frac{3}{2}(v_1^2 + v_2^2)-\frac{7}{2}{\bf v}_1.{\bf v}_2 -
\frac{1}{2}(\widehat{\bf r}_{12}.{\bf v}_1)(\widehat{\bf r}_{12}.{\bf v}_2)\Big]\,.\nonumber\\
\end{eqnarray}
 After substituting for ${\bf v}_1$ and ${\bf v}_2$ and integrating over the mass densities $\rho({\bf x})$ and $\rho_E({\bf x}_E)$, the terms containing factors of both ${\pmb \omega}$ and ${\pmb \omega}_E$ are
%\begin{eqnarray}\label{comb}
% \delta L&=& \frac{3G}{2c^2}\int\!\int d^3{\bf x}\,d^3{\bf x}_E \,
% \rho({\bf x})\rho_E({\bf x_E})\nonumber\\
% && \hspace*{1cm}\times\, \frac{|{\bf V}+({\pmb \omega}\times
% {\bf x})-({\pmb \omega}_E\times {\bf x}_E)|^2}
% {|{\bf r}+{\bf x}-{\bf x}_E|}.
%\end{eqnarray}
% The terms proportional to $|{\bf V}|^2$, $|{\pmb \omega}\times
% {\bf x}|^2$ and $|{\pmb \omega}_E\times {\bf x}_E|^2$ represent the
% corrections to the mass energies of the gyroscope and the Earth
% corresponding to the translational kinetic energy of the gyroscope,
% the rotational kinetic energy of the gyroscope and the rotational
% kinetic energy of the Earth, respectively. The term proportional to
% ${\bf V}.({\pmb \omega}\times {\bf x})$ correpsonds to the geodetic
% precession of the gyroscope calculated in Section~\ref{geo} and the
% term proportional to ${\bf V}.({\pmb \omega}_E\times {\bf x}_E)$
% corresponds to the (much smaller) geodetic precession of the Earth.
% The remaining contribution to equation~(\ref{comb}) is
\begin{eqnarray}\label{rem}
 \delta L &=& -\frac{G}{2c^2}\int\!\int d^3{\bf x}\,d^3{\bf x}_E \,
 \rho({\bf x})\rho_E({\bf x}_E)\, \nonumber\\
 &&\hspace*{0.2cm} \left[ 7 \frac{({\pmb \omega}\times {\bf x}).
 ({\pmb \omega}_E\times {\bf x}_E)}{r_{12}} \right. \nonumber\\
 && \hspace*{0.5cm}\left. +~ \frac{{\bf r}_{12}.({\pmb \omega}\times {\bf x})\,{\bf r}_{12}.({\pmb \omega}_E\times {\bf x}_E)}{r_{12}^3}\right]\nonumber\\
 &=& -\frac{G}{2c^2}\int d^3{\bf x}\,\rho({\bf x})I({\bf x})
\end{eqnarray}
 where, after defining ${\bf R}= {\bf r}+{\bf x}$, the integral $I({\bf x})$ may be written in the form
\begin{eqnarray}\label{i}
 &&\hspace*{-0.4cm} I({\bf x})\,=\, \int d^3{\bf x}_E \,\rho_E({\bf x}_E) \left[ 7 \frac{({\pmb \omega}\times {\bf x}). ({\pmb \omega}_E\times {\bf x}_E)}{r_{12}} \right. \nonumber\\
 && \hspace*{-0.3cm}\left. +~ \frac{{\bf R}.({\pmb \omega}\times {\bf x})\,{\bf R}.({\pmb \omega}_E\times {\bf x}_E)}{r_{12}^3}- \frac{{\bf x}_E.({\pmb \omega}\times {\bf x})\,{\bf R}.({\pmb \omega}_E\times {\bf x}_E)}{r_{12}^3}
 \right]\!.\hspace*{-0.5cm}\nonumber\\
\end{eqnarray}
 The integral~(\ref{i}) may be evaluated as a power series in $1/R$ using the expansions
\begin{eqnarray}
 &&\hspace*{-0.3cm} \frac{1}{r_{12}}\,=\, \frac{1}{R} \,+\, \frac{\widehat{\bf R}.{\bf x}_E}{R^2}\,+\,\frac{3(\widehat{\bf R}.{\bf x}_E)^2-x_E^2}{2R^3} \,+\,\dots \nonumber\\
 \mbox{and}\hspace*{0.3cm}&& \frac{1}{r_{12}^3}\,=\, \frac{1}{R^3} \,+\, \frac{3\widehat{\bf R}.{\bf x}_E}{R^4}\,+\,\dots\,.
\end{eqnarray}
 Assuming the Earth to be spherical, the integral of a term containing two factors of ${\bf x}_E$ may be evaluated using the formula analagous to~(\ref{sec}) and the integral of a term containing an odd number of factors of ${\bf x}_E$ is zero.
% and the formula
%\begin{eqnarray}\label{fi1}
% \int d^3{\bf x}_E\,\rho({\bf x}_E)({\bf a}.{\bf x}_E)
%  ({\bf b}.{\bf x_E})= \frac{1}{2}I_E({\bf a}.{\bf b})\,,
%\end{eqnarray}
% where $I_E$ is the moment of inertia of the Earth. Noting that the integral of a term containing an odd number of powers of ${\bf x}_E$ is zero,
 The expansion of~(\ref{i}) reduces to
\begin{eqnarray}
 I({\bf x})\,=\, 4I_E\frac{\widehat{\bf R}.[({\pmb \omega}\times {\bf x})\times {\bf \omega}_E]}{R^2}\,,
\end{eqnarray}
 to order $1/R^2$, and substituting back into~(\ref{rem}) then gives
\begin{eqnarray}\label{rem2}
 \delta L\,=\, -\frac{2GI_E}{c^2}\int d^3{\bf x}\,\rho({\bf x})
 \frac{\widehat{\bf R}.[({\pmb \omega}\times {\bf x})\times {\bf \omega}_E]}{R^2}\,.~~
\end{eqnarray}
 The integral~(\ref{rem2}) may similarly be evaluated in powers of $1/r$ using the expansion
\begin{eqnarray}
 \frac{\widehat{\bf R}}{R^2}\,=\,\frac{\widehat{\bf r}}{r^2}\,+\,
 \frac{{\bf x}- 3\widehat{\bf r}(\widehat{\bf r}.{\bf x})}{r^3}\,+\,\dots\,,
\end{eqnarray}
 which gives
\begin{eqnarray}\label{rem3}
 && \hspace*{-1cm}\delta L\,=\, \frac{2GI_E}{r^3c^2}\int d^3{\bf x}\,\rho({\bf x})\,\Big\{[{\pmb \omega}({\pmb \omega}_E.{\bf x})- {\bf x}({\pmb \omega}.{\pmb \omega}_E)]\nonumber\\
&&\hspace*{3.2cm} .[{\bf x}- 3\widehat{\bf r}(\widehat{\bf r}.{\bf x})]\Big\}\,.
\end{eqnarray}
 After applying~(\ref{sec}), equation~(\ref{rem3}) reduces to
\begin{eqnarray}
 \delta L&=& \frac{GII_E}{r^3c^2}[{\pmb \omega}.{\pmb \omega}_E - 3(\hat{\bf r}.{\pmb \omega})(\hat{\bf r}.
 {\pmb \omega}_E)]\nonumber\\
 &=& \frac{G}{r^3c^2}{\bf S}.[{\bf J} - 3\hat{\bf r}(\hat{\bf r}.{\bf J})],
\end{eqnarray}
 where ${\bf S}$ is the angular momentum of the gyroscope and ${\bf J}$ is the angular momentum of the Earth. The rate of gravitomagnetic precession is therefore
\begin{eqnarray}
 {\bf\Omega}_{gm}= \frac{G}{r^3c^2}[3\hat{\bf r}(\hat{\bf r}.{\bf J})
  -{\bf J}],
\end{eqnarray}
 in agreement with the standard result.
%\begin{eqnarray}
% &&\int d^3{\bf x}\,\rho({\bf x})({\bf a}.{\bf x})({\bf b}.{\bf x})
% = \frac{1}{2}I({\bf a}.{\bf b}) \nonumber\\
% \mbox{and}\hspace*{0.5cm}&&
% \int d^3{\bf x}_E\,\rho({\bf x}_E)({\bf a}.{\bf x}_E)
%  ({\bf b}.{\bf x_E})= \frac{1}{2}I_E({\bf a}.{\bf b}),\nonumber
%\end{eqnarray}
% for arbitrary constant vectors ${\bf a}$ and ${\bf b}$, where $I$ and
% $I_E$ are the moments of inertia of the gyroscope and the Earth,
% respectively, the expression~(\ref{rest}) reduces to 

%% file: applight2.tex
\section{The propagation of light}\label{applight2}
 Consider a test mass $m$ moving in the gravitational field of a mass $M$. After setting the velocity of the mass $M$ equal to zero, the Lagrangian~(\ref{leih}) for the mass $m$, in a frame at rest relative to distant matter is
\begin{eqnarray}\label{llight}
  L\,= \, \frac{1}{2}m v^2  + \frac{1}{8}\frac{mv^4}{c^2} \,+\, \frac{GmM}{r}\Big(1+ \frac{3v^2}{2c^2}\Big)\,,
\end{eqnarray}
 to lowest order in $GM/rc^2$.
% The corresponding Hamiltonian is
%\begin{eqnarray}\label{hlight}
%  H\,= \,\frac{1}{2}m v^2  + \frac{3}{8}\frac{mv^4}{c^2} \,-\, \frac{GmM}{r}\Big(1- \frac{3v^2}{2c^2}\Big)\,.
%\end{eqnarray}
 After subtracting the constant $mc^2$, the Lagrangian~(\ref{llight}) may be written in the form
\begin{eqnarray}\label{llight2}
  L\,= \, -\Big[1-\frac{v^2}{c(r)^2}\Big]^{1/2}m(r)c(r)^2\,,
\end{eqnarray}
 for suitable functions $m(r)$ and $c(r)$, which is a generalisation of the Lagrangian $-mc^2/\gamma$ for a free particle. Expanding the gamma factor in~(\ref{llight2}) and comparing with~(\ref{llight}) gives
\begin{eqnarray}
  && \hspace*{-0.5cm}m(r)c(r)^2\,=\,mc^2\Big(1-\frac{GM}{rc^2}\Big)\label{lm} \hspace*{0.5cm}\\
  \mbox{and}\hspace*{0.5cm}&& m(r)\,=\,m\Big(1+\frac{3GM}{rc^2}\Big)\,,\label{lmc}
\end{eqnarray}
 where~(\ref{lm}) comes from equating the terms independent of $v$ and~(\ref{lmc}) comes from equating the terms proportional to $v^2$. Combining~(\ref{lm}) and(\ref{lmc}) gives, to lowest order in $GM/rc^2$,
\begin{eqnarray}
 c(r)\,=\,c\Big(1-\frac{2GM}{rc^2}\Big)\label{lc}\,.
\end{eqnarray}
 The momentum corresponding to the Lagrangian~(\ref{llight2}) is
\begin{eqnarray}\label{mom}
 {\bf p}\,=\, m(r){\bf v}\Big[1-\frac{v^2}{c(r)^2}\Big]^{-1/2}
\end{eqnarray}
 and the Hamiltonian is
\begin{eqnarray}\label{hlight2}
  H\,= \, \Big[1-\frac{v^2}{c(r)^2}\Big]^{-1/2}m(r)c(r)^2\,.
\end{eqnarray}

 A photon may be considered as the zero-mass limit of a massive particle. If the limit is taken so that the energy remains constant then equation~(\ref{hlight2}) implies that the velocity $v$ tends to $c(r)$, so $c(r)$ is identified as the speed of light in a gravitational field. The variational principle needed to calculate the path of a photon may be derived by considering the zero-mass limit of Hamilton's principle. Since the speed of the photon is given by $c(r)$ on the varied path as well as on the actual path, the times at the endpoints are not fixed during the variation. It is therefore necessary to rewrite Hamilton's principle as a principle of least action~\cite{yourgrau}, which states that
\begin{eqnarray}\label{pla}
 \delta \int {\bf p}.d{\bf x}\,=\,0
\end{eqnarray}
 when the path is varied so as to keep the endpoints fixed and the energy, $H$, constant. Now ${\bf p}.d{\bf x}= {\bf p}.\widehat{\bf v}ds$ and, in the limit $v\rightarrow c(r)$, equation~(\ref{mom}) gives
\begin{eqnarray}\label{pv}
 {\bf p}.\widehat{\bf v}\,=\, \frac{Hv}{c(r)^2} \rightarrow \frac{H}{c(r)}\,.
\end{eqnarray}
 The variational principle therefore becomes
\begin{eqnarray}\label{flt}
 \delta \int \frac{ds}{c(r)}\,=\,0\,,
\end{eqnarray}
 where the photon frequency is constant along the path, which is Fermat's principle of least time.

 A photon may also be considered as a localised quantum of the electromagnetic field since, as is well known~\cite{born}, Fermat's principle is the geometric optics limit of Maxwell's equations for an electromagnetic wave in a dielectric medium in which the speed of light is $c(r)$. However, the propagation of electromagnetic waves through the entire coordinate space is not consistent with the string model, since physical space is restricted to the strings. For consistency, electrodynamics should be formulated as a direct action theory, in which charged particles interact directly with each other via the strings and there are no independent degrees of freedom associated with the electromagnetic field. Classical electrodynamics can in fact be reformulated as a direct action theory by eliminating the electromagnetic field variables from the action~\cite{darwin}. For two particles, with charges $q_1$ and $q_2$, the resulting Lagrangian to ${\mathcal O}(v^2/c^2)$ is
\begin{eqnarray}\label{darwin}
  && \hspace*{-1cm}L = \sum_i \Big[\frac{1}{2}m_iv_i^2 + \frac{1}{8}m_i\frac{v_i^4}{c^2}\Big]-\frac{q_1q_2}{r}
 \nonumber\\ &&\hspace*{0.8cm} +~\frac{q_1q_2}{2rc^2}\Big[{\bf v}_1.{\bf v}_2 + (\widehat{\bf r}.{\bf v}_1)
 (\widehat{\bf r}.{\bf v}_2)\Big]\,,
\end{eqnarray}
 which is known as the Darwin Lagrangian. The Lagrangian~(\ref{darwin}) can be included in the formalism of Appendix~\ref{applorentz2} by taking $\Gamma= -q_iq_j/4\pi\epsilon_0$, $a_1= 3/2$ and $b_1= -3$ in the Lagrangian~(\ref{ld5}) .

% According to Evans {\it et al}, AJP {\bf 64} 1404 (96), zero variation of $\int L \,dt$ with fixed positions and times at the endpoints is equivalent to zero variation of $\int {\bf p}.{\bf v} \,dt$ with fixed positions at the endpoints and $H$ held constant, i.e. to the zero variation of $\int {\bf p}.\widehat{\bf v} \,ds$. From~(\ref{mom}), we have
%\begin{eqnarray}\label{pv}
% {\bf p}.\widehat{\bf v}\,=\, mv\Big(1 + \frac{v^2}{2c^2} + \frac{3GM}{rc^2}\Big)\,.
%\end{eqnarray}
% Now consider the zero mass limit. To keep the Hamiltonian~(\ref{hlight2}) constant we require $v\rightarrow c(r)$, so
%\begin{eqnarray}\label{pv2}
% {\bf p}.\widehat{\bf v}\,\rightarrow\, \frac{H}{c}\Big(1+\frac{2GM}{rc^2}\Big)
%\end{eqnarray}
% and the variational principle becomes
%\begin{eqnarray}\label{flt}
% \delta \int \frac{ds}{c(r)}\,=\,0\,,
%\end{eqnarray}
% where $ds$ is the Euclidean path length, which is Fermat's principle of least time. Fermat's principle is known to be the geometric optics limit of Maxwell's equations~\cite{born}, so the Maxwell equations are those for a dielectric medium in which the speed of light is $c(r)$.

\subsection{Gravitational redshift}\label{gred}
 Equation~(\ref{lm}) shows that the mass-energy of a test mass at a distance $r$ from a mass $M$ is multiplied by a factor $1-GM/rc^2$. If the test mass is an atom, all the energy levels of the atom are multiplied by the same factor and the frequency of light emitted by the atom for a given atomic transition therefore has the form $f=f_0(1-GM/rc^2)$, where $f_0$ is the frequency of light for to the same atomic transition in the absence of the mass $M$. A photon emitted by an atom at a distance $r$ from the mass $M$ therefore propagates with a constant frequency $f_0(1-GM/rc^2)$. Compared to the frequency corresponding to the same transition in an atom receiving the photon at a distance $r+\Delta r$, the frequency of the photon is smaller by an amount $\Delta f$, given by
\begin{eqnarray}
 \frac{\Delta f}{f_0}&=&\frac{GM}{rc^2}-\frac{GM}{(r+\Delta r)c^2} \nonumber\\
 &\approx& \frac{GM}{rc^2}\frac{\Delta r}{r}\,,
\end{eqnarray}
 in agreement with the usual formula. The explanation for the gravitational redshift is essentially the same as in the dielectric model for gravity considered by Dicke~\cite{dicke2}.

\subsection{The bending of light}\label{lbend}
 If $\eta$ denotes an arbitrary parameterisation of the photon path then, in plane polar coordinates, the path length in equation~(\ref{flt}) is given by $ds= \sqrt{\dot r^2 + r^2\dot\theta^2}\,d\eta$. The corresponding Lagrangian, $L$, is independent of $\dot{\theta}$ and $\partial L/\partial \dot{\theta}=$ constant\, gives
\begin{eqnarray}
 \frac{r^2\dot{\theta}}{\sqrt{\dot{r}^2+r^2\dot{\theta}^2}}
  =A \Big(1 - \frac{2GM}{rc^2}\Big),\nonumber
\end{eqnarray}
 where $A$ is a constant. Thus, to lowest order in $GM/rc^2$,
\begin{eqnarray}
 &&\Big(\frac{dr}{d\theta}\Big)^2 + r^2 =
 \frac{r^4}{A^2}\Big(1 + \frac{4GM}{rc^2}\Big)~~~\nonumber\\
 \Rightarrow\hspace*{0.5cm}&&\Big(\frac{du}{d\theta}\Big)^2 + u^2
 = \frac{1}{A^2}\Big(1 + \frac{4GM}{rc^2}\Big),~~~
\end{eqnarray}
 where $u=1/r$. The zeroth order solution is $u= u_0 =
 A^{-1}\sin(\theta-\theta_0)$, so the constant $A$ is identified as the distance of
 closest approach. The first order solution is $u=u_0+u_1$ where $u_1$
 satisfies
\begin{eqnarray}
 \frac{du_1}{d\theta} \cos (\theta-\theta_0) + u_1 \sin
 (\theta-\theta_0) = \frac{2GM}{A^2c^2} \sin (\theta-\theta_0).\nonumber
\end{eqnarray}
 Dividing by $\cos^2 (\theta-\theta_0)$ and integrating gives
\begin{eqnarray}
 u = \frac{1}{A}\sin (\theta-\theta_0) + \frac{2GM}{A^2c^2}[1 + a
 \cos (\theta-\theta_0)],~~
\end{eqnarray}
 where $a$ is a constant of integration. The values of
 $\theta-\theta_0$ corresponding to $u=0$ are, approximately,
 $-2GM/Ac^2$ and $\pi+2GM/Ac^2$. The angle of deflection of the
 light ray is $\Delta \theta = 4GM/Ac^2$, which is twice the Newtonian
 value, in agreement with observation.

\subsection{Radar time delay}\label{radar}
 Experiments have shown that the time taken for a radar beam to travel
 from the Earth to a satellite is slightly increased if the beam passes
 close to the Sun. Let the distance of closest approach of the radar
 beam to the Sun be $d$ and let the distances of the Earth and the
 satellite from the point of closest approach be $r_e$ and $r_s$,
 respectively. If the beam travels to the satellite and is reflected
 back to Earth, the time delay due to the presence of the Sun is found,
 experimentally, to agree with the formula
\begin{eqnarray}\label{del}
 \delta t \approx \frac{4GM}{c^3}\,\ln\Big(\frac{4r_er_s}{d^2}\Big)
 \hspace*{0.5cm} (d\ll r_e,r_s),
\end{eqnarray}
 where $M$ is the mass of the Sun. The total time delay due to the reduction in the speed of light from $c$ to $c(r)$
 is given, from~(\ref{flt}), by
\begin{eqnarray}\label{delt}
 \delta t &=&  \int \Big[\frac{1}{c(r)} - \frac{1}{c}\Big]\,ds \nonumber\\
 &=&\frac{2}{c}\int_0^{\,r_e} \frac{2GM}{rc^2}\,ds +
 \frac{2}{c}\int_0^{\,r_s}\frac{2GM}{rc^2}\,ds\,,~~~
\end{eqnarray}
 to lowest order in $GM/rc^2$, where $s$ is the path length measured from the point of closest
 approach to the Sun. If $\theta$ is the angle between the direction of
 propagation of the radar beam and the direction of the Sun then
 $s=d\cot\theta$ and $r=d\,\mbox{cosec}\,\theta$ and the first integral becomes
\begin{eqnarray}
 && \frac{4GM}{c^3} \int_{0}^{\,r_e} \frac{ds}{r}\nonumber\\
 =&& 4\frac{GM}{c^3} {\Big[\ln(\cot\theta +
  \mbox{cosec}\,\theta)\Big]\,}_{\pi/2}^{\cot^{-1}(r_e/d)}\nonumber\\
 \approx&& \frac{4GM}{c^3}\,\ln\Big(\frac{2r_e}{d}\Big).
\end{eqnarray}
 The total time delay of the radar beam is therefore
\begin{eqnarray}
 \delta t \approx \frac{4GM}{c^3}\,\ln\Big(\frac{4r_er_s}{d^2}\Big),
\end{eqnarray}
 in agreement with~(\ref{del}).

%% file: appke3.tex
\section{The expansion history of the Universe}\label{appke3}
 The expansion of the universe may be calculated by taking into account the dependence of the string energies on the Hubble parameter. It is reasonable to suppose that, if the peculiar velocities of the particles are ignored, the energy in the string joining a particle of mass $m_i$ to a distant particle of mass $m_j$ has the form
\begin{eqnarray}\label{estring}
 \frac{Gm_im_j}{r_{ij}}f\Big(\frac{Hr_{ij}}{c}\Big)\,,
\end{eqnarray}
 for some function $f$, where $H$ is the Hubble parameter and $r$ is the length of the string. For the tests of the theory of gravity considered in Appendix~\ref{apprelgrav}, the time scales are much smaller than the time scale for the expansion of the Universe and the effect of the function $f$ is simply to rescale the masses $m_j$ of distant matter by a constant factor. The calculations in the previous Appendices therefore remain valid provided the factor $f$ is understood to be included in the masses $m_j$. In particular, equation~(\ref{ffrac}) for the total energy in the strings of a space quantum is replaced by
\begin{eqnarray}\label{ffrac2}
 \sum_{j\in S_d}\frac{Gm_im_j}{r_{ij}}f\Big(\frac{Hr_{ij}}{c}\Big)\,=\,\kappa m_ic^2\,,
\end{eqnarray}
where $\kappa= 3/40$.

 To calculate the function $f$, consider the kinetic energy in the strings associated with the Hubble flow. Since the Hubble velocity may exceed the speed of light, the kinetic energy will be calculated using the Newtonian formula. The recessional velocity of a point on the string at a distance $x$ from $m_i$ is $Hx$ and the energy of an element of string of length $dx$ is $Gm_im_jdx/r_{ij}^2$, assuming the mass of a string is distributed uniformly along its length, so the total kinetic energy in the string is
\begin{eqnarray}\label{kecalc}
 \int_0^{r_{ij}}\frac{1}{2}\frac{Gm_im_j}{r_{ij}^2c^2}(Hx)^2\,dx\,=\,\frac{Gm_im_j}{6r_{ij}}
 \Big(\frac{Hr_{ij}}{c}\Big)^{\!2}
\end{eqnarray}
 and equation~(\ref{ffrac2}) has the form
\begin{eqnarray}\label{ffrac3}
 \sum_{j\in S_d}\frac{Gm_im_j}{r_{ij}}\Big[1 + \frac{1}{6}\Big(\frac{Hr_{ij}}{c}\Big)^{\!2}\,\Big]\,=\,\kappa m_ic^2\,.
\end{eqnarray}
 The precise formula for the kinetic energy in the strings connected to distant matter may be more complicated since the nature of the instantaneous interaction transmitted by the strings is not yet understood. It is therefore of interest to consider a more general function $f$ of the form $f(x)= 1 + ax + bx^2$. Then
\begin{eqnarray}\label{ffrac4}
 \sum_{j\in S_d}\frac{Gm_im_j}{r_{ij}}\Big[1 + a\Big(\frac{Hr_{ij}}{c}\Big) + b\Big(\frac{Hr_{ij}}{c}\Big)^{\!2}\,\Big]\,=\,\kappa m_ic^2\,.\nonumber\\
\end{eqnarray}
 The sums in~(\ref{ffrac4}) may be evaluated in the continuum approximation, which gives
\begin{eqnarray}\label{gpe}
 && \sum_{j\in S_d} \frac{m_j}{r_{ij}}\,\,\approx\,\int_0^{R_U}\!\frac{\rho}{r}\,dV\,=\,\frac{3M_U}{2R_U}\nonumber\\
 \mbox{and}\hspace*{0.3cm}&& \sum_{j\in S_d} m_j r_{ij} \,\,\approx\,\int_0^{R_U}\!\rho r\,dV\,=\,\frac{3}{4}M_UR_U,\hspace*{0.5cm}
\end{eqnarray}
 where $M_U$ is the mass of the observable Universe, $R_U$ is the radius of the observable Universe and $\rho$ is the average mass density, and~(\ref{ffrac4}) then reduces to the equation
\begin{eqnarray}\label{ffrac5}
 \hspace*{-0.2cm}\frac{3GM_U}{2R_U}\Big[1 + \frac{2a}{3}\Big(\frac{HR_U}{c}\Big) + 2b\Big(\frac{HR_U}{c}\Big)^{\!2}\,\Big]\,=\,\kappa c^2\,.
\end{eqnarray}

% The cosmological string model
% therefore gives $R\sim t^{1/2}$ in the radiation era and suggests a transition from $R\sim t^{2/3}$ in the
% matter era to an accelerating phase with $R\sim t^2$. Could the transition be caused by the contraction
% in the short strings associated with galaxy formation\,?
% In the radiation era, almost all the mass in the photon space quanta is in the strings whereas almost all
% the mass in the baryon space quanta is at the centres. In the matter era, the energy in the photon space
% quanta is transferred from the strings to the centres and the energy in the baryon space quanta is
% transferred from the centres to the strings. In the acceleration era, the energy in the baryon strings is
% transferred from potential energy into kinetic energy. In all three eras, almost all the energy in the
% universe is in the strings, with the photon strings dominating in the radiation era and the baryon strings
% dominating in the matter and acceleration eras.

 Equation~(\ref{ffrac5}) may be used to calculate the expansion history of the universe if it is assumed that the constant $\kappa$, which represents the fraction of the energy of a space quantum in the strings, is time-independent. The expansion history is given by the scale factor, $R(t)$, is related to the radius $R_U$ by the usual formula
\begin{eqnarray}\label{ru}
 R_U(t)\,=\,R(t)\int_0^t \frac{c\,ds}{R(s)}\,.
\end{eqnarray}
 The total mass $M_U$ in the observable universe is given by
\begin{eqnarray}\label{mu}
 M_U\,=\,\frac{4\pi}{3}R_U^3(\rho_b\,+\,\rho_r)\,,
\end{eqnarray}
 where $\rho_b$ and $\rho_r$ are the densities of baryons and radiation, respectively. The parameters $\Omega_b$ and $\Omega_r$ may be defined in the same way as in the $\Lambda$CDM model, namely
\begin{eqnarray}\label{ob}
 \Omega_b\,=\,\frac{8\pi G\rho_b^0}{3H_0^2}\hspace{0.5cm}\mbox{and}\hspace{0.5cm}
 \Omega_r\,=\,\frac{8\pi G\rho_r^0}{3H_0^2}\,,
\end{eqnarray}
 where $\rho_b^0$ and $\rho_r^0$ are the densities at the present time and $H_0$ is the present value of the
 Hubble parameter.
% The conventional value $\Omega_b=0.04$ is also adopted since it is consistent with the
% observed amount of visible matter.
 Substitution into~(\ref{ffrac5}) gives
% Equation~(\ref{accel3}) may be solved for $\dot R$ by first completing the square to give
%\begin{eqnarray}\label{accel3}
% &&\hspace*{-0.3cm}\frac{3}{4}\Big(\frac{H_0R_U}{c}\Big)^2\Big(\frac{\Omega_b}{y^3}\,+\,\frac{\Omega_r}{y^4}\Big)
% \Big[1\,+\,\frac{2a}{3}\Big(\frac{HR_U}{c}\Big)\,+\,2b\Big(\frac{HR_U}{c}\Big)^{\!2}\,\Big]\nonumber\\
% && \hspace*{6cm}=\,\kappa\,,
%\end{eqnarray}
\begin{eqnarray}\label{accel4}
 &&\hspace*{-0.5cm}\frac{3}{4}\Big(\frac{H_0R_U}{c}\Big)^2\Big(\frac{\Omega_b}{y^3}\,+\,\frac{\Omega_r}{y^4}\Big)
\Big[1 + \frac{2a}{3}\Big(\frac{HR_U}{c}\Big) + 2b\Big(\frac{HR_U}{c}\Big)^{\!2}\,\Big]
% \Big[1\,-\frac{a^2}{18b}\,+\,2b\Big(\frac{HR_U}{c} +
% \frac{a}{6b}\Big)^{\!2}\,\Big]
\nonumber\\
 && \hspace*{6.5cm}=\,\kappa\,,\nonumber\\
\end{eqnarray}
 where $y=R/R_0$ and $R_0$ is the value of the scale factor at the present time. It is convenient to define the dimensionless time, $\tau$, by $\tau= H_0t$ and the dimensionless conformal time, $\eta$, by the equation
\begin{eqnarray}\label{edef}
 \eta(t)\,=\,H_0\int_0^t \frac{ds}{y(s)}\,.
\end{eqnarray}
 Then $H_0R_U/c= y \eta$ and $HR_U/c= \dot y \eta$, where the dot denotes differentiation with respect to $\tau$.
 Substituting into~(\ref{accel4}) and completing the square gives
\begin{eqnarray}\label{accel4b}
 \frac{\eta^2}{y}\Big(1+\frac{y_{eq}}{y}\Big)\Big[1\,-\frac{a^2}{18b}\,+\,2b\Big(\dot y \eta +
 \frac{a}{6b}\Big)^{\!2}\,\Big]\,=\,\frac{4\kappa}{3\Omega_b}\,,\nonumber\\
\end{eqnarray}
 where $y_{eq}= \Omega_r/\Omega_b$ is the value of $y$ at which the baryon and radiation densities are equal,
 which may be solved for $\dot y$ to give
\begin{eqnarray}\label{accel5}
 \dot y \eta\,=\,\Big[\frac{2\kappa}{3b\Omega_b}\frac{y}{\eta^2}\Big(1+\frac{y_{eq}}{y}\Big)^{\!-1} \!+
 \frac{a^2}{36b^2} - \frac{1}{2b}\Big]^{1/2}\,-\,\frac{a}{6b}\,.\nonumber\\
\end{eqnarray}

% which replaces the conventional Friedman equation for the evolution of the scale factor,
 To integrate numerically over several orders of magnitude in time, it is convenient to change the independent variable from $\tau$ to $u= \ln\eta$. Equation~(\ref{accel5}) then becomes
\begin{eqnarray}\label{accel6}
 \frac{1}{y}\frac{dy}{du}=\Big[\frac{2\kappa}{3b\Omega_b}\frac{y}{e^{2u}}\Big(1+\frac{y_{eq}}{y}\Big)^{\!-1} \!+ \frac{a^2}{36b^2} - \frac{1}{2b}\Big]^{1/2}\,-\,\frac{a}{6b}\,.\hspace*{-0.5cm}\nonumber\\
\end{eqnarray}
 The value of $\Omega_r$ is taken to be $8.2\times 10^{-5}$ and $\Omega_b$ is treated as a free parameter. The integration begins at the current epoch, $y=1$, and continues back to the singularity at $y=0$. The value of $\eta$ at the current epoch, $\eta=\eta_0$, is chosen so that the Hubble parameter is equal to $H_0$ at the present time, i.e. so that $\dot y/y= 1$. From~(\ref{accel4b}), it may be seen that the required value of $\eta_0$ satisfies the equation
\begin{eqnarray}\label{eeta}
 \eta_0^2(1+y_{eq})\Big(1\,+\,\frac{2a}{3}\eta_0\,+\,2b\eta_0^2\Big)\,=\,\frac{4\kappa}{3\Omega_b}\,,\nonumber\\
\end{eqnarray}
 which has a unique positive solution for $\eta_0$ when $a$ and $b$ are non-negative.

\begin{figure}[h]
\vspace*{0.5cm}
\includegraphics[height=5.5cm,width=7.5cm]{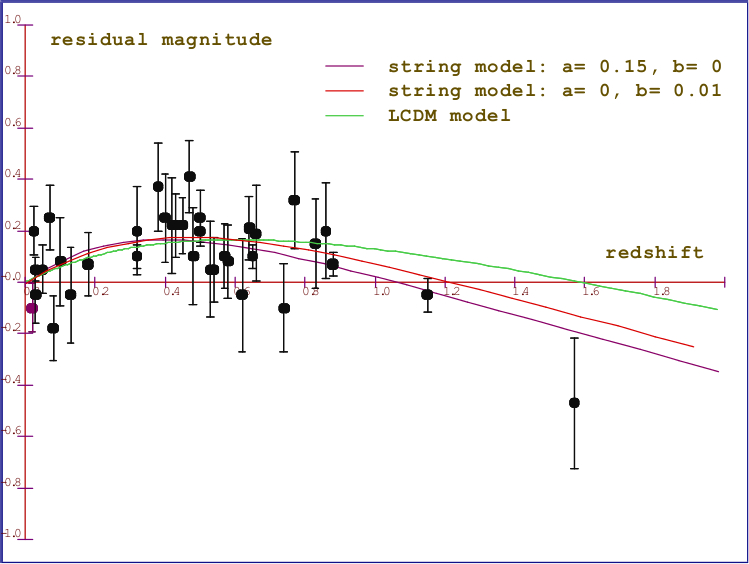}
\caption{\label{mz} Plot of residual magnitude against redshift for the scale factor in
 Figure~\ref{scalefactor} with parameters $a=0$, $b= 0.01$ and for the scale factor with parameters $a=0.15$, $b= 0$, together with together with the experimental data points and error bars from the Supernova Cosmology Project and the High-z Supernova Search Team. Both sets of parameters give as good a fit to the data as the $\Lambda$CDM model.}
\end{figure}

The integration was performed for different values of the parameters $a$, $b$ and $\Omega_b$ and  
a good fit to the observed expansion history, given experimentally as a plot of residual magnitude against redshift for 
distant supernovae, was found for a range of values of $a$ and $b$ with $\Omega_b\lesssim 0.01$. The time evolution 
of the scale factor in the string model for $a=0$, $b= 0.01$ and $\Omega_b= 0.01$ is shown in Figure~\ref{scalefactor}. 
The deceleration parameter, $q= -R\ddot R/\dot R^2$, is equal to $-0.76$ at the present time and the value of $HR_U/c$ at
 the present time is $2.91$. A similar time evolution can be obtained with $\Omega_b= 0.01$ for other values of $a$ and 
$b$, for example $a= 0.15$ and $b=0$, for which $q= -0.95$ and $HR_U/c= 2.79$. The magnitude-redshift relations for both 
sets of parameters are shown in Figure~\ref{mz}, together with the data from the Supernova Cosmology Project~\cite{knop} 
and the High-z Supernova Search Team~\cite{riess2}.

% The values $a= 1/9$, $b= 1/36$ and $B= 30$ are found to give good agreement with the
% experimental magnitude-redshift relation, with the time unit defined so the present time is $t=1$.
% Equation~(\ref{accel5}) is integrated from $t=1$ back to $t=0$ and the initial value of $\tau$ is adjusted
% to ensure that the graph of $\log y$ against $\log t$ is exactly linear for very small values of $t$, i.e.
% that the graph passes exactly through the origin.
% The value of $H_0$ in computer units is $1/\sqrt{20\Omega_b}=1.12$. Since $H_0= 70$\,km/s/% Mpc, the age of the universe at the present time is 1.12/(70\,km/s/Mpc) = 15.6\,Gyr.

\begin{figure}[h]
\vspace*{0.5cm}
\includegraphics[height=5.5cm,width=7.5cm]{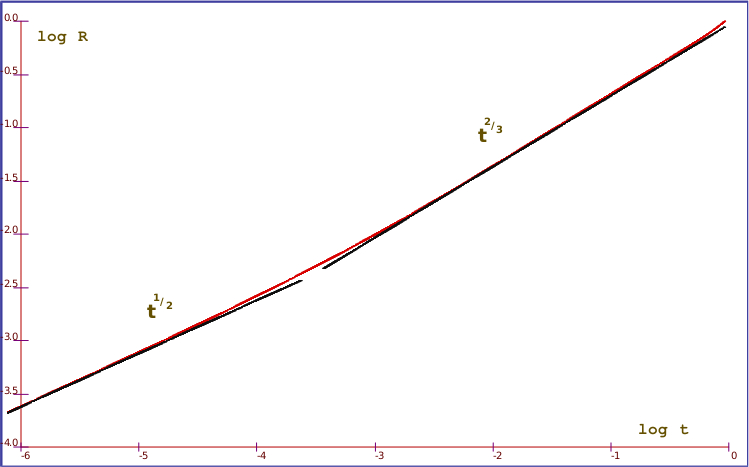}
\caption{\label{slopes} Plot of $\log R$ against $\log t$ for the best-fit scale factors $R(t)$ in the string model. The scale factor is proportional to $t^{1/2}$ in the radiation era and $t^{2/3}$ in the matter era, exactly as in $\Lambda$CDM.}
\end{figure}

 The graph of $\log R$ against $\log t$ in Figure~\ref{slopes} shows a transition from $R\sim t^{1/2}$ to $R\sim t^{2/3}$ at $\log y \approx -2.4$, corresponding to the transition from radiation domination to matter domination at redshift $z_{eq}\approx 250$.

It is remarkable that the evolution of the scale factor has the same form in both the radiation era and the matter era as in the conventional $\Lambda$CDM model, for all values of $a$ and $b$. The result may be understood by referring to equation~(\ref{accel4}). The quantity in square brackets varies very slowly compared with the factors of $1/y^3$ and $1/y^4$ so, to a good approximation, $R_U^2/R^3$ is constant in the matter era and $R_U^2/R^4$ is constant in the radiation era. It follows from the definition~(\ref{ru}) that $R_U$ is proportional to $t$ when both $R(t)\sim t^{1/2}$ and when $R(t)\sim t^{2/3}$, so $R_U^2/R^4$ is constant when $R(t)\sim t^{1/2}$ and $R_U^2/R^3$ is constant when $R(t)\sim t^{2/3}$. The string model therefore gives $R(t)\sim t^{1/2}$ in the radiation era and $R(t)\sim t^{2/3}$ in the matter era, as claimed.

It is also remarkable that, at least for some values of $a$ and $b$, the string model explains the transition from deceleration to acceleration. When the transition does occur, it is easy to show from equation~(\ref{accel4}) that the scale factor has the form $R\sim t^2$ at late times when $b\ne 0$ and grows exponentially when $b=0$. At late times, the radiation term in~(\ref{accel4}) is neglegible and $HR_U/c$ increases with time, so it follows that $R^3/R_U^2$ is proportional to $(HR_U/c)^2$ when $b\ne 0$ and proportional to $HR_U/c$ when $b=0$. Moreover, if $R$ accelerates at late times then the integral in~(\ref{ru}) converges, so $R_U$ has the same time dependence as $R$. Thus $H\sim 1/\sqrt R$ when $b\ne 0$, giving $R\sim t^2$, and $H \sim $\,const. when $b=0$, giving exponential growth.

The best-fit value of $\Omega_b$ in the string model is rather less than the value suggested by experiment, although the values suggested by different experiments are not yet in agreement. The smallest value of $\Omega_b$ consistent with nucleosynthesis calculations in the conventional model is $\Omega_b= 0.015$~\cite{kam} and the model in~\cite{mcgaugh} with no dark matter gives $\Omega_b= 0.016$.

If the value of $H$ at the present time is taken to be $70$\,km/s/Mpc then the value of $R_U$ corresponding to a
best-fit value of $HR_U/c$ at the present time of $2.8$ is about $12$\,Gpc, or about $3.6\times 10^{26}$\,m. Neglecting the contribution from radiation, equations~(\ref{mu}) and~(\ref{ob}) give the relation
\begin{eqnarray}
 GM_U\,=\,\frac{\Omega_b}{2}\Big(\frac{HR_U}{c}\Big)^2R_Uc^2\,.
\end{eqnarray}
 Taking $\Omega_b= 0.01$ gives $GM_U= 0.04 R_Uc^2$ and a corresponding value of $M_U$ of about $10^{22}$ Solar masses.

%% file: appdarkmatter.tex
\section{\lq Dark matter' in the string model}\label{appdarkmatter}

 Consider one of the Machian strings of a test mass $m$ connected to a distant mass $\overline m$. In the absence of any interactions between the Machian strings, the string is straight and has total energy $Gm\overline m/R$, where $R$ is the length of the string. The energy per unit length is $T(R)= Gm\overline m/R^2$, assuming the energy to be distributed uniformly along the string. Now consider the interaction between the Machian strings of $m$ and the Machian strings of a larger mass $M$. Let $f$ be the change the energy per unit length of the Machian strings of $m$ and suppose, as discussed in Section~\ref{dm}, that the value of $f$ at a given point depends on the ratio of the density of Machian strings of $M$ at that point to the density of background strings. If $\lambda$ denotes the mass per unit length in a Machian string joining two particles of unit mass, the Machian strings within a distance $r$ of the $M$ contain a mass $\lambda r MM_U$. The corresponding mass density, $\rho_M$, in the Machian strings of $M$ is therefore given by $4\pi \rho_M r^2dr = \lambda dr M M_U$, so that $\rho_M= \lambda MM_U/4\pi r^2$. The mass density in the background Machian strings, $\rho_b$, is given by $\rho_b=(\lambda M_U^2R_U/2)/(4\pi R_U^3/3)$, so that $\rho_M/\rho_b \sim (M/M_U)(R_U^2/r^2)$. The ratio of densities is therefore given by the variable
\begin{eqnarray}\label{u}
 u\,=\,\frac{M}{|{\bf x}-{\bf x}_M|^2}\Big(\frac{M_U}{R_U^2}\Big)^{\!-1}\,.
\end{eqnarray}
% since the density of  strings of $M$ at the point ${\bf x}$ is proportional to $M/|{\bf x}-{\bf x}_M|^2$, where ${\bf x}_M$ is the position of the centre of the mass $M$, and the density of background strings is proportional to $M_U/R_U^2$.
 The function $f$ is assumed to be a function of $u$, so the energy per unit length of the Machian strings of $m$ at the point ${\bf x}$ may be written in the form
%\begin{eqnarray}\label{dens2}
% {\mathcal E}({\bf x})\,=\,\frac{\kappa mc^2}{NR_U}\Big\{1 + f[u({\bf x})]\Big\}.
%\end{eqnarray}
\begin{eqnarray}\label{dens}
 {\mathcal E}(R,{\bf x})\,=\,T(R)\Big\{1 + f[u({\bf x})]\Big\}\,.
\end{eqnarray}

 At the position of the test mass $m$, the ratio of Machian string densities~(\ref{u}) is equal to
\begin{eqnarray}\label{adens}
 \frac{M}{r^2}\Big(\frac{M_U}{R_U^2}\Big)^{\!-1}\,\approx\,\frac{GM}{r^2}\Big/0.1\,a_0\,,
\end{eqnarray}
 where $r$ is the distance between $m$ and $M$, since $a_0\approx 0.4 c^2/R_U$ and, from Appendix~\ref{appke3}, $GM_U\approx 0.04 R_Uc^2$. Interaction between the Machian strings is therefore expected to lead to corrections to the Newtonian law of gravitation with an acceleration scale of order $0.1 a_0$.

 The magnitude of the expected additional acceleration can also be estimated by considering the limiting case of 
maximum asymmetry in the Machian strings of $m$, where all the strings point in the same direction at the centre. The 
tension $T(R)= Gm\overline m/R^2$ in a string of length $R$ connected to a distant particle of mass $m$ makes a 
contribution to the total acceleration of magnitude $G\overline m/R^2$. If all the Machian strings around $m$ were 
arranged to point in the same direction at $m$, without changing their tensions, then the total acceleration would be 
given by integrating $G\overline m/R^2$ over all distant matter. The resulting acceleration would be
\begin{eqnarray}\label{amax}
 && \hspace*{-1cm}a_{max}\,=\, \int \frac{G\rho}{r^2}\,dV\,=\,4\pi G\rho R_U\nonumber\\
 \Rightarrow\hspace*{0.5cm} && a_{max}\,=\,\frac{2\kappa c^2}{R_U}\,,
\end{eqnarray}
 since~(\ref{kd}) gives
\begin{eqnarray}\label{kapro}
 && \kappa c^2\,=\, \int \frac{G\rho}{r}\,dV\,=\,2\pi G\rho R_U^2\,.
\end{eqnarray}
 In terms of the MOND acceleration scale $a_0\approx 0.4 c^2/R_U$, equation~(\ref{amax}) gives, with $\kappa= 3/40$ from~(\ref{kapsol}),
\begin{eqnarray}\label{amax}
 a_{max}\,\approx\, 0.4 a_0\,.
\end{eqnarray}

\begin{figure}[h]
\vspace*{0.5cm}
\includegraphics[height=6cm,width=6cm]{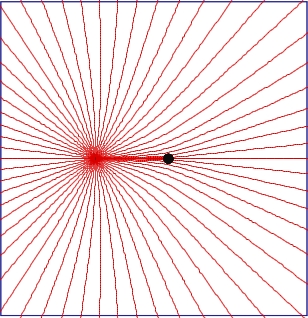}
%\vspace*{-0.8cm}
\caption{\label{asym} Asymmetry of Machian strings for a test mass at the edge of a galaxy. The centre of the galaxy is represented by the black dot and the centre of the test mass is on the left, at the distance $R_g$ defined by equation~(\ref{rg}). All the strings shown are Machian strings of the test mass.}
\end{figure}

 To calculate the additional acceleration explicitly, a particular form must be chosen for the function $f(u)$. The limit $u\rightarrow 0$ corresponds to the limit of no interaction, so clearly $f(u)\rightarrow 0$ as $u\rightarrow 0$. Appendix~\ref{appdarkmatter1b} shows that the condition $f<1$ must be applied to ensure that the string tension remains positive, so $f(u)$ must tend to some constant $A<1$ in the limit $u\rightarrow \infty$. To obtain flat galaxy rotation curves, the additional acceleration is required to be proportional to $1/r$ at large distances from the centre of the galaxy. It turns out that a $1/r$ acceleration is obtained if $f(u)\sim \sqrt u$ as $u\rightarrow 0$, so consider the function $f_1(u)$ defined by
\begin{eqnarray}\label{fu}
 f_1(u)\,=\, \frac{A\sqrt{u}}{1+\sqrt{u}}\,.
\end{eqnarray}
 The corresponding Machian string paths were calculated using the method described in Appendix~\ref{appdarkmatter1}. The pattern of Machian strings for a test mass at the edge of a typical galaxy is shown in Figure~\ref{asym}, where the edge of the galaxy is defined as the distance, $R_g$, at which the Newtonian acceleration is equal to $a_0$. Since $a_0\approx 0.4 \,c^2/R_U$, $R_g$ is given by
\begin{eqnarray}\label{rg}
 &&\frac{GM}{R_g^2}\,\approx\,0.4\frac{c^2}{R_U}\hspace*{0.5cm} \nonumber\\
 \Rightarrow\hspace*{0.5cm} && R_g \,\approx \,0.32 \,R_U\sqrt{\frac{M}{M_U}}\,.\hspace*{0.5cm}
\end{eqnarray}

%A specific model for the interaction between the field strings of two
% masses is suggested

 In Appendix~\ref{appdarkmatter1b}, the tensions in the Machian strings at $m$ are calculated and integrated over the asymmetrical Machian string distribution to find the total additional acceleration of the centre. In Appendix~\ref{appdarkmatter2}, the additional force between the two masses is found independently by calculating the change in total energy of the Machian strings as a function of the distance between the masses and the two methods are found to be in agreement. For the function $f_2(u)$, defined by
\begin{eqnarray}\label{fu2}
 f_2(u)\,=\, \left\{ \begin{array}{ll}
         A\sqrt u& \hspace*{0.3cm} u \le 1\\
        ~A&  \hspace*{0.3cm} u > 1\,,\end{array} \right.
\end{eqnarray}
 which has the same limits as $f_1$ for both $u\ll 1$ and $u\gg 1$, the formula for the acceleration calculated in Appendix~\ref{appdarkmatter2} can be evaluated analytically.

 Figure~\ref{ffu} shows the accelerations corresponding to the interaction functions $f_1$ and $f_2$ calculated in Appendix~\ref{appdarkmatter2} in the limit $A\rightarrow 1$.
\begin{figure}[h]
%\vspace*{0.5cm}
\includegraphics[height=5cm,width=7cm]{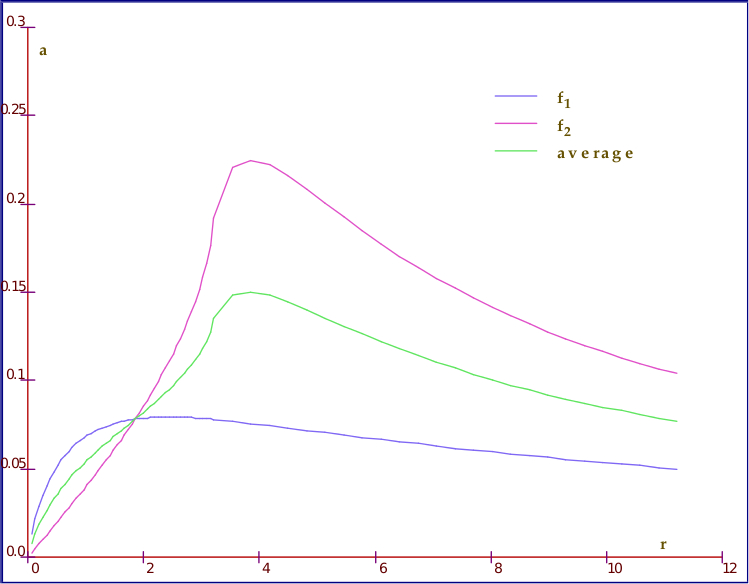}
%\vspace*{-0.8cm}
\caption{\label{ffu} The additional accelerations, in units of $a_0$, for the interactions between Machian strings corresponding to the functions $f_1$ and $f_2$ in equations~(\ref{fu}) and~(\ref{fu2}), respectively, for $A= 1$. The green curve shows the average of the two accelerations. The radial distance $r$ is in units of $R_g$.}
\end{figure}
 For both curves, the additional acceleration for $r \gg R_g$ is approximately $a_0R_g/r$, which is exactly the asymptotic behaviour in MOND. The average of the two additional acceleration curves, shown in green, is the curve used to plot Fig~\ref{atotal}.

 It is important to check that the additional acceleration is negligible in the Solar System since no dark matter effects on the motion of planets or asteroids have been observed. Indeed, the total mass of any dark matter within the orbit of Saturn is estimated~\cite{pitjev} to be less than $1.7\times 10^{-10}$ Solar masses, or about $3\times 10^{20}\,$kg. The corresponding Newtonian acceleration is $8.9\times 10^{-15}$\,m/s$^2\approx 7.5\times 10^{-5}\,a_0$, taking the orbit of Saturn to have radius $r\sim 1.5\times 10^{12}\,$m, which provides an upper limit for the magnitude of the additional acceleration. The value of $u$ at the radius of Saturn is about  $4.4\times 10^5$, taking $M_U=10^{22}$ Solar masses and $R_U= 3.6\times 10^{26}\,$m. According to the calculations in Appendices~\ref{appdarkmatter1b} and~\ref{appdarkmatter2} below, the additional acceleration on a test mass $m$ for the functions $f_1$ and $f_2$ is approximately $(3\kappa/\sqrt{u_0})(c^2/R_U)\approx 0.6a_0/\sqrt{u_0}$ for $u_0\gg 1$, where $u_0$ is the value of $u$ at the centre of $m$.  The corresponding additional acceleration at the radius of the orbit of Saturn is $9.0\times 10^{-4}\,a_0$, which is too large by an order of magnitude, so it appears that a different function $f(u)$ is needed.

 The functions $f_1$ and $f_2$ defined above can easily be generalised by changing the value of $u$ at which the transition from $f\sim A\sqrt u$ for $u\ll 1$ to $f\sim A$ for $u\ll 1$ occurs.
% The magnitude of the additional acceleration for $r \gg R_g$ does of course depend on the precise form of the function $f$ as well as on the value of $A$.
 Suppose, for example, that the length scale of $f(u)$ is decreased by a factor $\lambda$. The function $f_2(u)$ in~(\ref{fu2}) then becomes
\begin{eqnarray}\label{fu3}
 f_3(u)\,=\, \left\{ \begin{array}{ll}
         A\sqrt {\lambda u}& \hspace*{0.3cm} u \le 1/\lambda\\
        ~A&  \hspace*{0.3cm} u > 1/\lambda\,.\end{array} \right.
\end{eqnarray}
 The function $f_3(u)$ is larger by a factor of $\sqrt\lambda$ for $u\ll 1$, so the additional acceleration for $u_0 \ll 1$ is also expected to be larger by a factor of $\sqrt\lambda$. The effect on the additional acceleration for $u_0 \gg 1$ may be found by repeating the analytic calculation in Appendix~\ref{appdarkmatter2}. When the function $f_3$ is substituted into equation~(\ref{change4}), the function $I(\rho)$ in equation~(\ref{I}) becomes
\begin{eqnarray}\label{I}
 && \hspace*{-1.5cm}I(\rho)=\frac{3}{2}\sqrt\lambda +\Big(\frac{\sqrt\lambda}{2}-\frac{\rho}{4}-\frac{\lambda}{4\rho}\Big)\ln
 \Big|\frac{\rho+\sqrt\lambda}{\rho-\sqrt\lambda}\Big|\nonumber\\
 && \hspace*{2cm}+~\sqrt\lambda\ln\Big(\!\frac{\sqrt{M_U/M}\,}{\rho+\sqrt\lambda}\!\Big)\,,
\end{eqnarray}
 where $\rho= 1/\sqrt{u_0}$, which has the limits $I(\rho) = \sqrt\lambda(1+\ln\sqrt{\frac{M_U}{\!M}} -\ln \rho + \dots)$ for $\rho\gg 1$ and $I(\rho) = \sqrt\lambda(1+\ln\sqrt{\frac{M_U}{\lambda M}}) -\rho^2/(6\sqrt\lambda) + \dots$ for $\rho\ll 1$. The additional acceleration is therefore multiplied by $\sqrt\lambda$ for $\rho\gg 1$, i.e. for $u_0\ll 1$, as expected, and divided by $\sqrt\lambda$ for $u_0 \gg 1$. Thus, for $\lambda>1$, the function $f_3(u)$ gives a larger additional acceleration in the MOND regime and a smaller additional acceleration in the Solar System. The exact form of the function $f(u)$ remains to be derived by considering more carefully the nature of the interaction between the Machian strings.

%\begin{figure}[h]
%%\vspace*{0.5cm}
%\includegraphics[height=5cm,width=7cm]{acombine.jpg}
%%\vspace*{-0.8cm}
%\caption{\label{ffc} The additional accelerations compared to the MOND acceleration. The acceleration is in units of $a_0$ and $r$ is in units of $R_g$.}
%\end{figure}

\input{appdarkmatter1.tex}

\input{appdarkmatter2.tex}

%% file: appdarkmatter1.tex
\subsection{Calculation of Machian string paths}\label{appdarkmatter1}
 Let $s$ be the path length along one of the Machian strings of $m$ connected to a distant mass $\overline m$, with $s= s_i$ at $m$ and $s= s_f$ at $\overline m$. If ${\bf x}(s)$ denotes the position of the point along the string at path length $s$ then ${\bf x}(s_i)={\bf X}_i$ and ${\bf x}(s_f)={\bf X}_f$, where ${\bf X}_i$ and ${\bf X}_f$ are the positions of the masses $m$ and $\overline m$, respectively. The total energy in the string is
\begin{eqnarray}\label{e2}
 E=\int_{s_i}^{s_f}\! {\mathcal E}(R,{\bf x})~ds\,,
\end{eqnarray}
 where ${\mathcal E}(R,{\bf x})$ is the energy per unit length defined in~(\ref{dens}) and $R=s_f-s_i$ is the length of the string. According to the IMC, the total energy of all the strings connected to the masses $m$ and $\overline m$ is $E_S\,=\,mc^2 + \overline mc^2- E$. To minimise the total energy of the system it is therefore necessary to find a string path for which the energy~(\ref{e2}) is a maximum. Since the energy~(\ref{e2}) tends to zero in the limit that the string becomes infinitely long, string paths that maximise~(\ref{e2}) do exist.

 Consider a variation $\delta {\bf x}$ of the string path with the string held fixed at the distant mass
 $\overline m$, so that $\delta{\bf x}= \delta {\bf X}_i$, say, at $s=s_i$ and $\delta{\bf x}= {\bf 0}$ at
 $s=s_f$. Since the total length of the string changes it is convenient to introduce the parameter $\sigma$ along
 the string path so that $\sigma$ is fixed at $m$ and $\overline m$, with $\sigma=0$ at $m$ and $\sigma=1$ at $\overline m$. The energy $E$ is then given by
\begin{eqnarray}\label{t1}
 E=\int_0^1\! {\mathcal E}(R,{\bf x})\,|{\bf x}^\prime|~d\sigma\,,
\end{eqnarray}
 where the prime denotes differentiation with respect to $\sigma$, and the string length $R$ is given by
\begin{eqnarray}\label{r1}
 R=\int_0^1\! |{\bf x}^\prime|~d\sigma\,.
\end{eqnarray}
 The variation of~(\ref{t1}) is
\begin{eqnarray}\label{t2}
 &&  \hspace*{-0.8cm}\delta E= \!\int_0^1\! \Big\{\Big(\frac{\partial{\mathcal E}}{\partial R}\delta R +
 {\pmb\nabla}{\mathcal E}.\delta{\bf x}\Big)\,|{\bf x}^\prime| + {\mathcal E}\frac{{\bf x}^\prime.\delta{\bf
 x}^\prime}{|{\bf x}^\prime|}\Big\}~d\sigma\,,~~ \\ &&  \hspace*{-0.3cm}\mbox{where} \hspace*{1.3cm} \delta R\,=
 \int_0^1\!\frac{{\bf x}^\prime.\delta{\bf x}^\prime}{|{\bf x}^\prime|}~d\sigma\,.\label{r2}
\end{eqnarray}
 Integrating~(\ref{r2}) by parts gives
\begin{eqnarray}\label{r3}
 \delta R= \Big(\frac{{\bf x}^\prime.\delta{\bf x}}{|{\bf x}^\prime|}\Big)_0^1\,-
  \int_0^1\! \Big\{\frac{{\bf x}^{\prime\prime}}{|{\bf x}^\prime|}- \frac{{\bf x}^\prime({\bf x}^\prime.
  {\bf x}^{\prime\prime})}{|{\bf x}^\prime|^3}\Big\}.\delta{\bf x}~d\sigma\,.~
\end{eqnarray}
 After changing back to the path length parameterisation, for which $|{\bf x}^\prime|=1$ and ${\bf x}^\prime.
 {\bf x}^{\prime\prime}=0$,~(\ref{r3}) becomes
\begin{eqnarray}\label{r4}
 \delta R\,=\, -{\bf x}^\prime(s_i).\delta {\bf X}_i\, -\int_{s_i}^{s_f}\! {\bf x}^{\prime\prime}.\delta{\bf
 x}\,\,ds\,.
\end{eqnarray}
 Similarly,~(\ref{t2}) becomes
\begin{eqnarray}\label{t3}
 && \hspace*{-0.8cm}\delta E\,=\, -\,{\mathcal E}(R,{\bf X}_i)\,{\bf x}^\prime(s_i).\delta {\bf X}_i\,
 \nonumber\\ && \hspace*{-0.8cm}+\int_{s_i}^{s_f}\!\!  \Big\{\frac{\partial{\mathcal E}}{\partial R}\delta R
 + \!\Big[{\pmb\nabla}{\mathcal E} - {\bf x}^\prime ({\bf x}^\prime.{\pmb\nabla}{\mathcal E}) -
 {\mathcal E}{\bf x}^{\prime\prime}\Big].\delta{\bf x}\Big\}\,ds.~
\end{eqnarray}
 The variation $\delta R$ is independent of $s$ and may be taken outside the integral. Substituting for
 $\delta R$ from~(\ref{r4}) then gives
\begin{eqnarray}\label{t4}
 && \hspace*{-0.8cm}\delta E\,=\, -\,\Big[{\mathcal E}(R,{\bf X}_i) + I(R)\Big]{\bf x}^\prime(s_i).\delta {\bf
 X}_i\, \nonumber\\ && \hspace*{-0.8cm}+\int_{s_i}^{s_f} \!\Big\{{\pmb\nabla}{\mathcal E} - {\bf x}^\prime
 ({\bf x}^\prime.{\pmb\nabla}{\mathcal E}) - [{\mathcal E}+I(R)]{\bf x}^{\prime\prime}\Big\}.\delta{\bf x}\,ds,~
 \\ && \hspace*{-0.3cm}\mbox{where} \hspace*{1cm} I(R)\,=\, \int_{s_i}^{s_f}\!\frac{\partial{\mathcal E}}
 {\partial R}\,ds\,.\label{ir}
\end{eqnarray}

 The requirement that $E$ is stationary for all variations of the string path for which the string is fixed at
 both ends, i.e. for which $\delta{\bf X}_i=0$, gives the path equation
\begin{eqnarray}\label{pe}
 {\pmb\nabla}{\mathcal E} - {\bf x}^\prime
 ({\bf x}^\prime.{\pmb\nabla}{\mathcal E}) - [{\mathcal E}+I(R)]{\bf x}^{\prime\prime}\,=\,0\,.
\end{eqnarray}
 After substituting ${\bf x}^{\prime\prime}=\kappa{\bf n}$, where $\kappa$ is the curvature and ${\bf n}$ is a
 unit vector normal to the tangent vector ${\bf x}^\prime$, the component of equation~(\ref{pe}) along ${\bf x}^\prime$ is found to be identically zero and the component along ${\bf n}$ gives the equation for the
 curvature,
\begin{eqnarray}\label{curv}
 \kappa({\bf x})\,=\,\frac{{\bf n}.{\pmb\nabla}{\mathcal E}}{{\mathcal E}(R,{\bf x})+I(R)}\,.
\end{eqnarray}
 A computer program was written to calculate the paths of Machian strings around a test mass $m$, with the
 energy per unit length of the Machian strings defined by~(\ref{dens}) and the interaction function $f_1(u)$ defined by~(\ref{fu}). If $\psi$ denotes the angle that a Machian string makes with the line of centres
 joining $m$ and $M$, the path equations are $x^\prime= \cos\psi$, $y^\prime= \sin\psi$ and $\psi^\prime=
 \kappa$, where the prime denotes differentiation with respect to path length and $\kappa$ is given by
 equation~(\ref{curv}). The value of $A$ was taken to be $0.5$ to avoid possible numerical problems associated with vanishing string tension when $A$ is close to $1$. Machian string paths were calculated for different values of $M$, for a given separation, $r$, of the two masses. Figure~\ref{asym} shows a two-dimensional set of string paths, lying in a plane containing the line of centres, for the case $r= R_g$. Since the effect of the
 mass $M$ is negligible at large distances, the initial directions of the strings were adjusted to ensure that the distribution of strings at large distances is uniform. The two-dimensional set of string paths then represents a cross-section of the actual three-dimensional Machian string distribution. The Machian strings around $m$ are seen to be asymmetric, with a higher density of strings on the side nearest to $M$.

\subsection{Direct calculation of the additional acceleration}\label{appdarkmatter1b}
 For paths satisfying the path equation~(\ref{pe}), it follows from~(\ref{t4}) that the change in the total energy of the string when mass $m$ is displaced by $\delta{\bf X}_i$ is
\begin{eqnarray}\label{c1}
 \delta E\,=\, -\,\Big[{\mathcal E}(R,{\bf X}_i) + I(R)\Big]{\bf x}^\prime(s_i).\delta {\bf X}_i\,.
\end{eqnarray}
 The total energy of the system is $E_S\,=\,mc^2 + \overline mc^2- E$, so
\begin{eqnarray}\label{c2}
 \delta E_S\,=\, \Big[{\mathcal E}(R,{\bf X}_i) + I(R)\Big]{\bf x}^\prime(s_i).\delta {\bf X}_i\,.
\end{eqnarray}
 If ${\bf F}$ denotes the force exerted by the string on the mass $m$ then the work done by the system is
 ${\bf F}.\delta{\bf X}_i$, so $\delta E_S\,=\,-{\bf F}.\delta{\bf X}_i$. The force exerted on the mass $m$
 is therefore
\begin{eqnarray}\label{force}
 {\bf F}\,=\, -\Big[{\mathcal E}(R,{\bf X}_i) + I(R)\Big]{\bf x}^\prime(s_i)\,,
\end{eqnarray}
 where ${\bf x}^\prime$ is the unit vector from $m$ to $\overline m$. The string tension at a general point ${\bf x}$ along the string is given by
\begin{eqnarray}\label{tens}
 T(R,{\bf x})\,=\,-[{\mathcal E}(R,{\bf x})+I(R)]\,.
\end{eqnarray}
 Note that, in the absence of any interactions between the strings, ${\mathcal E}= T(R)\,=\,Gm\overline m/R^2$
 and~(\ref{force}) reduces to the Newtonian gravitational force
\begin{eqnarray}
 {\bf F}\,=\, \frac{Gm\overline m}{R^2}\,{\bf x}^\prime\,.
\end{eqnarray}

 Consider the energy per unit length ${\mathcal E}(R,{\bf x})$ defined by equation~(\ref{dens}). After substituting~(\ref{dens}) into~(\ref{ir}), noting that $T(R)$ is proportional to $1/R^2$, the string
 tension~(\ref{tens}) becomes
% The string paths around a test mass $m$ at the origin and the tension forces exerted by the strings at the centre, in the presence of a mass $M$ at ${\bf x}_M$, can now be calculated for the model considered in Appendix~\ref{appdm1}. When the dependence on the length of the string is included, equation~(\ref{dens2}) for the energy per unit length becomes
%\begin{eqnarray}\label{dens}
% {\mathcal E}(R,{\bf x})\,=\,T(R)\Big\{1 + f[u({\bf x})]\Big\}\,.
%\end{eqnarray}
% where ${\bf x}$ is defined by equation~(\ref{u}).
%% Equation~(\ref{curv}) for the curvature of the string path and equation~(\ref{f1}) for the string tension
%%both %depend on the quantity ${\mathcal E}(R,{\bf x})+I(R)$.
% After substituting~(\ref{dens}) into~(\ref{ir}), noting that $T(R)$ is proportional to $1/R^2$, the string
% tension~(\ref{tens}) becomes
%\begin{eqnarray}\label{ei}
%% {\mathcal E}(R,{\bf x})+I(R)=T(R)[-1 + f(u)] - %\frac{2T(R)}{R}\int_{s_i}^{s_f}\!\!\!f(u)\,ds\,.\hspace*{-0.5cm}\nonumber\\
% {\mathcal E}(R,{\bf x})+I(R)\,=\,-T(R)\left[1 - f(u) + \frac{2}{R}\int_{s_i}^{s_f}\!\!\!f(u)\,ds\right].\hspace*{-0.5cm}\nonumber\\
%\end{eqnarray}
\begin{eqnarray}\label{ei}
 T(R,{\bf x})\,=\,T(R)\,F(R,{\bf x})\,,
\end{eqnarray}
 say, where
\begin{eqnarray}\label{gu}
 F(R,{\bf x})\,=\,1 - f[u({\bf x})] + \frac{2}{R}\int_{s_i}^{s_f}\!\!f[u({\bf x}(s))]\,ds.~~
\end{eqnarray}
 To ensure that the curvature~(\ref{curv}) remains finite, the string tension must be positive everywhere along
 the string so the function $F(R,[\bf x])$ must be positive. For a string of length $R_U$, the change of variables
 $s=\sigma R_U\sqrt{M/M_U}$ gives
\begin{eqnarray}\label{gu2}
 F(R,{\bf x})\,=\,1 - f[u({\bf x})] + 2\sqrt{\frac{M}{M_U}}\int_0^{\sqrt{\frac{M_U}{\!M}}}\!\!f(u)
 \,d\sigma\,,\hspace*{-0.5cm}\nonumber\\
\end{eqnarray}
 where $u=(\rho^2+\sigma^2-2\rho\sigma\cos\theta)^{-1}$, $\rho$ is defined by $r=\rho R_U\sqrt{M/M_U}$ and $\theta$ is the angle between ${\bf x}$ and ${\bf x}_M$. For the functions $f(u)$ defined in~(\ref{fu}) and~(\ref{fu2}), $f(u)\sim \sqrt{u}\sim 1/\sigma$ for $\sigma\gg 1$. The contribution to the integral in~(\ref{gu2}) is of order unity from $\sigma\lesssim 1$ and of order $\ln(M_U/M)$ from $\sigma\gtrsim 1$, so the third term in~(\ref{gu2}) is of order $\sqrt{M/M_U}\ln(M_U/M)$ and is therefore negligible since $M_U/M$ is very large. It follows that $F(R,{\bf x})<1$, so the interaction with $M$ reduces the tension in the Machian strings of $m$. The condition needed to ensure that the string tension remains positive is $f<1$.
% which is the condition assumed in Appendix~\ref{appdm1}.

% To calculate the acceleration, the asymmetry must be multiplied by the tension in the string at $m$, given
% by~(\ref{f1}). For a field string with length of order $R_U$, the energy density~(\ref{dens}) is
% approximately equal to $T(R)$ along almost all of the string path so $I(R)$, given by~(\ref{ir}), is
% approximately $-2T(R)$.
% The tension at $m$ in a field string connected to a distant mass $\overline m$ is
% therefore
%\begin{eqnarray}\label{tension}
% F\,=\,\frac{Gm\overline m}{R_U^2}[1-f(u)]\,.
%\end{eqnarray}
% The function $1-f(u)$ is approximately $1$ when $u \ll 1$ and $1/\sqrt u$ when $u \gg 1$.
% The additional acceleration on the centre corresponding to a two-dimensional pattern of field strings,
%such as %those in Figures~\ref{asym1}-\ref{asym3},

 For a given two-dimensional set of Machian string paths, such as those shown in Figure~\ref{asym}, the additional acceleration acting on the test mass $m$ may be calculated as follows. Since the distribution of strings at large distances is uniform, the total force acting on the centre is given by
\begin{eqnarray}\label{f1}
 {\bf F}\,=\, \frac{1}{4\pi}\int {\bf T}(\theta)\sin\theta\,d\theta d\phi\,=\, \frac{1}{2}\int
 {\bf T}(\theta)\sin\theta\,d\theta\,,\nonumber\\
\end{eqnarray}
 where ${\bf T}(\theta)$ is the tension at the origin in a string whose direction at large distances is
 at an angle $\theta$ to the line of centres. When $A<1$, the integral in~(\ref{gu2}) is negligible and the
 magnitude of the string tension at $m$ is $T(R)[1-f(u_0)]$ for all strings, where $u_0$ is the value of $u$ at $m$. The component of the total force along the line of centres, from $M$ to $m$, is then
\begin{eqnarray}\label{f2}
 F\,=\, -\frac{1}{2}T(R)[1-f(u_0)]\int \cos\psi\, d\cos\theta \,,
\end{eqnarray}
 where $\psi$ denotes the angle made by the initial direction of a string, at the centre of $m$, with
 the line of centres. According to equation~(\ref{amax}), the acceleration of $m$ due to the Machian strings
 is equal to $2\kappa c^2/R_U$ when the tension in all the strings is $T(R)$ and all the strings point from $m$ to $M$ at the centre of $m$, i.e. when $\psi=0$ for all strings. The acceleration due to the Machian strings corresponding to~(\ref{f2}) is therefore
\begin{eqnarray}\label{af}
 a\,=\, -\frac{2\kappa c^2}{R_U}[1-f(u_0)]\langle\cos\psi \rangle\,,
\end{eqnarray}
 where $\langle\cos\psi \rangle$ is the average value of $\cos\psi$, as a function of $\cos\theta$,
 when $\cos\theta$ uniformly distributed in the interval $[-1,1]$.

\begin{figure}[h]
\vspace*{0.5cm}
\includegraphics[height=5cm,width=7cm]{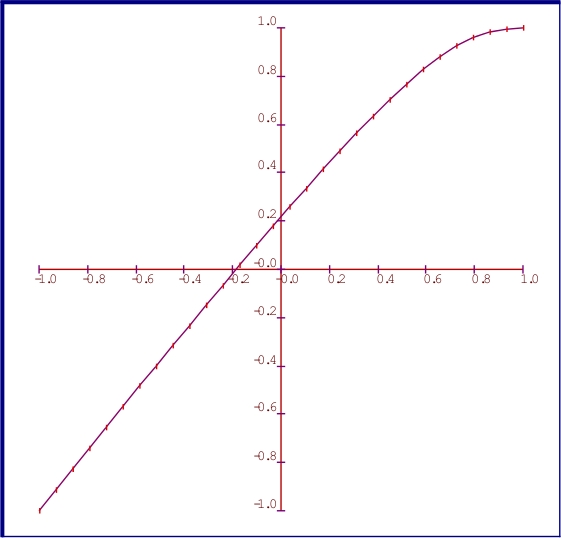}
%\vspace*{-0.8cm}
\caption{\label{cos} A plot of $\cos\psi$ as a function of $\cos\theta$ for the strings in~Figure~\ref{asym}, where $\psi$ is the angle that a string makes with the horizontal and $\theta$ is the angle of the string at a large distance from $M$. The asymmetry, given by averaging the function over the range $[-1,1]$, is equal to $0.16$.}
\end{figure}

 The quantity $\langle\cos\psi \rangle$ defines the asymmetry of the distribution of Machian strings. The asymmetry is zero for $M=0$, when the Machian strings of $m$ are spherically symmetric, and unity in the limit that all strings of $m$ initially point towards $M$.
% Let $\psi_i$ denote the angle made by the initial direction of the $i^{th}$ string, at the centre of
% $m$, with the line of centres.
% Numerical calculations show that, when $f(u)\lesssim 1$, the asymmetry in the strings at $m$
% is approximately proportional to $f(u)$, where the asymmetry is defined as the average value of
% $\cos\theta$ for all the field strings at $m$, where $\theta$ is the angle between the direction
% ${\bf x}^\prime$ along a given string and the line of centres from $m$ to $M$. For the purposes of
%numerical calculation, a simple smooth function $f$ is chosen to agree with the limits in~(\ref{fu}).
%\begin{eqnarray}\label{fn}
% f(u)\,=\,\frac{\sqrt u}{1 + \sqrt u}\,,
%\end{eqnarray}
% which is approximately equal to $\sqrt u$ when $u \ll 1$ and increases monotonically to $1$ as $u$ tends to
% infinity.
 A plot of $\cos\psi$ as a function of $\cos\theta$ for the set of Machian string paths in Figure~\ref{asym}
 is shown in Figure~\ref{cos}.
\noindent
 The distribution of strings and the resulting asymmetry was calculated for different values of $M$, corresponding to values of $u_0$ in the range $10^{-6} < u_0 < 10^6$.
\begin{figure}[h]
\vspace*{0.5cm}
\includegraphics[height=5cm,width=7cm]{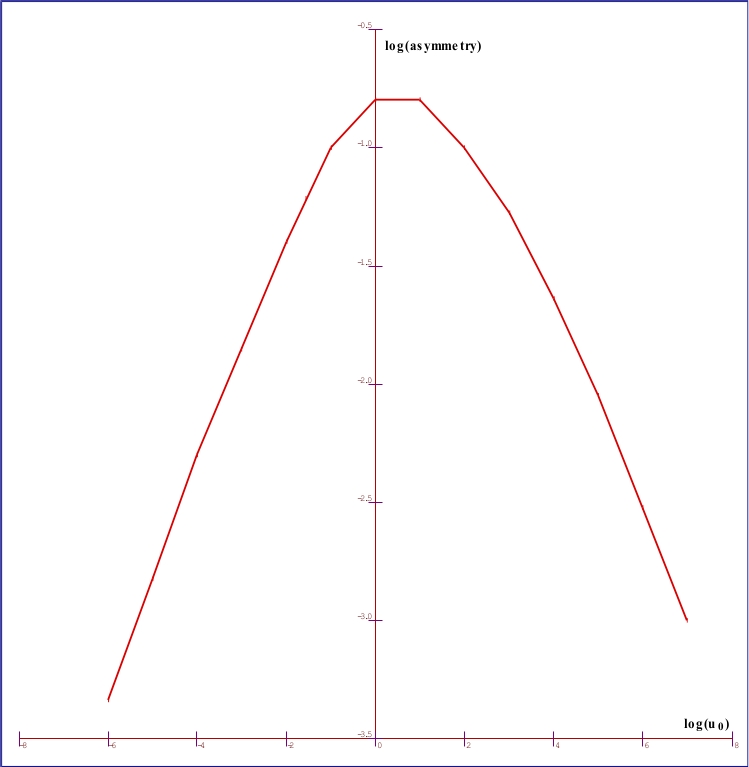}
%\vspace*{-0.8cm}
\caption{\label{log} A log-log plot of the asymmetry in the Machian strings of a test mass due to the presence of a mass $M$ at a distance $r$ as a function of $u_0$, where $u_0$ is the ratio of $M/r^2$ to $M_U/R_U^2$.}
\end{figure}
\noindent
 It may be seen from~Figure~\ref{log} that the asymmetry is largest for $u_0\sim 1$. The variation of the asymmetry for $u_0\ll 1$ and $u_0\gg 1$ is given by
\begin{eqnarray}\label{acc2}
 \langle\cos\psi \rangle\,\approx\, \left\{ \begin{array}{ll}
         0.5\sqrt u_0& \mbox{\hspace*{0.2cm}for \hspace*{0.2cm} $u_0 \ll 1$}\\
        3/\sqrt u_0 & \mbox{\hspace*{0.2cm}for \hspace*{0.2cm} $u_0 \gg 1$}\,.\end{array} \right.
\end{eqnarray}
 The corresponding acceleration is calculated for the data points in~Figure~\ref{log}, using equation~(\ref{af}), and is shown in~Figure~\ref{comp}.
\begin{figure}[h]
\vspace*{0.5cm}
\includegraphics[height=5cm,width=7cm]{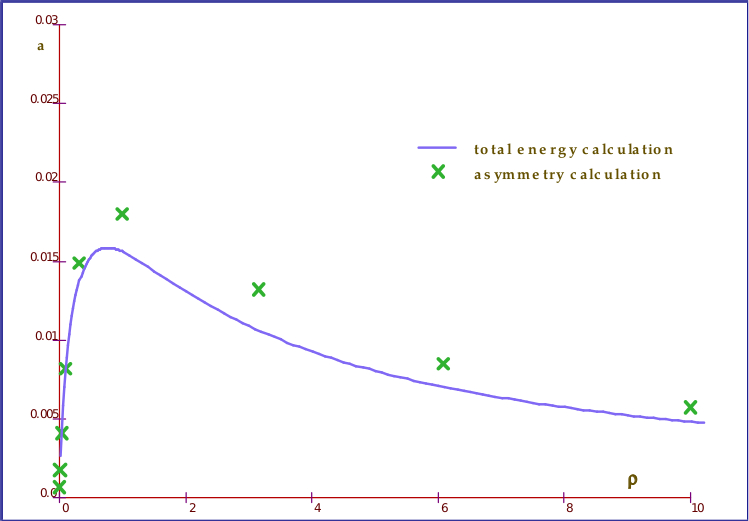}
%\vspace*{-0.8cm}
\caption{\label{comp} The magnitude of the additional acceleration, in units of $c^2/R_U$, calculated from the asymmetry of string paths (green crosses) using the data points in~Figure~\ref{log}, for the interaction defined by the function~(\ref{fu}) with $A= 0.5$. The blue curve is calculated in Appendix~\ref{appdarkmatter2} using an approximate formula for the total interaction energy.}
\end{figure}
 The additional acceleration for the function $f_1(u)$, with $A= 0.5$, therefore has the limits
\begin{eqnarray}\label{lim}
 a\,\approx\, -\frac{2\kappa c^2}{R_U}\left\{ \begin{array}{ll}
         0.5\sqrt u_0& \mbox{\hspace*{0.2cm}for \hspace*{0.2cm} $u_0 \ll 1$}\\
        1.5/\sqrt u_0 & \mbox{\hspace*{0.2cm}for \hspace*{0.2cm} $u_0 \gg 1$}\,.\end{array} \right.
\end{eqnarray}
 For comparison with the results of Appendix~\ref{appdarkmatter2}, it is convenient to rewrite the limits in terms of the variable $\rho$ defined by $r=\rho R_U\sqrt{M/M_U}$, which is related to $u_0$ by the simple equation $u_0=1/\rho^2$. In terms of $\rho$,~(\ref{lim}) becomes
\begin{eqnarray}\label{lim2}
 a\,\approx\, -\frac{2\kappa c^2}{R_U}\left\{ \begin{array}{ll}
         1.5\rho& \mbox{\hspace*{0.2cm}for \hspace*{0.2cm} $\rho \ll 1$}\\
        0.5/\rho & \mbox{\hspace*{0.2cm}for \hspace*{0.2cm} $\rho \gg 1$}\,.\end{array} \right.
\end{eqnarray}

%% file: appdarkmatter2.tex
\subsection{Calculation of the additional acceleration from the Machian string interaction energy}\label{appdarkmatter2}
 The change in the total energy of the Machian strings of $m$ corresponding to the interaction with the Machian strings of $M$ defined by equation~(\ref{dens}) is
\begin{eqnarray}\label{change}
% \Delta E\,=\,\frac{mc^2}{NR_U}\sum_i\int_0^{R_U} \!f (u)\,ds\,,
 \Delta E\,=\,\sum_i\frac{Gm\overline m_i}{R_i^2}\int_0^{R_i} \!f[u({\bf x})]\,ds\,,
\end{eqnarray}
 where $s$ denotes the path length along the $i^{th}$ string. Although the Machian strings around $m$ are asymmetrical, as illustrated in Figure~\ref{asym}, a symmetrical distribution of strings around $m$ may nevertheless be assumed in order to calculate an approximate formula for $\Delta E$. The total mass $\overline m$
 in an elemental solid angle $d\Omega$ and thickness $dR$ at radius $R$ is $\rho R^2 dRd\Omega$, where $\rho$ is the average matter density, so~(\ref{change}) then becomes, after integrating over the azimuthal angle,
% The sum over the strings may then be replaced by an integral over the angle $\theta$ that the string makes with the line of centres, so
\begin{eqnarray}\label{change2}
% \Delta E\,\approx\,\frac{mc^2}{2R_U}\int_0^{R_U}\!\int_0^\pi \!f(u)\sin\theta\,d\theta\,ds\,.
 \Delta E\,\approx\,2\pi Gm\rho\int_0^{R_U}\!\!\!dR\!\int_0^\pi \!\!\sin\theta\,d\theta \int_0^R\!\!f[u({\bf x})]\,ds\,.~~~
\end{eqnarray}
 Let $r$ be the distance between the centres of the two masses, so that $|{\bf x}-{\bf x}_M|^2=r^2 + s^2 -2rs\cos\theta$. Since $f\rightarrow 0$ as $s\rightarrow \infty$, the integral over $s$ in~(\ref{change2}) is insensitive to the value of $R$, so $R$ can be replaced by $R_U$ in the upper limit of the integral over $s$ and the integral over $R$ then simply gives a factor of $R_U$. If the dimensionless variables $\rho$ and $\sigma$ are defined by $r=\rho R_U\sqrt{M/M_U}$ and $s=\sigma R_U\sqrt{M/M_U}$ then $u=(\rho^2+\sigma^2-2\rho\sigma\cos\theta)^{-1}$. Equation~(\ref{change2}) then becomes, using~(\ref{kapro}),
\begin{eqnarray}\label{change3}
% \Delta E\,\approx\,\frac{mc^2}{2}\sqrt{\frac{M}{M_U}}\int_0^{\sqrt{\frac{M_U}{\!M}}}\!\!\int_{-1}^1 %\!f(u)\,d\cos\theta\,d\sigma\,.~
 \Delta E\,\approx\,\kappa mc^2 \sqrt{\frac{M}{M_U}}\int_0^{\sqrt{\frac{M_U}{\!M}}}\!\!d\sigma\int_{-1}^1 \!f[u({\bf x})]\,d\cos\theta\,. \hspace*{-0.5cm}~\nonumber\\
\end{eqnarray}
 Changing variables from $\cos\theta$ to $u$, using $du\,=\,2\rho\sigma u^2\,d\cos\theta$, gives
\begin{eqnarray}\label{change4}
 \Delta E\approx \frac{\kappa mc^2}{2r}\frac{MR_U}{M_U}\int_0^{\sqrt{\frac{M_U}{\!M}}}\!\frac{d\sigma}{\sigma}
 \int_{(\rho+\sigma)^{-2}}^{(\rho-\sigma)^{-2}} \!\frac{f(u)}{u^2}\,du\,.~\hspace*{-0.5cm}\nonumber\\
\end{eqnarray}
 When the IMC is taken into account, the corresponding change in energy of the whole system, including all the other masses other than $m$ and $M$, is equal to $-\Delta E$. The additional acceleration of the mass $m$ due to the interaction between the Machian strings is therefore given by
\begin{eqnarray}\label{a1}
 a\,=\,\frac{1}{m}\frac{d\,\Delta E}{dr}\,.
\end{eqnarray}

 For the function $f_1(u)$, the energy~(\ref{change4}) can be evaluated numerically and the result, for $A= 0.5$, is shown as the blue curve in Figure~\ref{comp}. The corresponding limits are
\begin{eqnarray}\label{a3}
 a\,\approx\, -\frac{2\kappa c^2}{R_U}\left\{ \begin{array}{ll}
        \! 1.5\rho  \hspace*{0.5cm}\rho \ll 1 ~~\\
        \! 0.4/\rho  \hspace*{0.5cm}\rho \gg 1\,,~~\end{array} \right.
\end{eqnarray}
 which is seen to be in good agreement with~(\ref{lim2}) and confirms the validity of the above calculations and the calculations in Appendix~\ref{appdarkmatter1b} with an accuracy of $\sim\! 10\%$.

\begin{figure}[h]
%\vspace*{0.5cm}
\includegraphics[height=5cm,width=7cm]{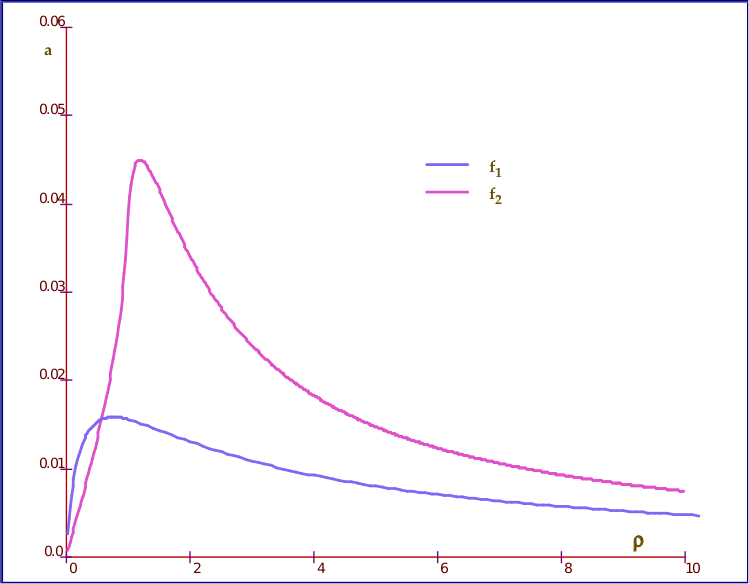}
%\vspace*{-0.8cm}
\caption{\label{ff} The magnitude of the additional acceleration, in units of $c^2/R_U$, for the interactions between Machian strings corresponding to the functions $f_1$ and $f_2$ in equations~(\ref{fu}) and~(\ref{fu2}), respectively, for $A= 0.5$. Both calculations are based on the approximate formula~(\ref{change2}) for the total interaction energy. The blue curve is the same as in Figure~\ref{comp}.}
\end{figure}

 If $f(u)$ is taken to be the function $f_2(u)$ defined in~(\ref{fu2}), the energy~(\ref{change4}) can be evaluated analytically. The result is
%\begin{eqnarray}\label{c5}
% &&\Delta E= -mc^2\sqrt{\frac{M}{M_U}}\left\{\frac{3}{2} +\Big(\frac{1}{2}-\frac{x}{4}-\frac{1}{4x}\Big)\ln
% \Big|\frac{x+1}{x-1}\Big|~\right.\nonumber\\
% &&\left.\hspace*{3.5cm}+~\ln\Big(\frac{\sqrt{M_U/M}\,}{x+1}\Big)\right\}
%\end{eqnarray}
\begin{eqnarray}\label{c5}
 \Delta E\,\approx\, 2\kappa A\, mc^2\sqrt{\frac{M}{M_U}}\,I(\rho)\,,
\end{eqnarray}
 where the function $I(\rho)$ is given by
\begin{eqnarray}\label{I}
% I(x)\,=\,\left\{\frac{3}{2} +\Big(\frac{1}{2}-\frac{x}{4}-\frac{1}{4x}\Big)\ln
% \Big|\frac{x+1}{x-1}\Big|~\right.\nonumber\\
% &&\left.\hspace*{3.5cm}+~\ln\Big(\frac{\sqrt{M_U/M}\,}{x+1}\Big)\right\}
 I(\rho)=\frac{3}{2} +\Big(\frac{1}{2}-\frac{\rho}{4}-\frac{1}{4\rho}\Big)\ln
 \Big|\frac{\rho+1}{\rho-1}\Big|+\ln\Big(\!\frac{\sqrt{M_U/M}\,}{\rho+1}\!\Big)\,,
 \hspace*{-0.5cm}\nonumber\\
\end{eqnarray}
 and the corresponding additional acceleration is
\begin{eqnarray}\label{a2}
 a\,\approx\,2\kappa A\frac{c^2}{R_U}\,I^\prime(\rho)\,.
\end{eqnarray}
For $\rho\gg 1$, $I(\rho) = 1+\ln\sqrt{\frac{M_U}{\!M}} -\ln \rho + \dots$ and the additional acceleration is therefore $A/\rho$ times $c^2/R_U$, which is indeed proportional to $1/r$. For $\rho\ll 1$, $I(\rho) = 1+\ln\sqrt{\frac{M_U}{\!M}} -\rho^2/6 + \dots$ so, for $A= 0.5$,~(\ref{a2}) has the limits
\begin{eqnarray}\label{lim3}
 a\,\approx\, -\frac{2\kappa c^2}{R_U}\left\{ \begin{array}{ll}
         \rho/6& \mbox{\hspace*{0.2cm}for \hspace*{0.2cm} $\rho \ll 1$}\\
        0.5/\rho & \mbox{\hspace*{0.2cm}for \hspace*{0.2cm} $\rho \gg 1$}\,.\end{array} \right.
\end{eqnarray}
 The accelerations corresponding to the two functions $f_1$ and $f_2$ are shown in Figure~\ref{ff}.

 To plot the curves shown in Figure~\ref{ffu}, the values on the y-axis were multiplied by a factor of $2.5$ to convert  from units of $c^2/R_U$ to units of $a_0$. The values on the x-axis were multiplied by a factor of $3.2$ to convert from $\rho$ to the distance $r$ measured in units of $R_g$ since, from equation~(\ref{rg}), $r/R_g= 3.2\rho$.

%\noindent
% The function $I(x)$ has the expansions
%\begin{eqnarray}\label{I2}
% ~~I(x)\,=\,\left\{ \begin{array}{ll}
%         \!1+\ln\sqrt{\frac{M_U}{\!M}} - x^2/6 + \dots & \hspace*{0.2cm} x \ll 1 \\
%  \!1+\ln\sqrt{\frac{M_U}{\!M}} -\ln x - 1/6x^2 + \dots&  \hspace*{0.2cm} x \gg 1\,,\end{array} \right.
%  \nonumber
%\end{eqnarray}
% so the acceleration~(\ref{a2}) has the limits
%\begin{eqnarray}\label{a3}
% a\,=\,-A\frac{c^2}{R_U}\left\{ \begin{array}{ll}
%%         \frac{x}{3} + \dots & \hspace*{0.3cm} x \ll 1 ~~\\
%%        \frac{1}{x} - \frac{5}{3x^3} + \dots&  \hspace*{0.3cm} x \gg 1\,,~~\end{array} \right.
%        \! x/3 + \dots & \hspace*{0.3cm} x \ll 1 ~~\\
%        \! 1/x - 1/3x^3 + \dots&  \hspace*{0.3cm} x \gg 1\,.~~\end{array} \right.
%\end{eqnarray}
% For definiteness, the limit $A\rightarrow 1$ is assumed. The corresponding acceleration is plotted in %Figure~\ref{adm}.

%% file: appgravrad.tex
\section{Gravitational radiation}\label{appgravrad}
 Consider a binary pulsar consisting of two masses, $m_1$ and $m_2$, in orbit around each other.
% In Newtonian gravity, the separation $r$ is given by
%\begin{eqnarray}\label{rel}
% r\,=\,\frac{a(1-e^2)}{1+e\cos\theta}\,,
%\end{eqnarray}
% where $a$ is the semi-major axis and $e$ is the eccentricity of the orbit. The orbital period, $T$, is given by $2\pi/T= \sqrt{GM/a^3}$, where $M= m_1+m_2$, in accordance with Kepler's third law.
 For the special case when the masses move in a circular orbits, the masses have constant angular velocity $\omega= \sqrt{GM/r^3}$, where $M= m_1+m_2$ and $r$ is the separation of the two masses.

 In the string model, each mass exerts a periodic distortion in the Machian strings of the other as it orbits around it. The distortions have angular frequency $\omega$ and are expected to generate gravitational waves of angular frequency $\omega$ in the Machian strings of each mass. The speed of the waves can be calculated from the Lagrangian~(\ref{ldd}) by considering perturbations in a Machian string of the form $X^\mu=(c\tau,{\bf \xi}\sigma+\delta{\bf x})$. Then $(\partial X^\mu/\partial \tau)(\partial X_\mu/\partial \sigma)= -{\pmb \xi}.\delta\dot{\bf x} - \delta\dot{\bf x}.\delta{\bf x}^\prime$, $(\partial X^\mu/\partial \sigma)^2= -\xi^2-2{\bf \xi}.\delta{\bf x}^\prime- {\delta{\bf x}^\prime}^2$ and $s= 1-\delta\dot{\bf x}^2/c^2$. For transverse perturbations, with ${\bf \xi}.\delta{\bf x}=0$, expansion of~(\ref{ldd}) to quadratic order gives
\begin{eqnarray}\label{lexp}
 {\mathcal L}&=& \Gamma\Big(1+a_1\frac{\delta\dot{\bf x}^2}{c^2}+\dots\Big)\Big(\xi^2 + {\delta{\bf x}^\prime}^2 +\dots \Big)^{-1/2}\nonumber\\
 &=& \frac{\Gamma}{\xi}\Big(1 + a_1\frac{\delta\dot{\bf x}^2}{c^2} - \frac{{\delta{\bf x}^\prime}^2}{2\xi^2}+\dots\Big)\,.
\end{eqnarray}
 The equation of motion for $\delta{\bf x}$ corresponding to~(\ref{lexp}) is $2a_1\delta\ddot{\bf x}/c^2= \delta{\bf x}^{\prime\prime}/\xi^2$, so that transverse gravitational waves propagate along the strings with speed
\begin{eqnarray}\label{cgrav}
 c_g\,=\,\frac{c}{\sqrt{2a_1}}\,.
\end{eqnarray}
 Waves with angular frequency $\omega$ and amplitude $A$ have energy $\Gamma A^2\omega^2/2\xi^2c_g^2$ per unit length along the string, so the energy flux in the string is $\Gamma A^2\omega^2/2\xi^2c_g$. If all the Machian strings of the mass $m_i$ are assumed to have waves of amplitude $A_i$, the total rate of gravitational radiation from $m_i$ is
\begin{eqnarray}\label{cgrav}
 \sum_j \frac{Gm_im_j A_i^2\omega^2}{2r_{ij}^2c_g}\,=\,\frac{3Gm_iM_UA_i^2\omega^2}{2c_gR_U^2}\,.
% \approx\,\frac{\kappa m_ic^2a_i^2\omega^2}{c_gR_U}\,,\nonumber\\
\end{eqnarray}
% where equation~(\ref{ffrac5}) has been used and the $HR_U/c$ terms have been neglected.
% The rate of radiation of energy in the strings may be estimated by multiplying the energy density per unit length in the strings by the speed of waves in the strings. The energy needed to give a string of tension $T$ a transverse displacement $y(x)$ is $T(\partial y/\partial x)^2$ per unit length. If the displacement is a wave with amplitude $a$, frequency $\omega$ and speed $c$, the energy per unit length is $T(a\omega/c)^2$ and the corresponding energy flux is $Ta^2\omega^2/c$. Integrating the magnitude of the string tension over all strings would give $\kappa mc^2/R_U$, so the energy flux is expected to be some fraction of $\kappa mca^2\omega^2/R_U$\,.
% The rate of energy loss in the strings connected to $m_1$ is
%\begin{eqnarray}
% \frac{dE_1}{dt}\,=\, -\frac{\kappa m_1ca_1^2\omega^2}{R_U}\,=\, -\frac{\kappa Gc}{R_U r^3}
%m_1a_1^2(m_1+m_2)\,,
%\end{eqnarray}
% where $a_1$ is the amplitude of the waves in the strings of $m_1$, with a similar formula for the rate of energy loss in the strings connected to $m_2$, so
 With $\omega^2=GM/r^3$, the total rate of energy loss due to gravitational radiation in the strings connected to the two masses is therefore
\begin{eqnarray}\label{trad}
 \frac{dE}{dt}\,=\, -\frac{3G^2M_U(m_1A_1^2 + m_2A_2^2)(m_1+m_2)}{2c_gR_U^2 r^3}\,.
\end{eqnarray}

 The amplitudes $A_1$ and $A_2$ of the gravitational waves in the strings of the two masses could be found by generalising the numerical calculation of the field string paths in Appendix~\ref{appdarkmatter}. The interaction between the Machian strings of $m_1$ and $m_2$ has so far been considered only for the case when $m_1$ is very much smaller than $m_2$, so the interaction must first be defined for the case when the masses are of comparable magnitude. The paths of the strings would also have to be calculated as a function of time and such calculations have not yet been attempted.

 The prediction of General Relativity~\cite{pm}, as confirmed experimentally by observations of the decay of the orbital period of binary pulsars~\cite{taylor,anton}, is
\begin{eqnarray}\label{grad}
 \frac{dE}{dt}\,=\, -\frac{32G^4(m_1m_2)^2(m_1+m_2)}{5r^5c^5}\,.
\end{eqnarray}
 Comparison with equation~(\ref{trad}) shows that the amplitude of the waves in the Machian strings of $m_1$ required for agreement with experiment has the form
% An order of magnitude estimate may be obtained using dimensional analysis.
% Equation~(\ref{u}) in Appendix~\ref{appdarkmatter} shows that the density of strings around $m_1$ has a length scale $R_U\sqrt{m_1/M_U}$, namely the distance around $m_1$ within which the density of Machian strings of $m_1$ is larger than the density of background strings, and it is reasonable to suppose that the waves in the strings
% are generated over a distance of order $L$. The amplitude of the waves is of order $L\theta$, where $\theta$ is the angle of deflection of the strings of $m_1$ due to $m_2$. The value of $\theta$ may be estimated by equating the gravitational force on the strings of $m_1$ exerted by $m_2$ with the restoring force due to the string tension. {\it But actually it can't because there is no direct gravitational action of $m_2$ on the strings of $m_1$. The only interaction is between the strings of $m_1$ and the strings of $m_2$.}
% A length $ds$ of string with tension $T$ has a mass $Tds/c^2$, so the gravitational force due to $m_2$ is $Gm_2Tds/r^2c^2$. The force falls off over a distance $r$ and the total force on a length of string within a distance $r$ from $m_1$ is therefore of order $Gm_2T/rc^2$. If $\theta$ is the total deflection angle of the length $r$ of string then the restoring force is $T\theta$, for $\theta\ll 1$, so $\theta\sim Gm_2/rc^2$.
% The amplitude of the waves in the strings of $m_1$ due to the motion of $m_2$ should therefore be of the form
\begin{eqnarray}\label{a1v}
 A_1 \,=\, \eta R_U\sqrt{\frac{m_1}{M_U}}\Big(\frac{Gm_2}{rc^2}\Big)\,,
\end{eqnarray}
 with a similar expression for $A_2$, where the dimensionless parameter $\eta$ is given by
\begin{eqnarray}\label{eta}
 \eta^2\,=\, \frac{32}{15\sqrt{2a_1}}\,.
\end{eqnarray}
 The value of $\eta$ corresponding to the value of $a_1$ given by~(\ref{psol2}) in Appendix~\ref{appmg4} is
 $\eta= 0.584$. In equation~(\ref{a1v}), the length scale $R_U\sqrt{m_1/M_U}$ is the distance around $m_1$ within which the density of Machian strings of $m_1$ is larger than the density of background strings and the dimensionless factor $Gm_2/rc^2$ is a measure of the strength of the gravitational interaction at $m_1$ due to $m_2$.

% Equation~(\ref{trad}) is in agreement with the formula~(\ref{grad}) if
%\begin{eqnarray}
% m_1a_1^2 + m_2a_2^2 \,=\, \frac{32G^3c_g(m_1m_2)^2R_U}{5\kappa r^2c^7}\,.
%\end{eqnarray}
% If the radiation is shared equally between the strings of the two masses, so that $m_1a_1^2= m_2a_2^2$, then
%\begin{eqnarray}\label{a1r}
% a_1^2 \,=\, \frac{16G^3m_1m_2^2R_U}{5\kappa r^2c^6}\,=\,\frac{16Gm_1R_U}{5\kappa c^2}
%\Big(\frac{Gm_2}{rc^2}\Big)^2\,.
%\end{eqnarray}
% According to equation~(\ref{ffrac5}) Appendix~\ref{appke3}, $GM_U= 2\kappa R_Uc^2/3$ if the effect of the expansion of the Universe is neglected. Equation~(\ref{a1r}) then requires
%\begin{eqnarray}
% a_1 \,=\, \sqrt{\frac{32}{15}}R_U\sqrt{\frac{m_1}{M_U}}\Big(\frac{Gm_2}{rc^2}\Big)\,.
%\end{eqnarray}
% Physically, $Gm_2/rc^2$ gives the strength of gravitational effects at $m_1$ due to $m_2$ and
% Is it reasonable to expect that the amplitude of waves in the strings of $m_1$ is the product of the two\,? One can argue that the amplitude should be proportional to $m_2$, since it is the orbit of $m_2$ around $m_1$ that produces the waves. One can also argue that the relevant lengthscale for the strings of $m_1$ is the distance around $m_1$ for which the strings of $m_1$ are dominant. Then it remains to do numerical simulations to check that the waves do indeed have this amplitude. The required calculations generalise the calculations in Appendix~\ref{appdarkmatter} because the masses are comparable and because they are time-dependent.